%% file: ms.tex
\documentclass[numberedappendix]{emulateapj}
\slugcomment{DRAFT \today}
\shortauthors{Covey et al. 2007}
\shorttitle{SDSS/2MASS Stellar SEDs} 
\usepackage{lscape}

\begin{document}

\title{Stellar SEDs from 0.3--2.5$\mu$\lowercase{m}: Tracing the Stellar Locus and Searching for Color Outliers in SDSS and 2MASS}

\author{Covey, K. R.\altaffilmark{1}, {Ivezi{\' c}}, {\v Z}\altaffilmark{2}, Schlegel, D.\altaffilmark{3}, Finkbeiner, D.\altaffilmark{1}, Padmanabhan, N.\altaffilmark{3}, Lupton, R.H.\altaffilmark{4}, Ag{\" u}eros, M.~A.\altaffilmark{5}, Bochanski, J.~J.\altaffilmark{2}, Hawley, S.~L.\altaffilmark{2}, West, A.~A.\altaffilmark{6}, Seth, A.\altaffilmark{1}, Kimball, A.\altaffilmark{2}, Gogarten, S.M.\altaffilmark{2},  Claire, M.\altaffilmark{2}, Haggard, D.\altaffilmark{2}, Kaib, N.\altaffilmark{2}, Schneider, D. \altaffilmark{6}, Sesar, B.\altaffilmark{2}}

\altaffiltext{1}{Harvard Smithsonian Center for Astrophysics, 60 Garden St., Cambridge, MA 02138; kcovey@cfa.harvard.edu}
\altaffiltext{2}{University of Washington, Department of Astronomy, Box 351580, Seattle, WA 98195.}
\altaffiltext{3}{Lawrence Berkeley National Laboratory, 1 Cyclotron Road, Mail Stop 50R5032, Berkeley, CA 94720}
\altaffiltext{4}{Princeton University Observatory, Princeton, NJ 08544.}
\altaffiltext{5}{NSF Fellow; Columbia University, Department of Astronomy, 550 West 120th St, New York, NY 10027}
\altaffiltext{6}{Univeristy of California Berkeley, Astronomy Department, 601 Campbell Hall, Berkeley, CA 94720-3411}

\begin{abstract}

The Sloan Digital Sky Survey (SDSS) and Two Micron All Sky Survey (2MASS)
are rich resources for studying stellar astrophysics and the structure and formation history of the Galaxy.  As new
surveys and instruments adopt similar filter sets, it is increasingly
important to understand the properties of the $ugrizJHK_s$ stellar locus, both
to inform studies of `normal' main sequence stars as well as for robust
searches for point sources with unusual colors. Using a sample of $\sim$ 
600,000 point sources detected by SDSS and 2MASS, we tabulate the position 
and width of the $ugrizJHK_s$ stellar locus as a function of $g-i$ color, and 
provide accurate polynomial fits. We map the Morgan-Keenan spectral type sequence to the median 
stellar locus by using synthetic photometry of spectral standards and by
analyzing 3000 SDSS stellar spectra with a custom spectral typing pipeline, described in full in an attached Appendix.

Having characterized the properties of `normal' main sequence stars, we develop an algorithm for identifying point sources whose colors differ significantly from those of normal stars.  This algorithm calculates a point source's minimum separation from the stellar locus in a seven-dimensional color space, and robustly identifies objects with unusual colors, as well as spurious SDSS/2MASS matches.
Analysis of a final catalog of 2117 color outliers
identifies 370 white-dwarf/M dwarf (WDMD) pairs, 93 QSOs, and 90 M giant/carbon star 
candidates, and demonstrates that WDMD pairs and QSOs can be distinguished on the basis of their $J-K_s$ and $r-z$ colors.  We also identify a group of objects with correlated offsets in the $u-g$ vs. $g-r$ and $g-r$ vs. $r-i$ color-color spaces, but subsequent follow-up is required to reveal the nature of these objects. Future applications of this 
algorithm to a matched SDSS-UKIDSS catalog may well identify additional classes of objects with unusual colors by probing new areas
of color-magnitude space.  
\end{abstract}


\section{Introduction \label{intro}}

The Two Micron All Sky Survey \citep[2MASS;][]{Skrutskie1997} and Sloan Digital Sky Survey \citep[SDSS;][]{York2000} are fertile grounds for identifying rare stellar objects with unusual colors, such as brown dwarfs \citep{Kirkpatrick1999,Strauss1999,Burgasser1999}, carbon stars \citep{Margon2002}, RR Lyrae stars \citep{Ivezic2005}, and white dwarf/M dwarf pairs \citep{Raymond2003,Smolcic2004,Silvestri2006}.  Relatively little attention has been devoted to understanding and documenting the detailed characteristics of `normal' main-sequence stars detected in these surveys, despite the fact that they represent an important source of contamination in the search for any set of rare objects.  

Recent studies have also demonstrated the power of combining survey data
at differing wavelengths to provide insights into the nature of
various astrophysical sources.  \citet{Finlator2000} showed that population
synthesis models of the Galaxy, based on \citet{Kurucz1979} models, can reproduce
the colors of the eight band SDSS/2MASS stellar locus (except for stars colder 
than $\sim$3000 K).  \citet{Agueros2005} analyzed a catalog of GALEX-SDSS sources, constraining the fraction of GALEX sources with optical counterparts and investigating the ability of GALEX and GALEX-SDSS catalogs to distinguish between various classes of UV-bright objects (white dwarfs, starburst galaxies, AGN, etc.), while \citet{Anderson2007} have combined the ROSAT and SDSS catalogs to construct the largest sample of X-ray luminous QSOs to date, including hundreds of rare objects such as X-ray emitting BL Lacs.  \citet{West2007} utilized a matched SDSS-HI dataset to investigate the relationship between the stellar population, neutral gas content, and star formation history of a homogeneous set of galaxies spanning a wide range of masses and morphologies.  By supplementing the `main' SDSS sample of galaxies with observations from surveys ranging from the X-ray (ROSAT) to the radio (FIRST, NVSS), \citet{Obric2006} demonstrated that the optical/near-infrared (NIR) SEDs of galaxies are remarkably uniform; the K-band flux of a galaxy can be predicted with an accuracy of 0.1 mags based soley on its $u-r$ color, redshift, and estimated dust content.  These studies show that the union of datasets derived from surveys covering differing wavelength ranges allow investigations that would be impossible using either dataset in isolation. 

While the main science drivers of SDSS-I were extragalactic in nature, the
SEGUE component of SDSS-II is extending the survey footprint
through the Galactic plane.  In addition, several other surveys are now or will
shortly be mapping much of the sky in the optical (the Panoramic
Survey Telescope \& Rapid Response System [Pan-STARRS], \citealp{Kaiser2002}, and SkyMapper,
\citealp{Keller2007}) and near infrared (the United Kingdom Infrared Deep Sky 
Survey [UKIDSS] \citealp{Warren2007}) using filter systems that closely resemble those
of SDSS and 2MASS. As a result, these surveys are increasingly useful for
anyone who wishes to identify and study stellar objects.

To aid in the study of point sources identified by these surveys, we present a detailed characterization of the optical/NIR properties of stars detected by both SDSS and 2MASS, as well as an algorithm for identifying point sources with optical/NIR colors that differ significantly from those of typical main sequence stars.  In \S 2 we describe the assembly of our matched SDSS/2MASS stellar catalog.  We provide a detailed characterization of the locus of main sequence stars in SDSS/2MASS color-color space in \S 3.  We give a robust algorithm for identifying color outliers from the stellar locus in \S 4, along with a description of the resultant catalog generated by the application of this algorithm to our dataset.   We summarize our work and highlight our conclusions in \S 5.

\section{Assembling a Catalog of Matched SDSS/2MASS Point Sources \label{matching}}

\subsection{The Sloan Digital Sky Survey \label{sdss}}

The Sloan Digital Sky Survey has imaged nearly a quarter of the sky, centered on the North Galactic Cap, at optical wavelengths \citep{York2000}.  The latest public Data Release (DR5) includes photometry for $2.1 \times 10^8$ unique objects over $\sim 8000$ deg$^2$ of sky \citep{Adelman-McCarthy2007}.  Since the completion of the survey's initial mission, new scientific studies (collectively known as SDSS-II) have begun to add new spatial and temporal coverage to the SDSS database\footnote{see http://www.sdss.org}. The SDSS camera \citep{Gunn1998}, populated by six CCDs in each of five filters \citep[{\it u, g, r, i, z}; ][]{Fukugita1996} observes in time delay and integrate (TDI) mode to generate near-simultaneous photometry along a strip 2.5 degrees in width; a second scan, slightly offset from the first, fills in the areas on the sky that fall in the gaps between CCDs.  The survey's photometric calibration strategy \citep{Hogg2001,Smith2002,Tucker2006} produces a final catalog which is 95\% complete to a depth of $r \sim 22.2$, and accurate to $0.02$ mags (both absolute and RMS error) for sources not limited by Poisson statistics \citep{Ivezic2004}. Sources with $r<20.5$ have astrometric errors less than $0.1\arcsec$ per coordinate \citep[rms ;][]{Pier2003}, and robust star/galaxy separation is achieved above $r\sim 21.5$ \citep{Lupton2001}.

SDSS photometry provides candidates for spectroscopic observation with the SDSS twin fiber-fed spectrographs \citep{Newman2004}, mounted on the same dedicated 2.5 m telescope \citep{Gunn2006} as the survey's imaging camera. The spectrographs give wavelength coverage from 3800 \AA\ to 9200 \AA\ with resolution $\lambda/\Delta\lambda \sim 1800$.  Each fiber has a 3 arcsec diameter and is plugged into a hole in a pre-drilled metal plate allowing observations across a 3 degree field.  A single plate accomodates 640 fibers (320 going to each spectrograph), of which $\approx 50$ are typically reserved for calibration purposes.  The SDSS spectroscopic catalog contains over $10^6$ objects, including $6 \times 10^5$ galaxies, $1.5 \times 10^5$ stars, and nearly $8 \times 10^4$ quasars \citep{Schneider2007}. 

\subsection{The Two Micron All Sky Survey \label{2mass}}

The Two Micron All Sky Survey used two 1.3 m telescopes to survey the entire sky in near--infrared light\footnote{See http://www.ipac.caltech.edu/2mass.} \citep{Skrutskie1997, Cutri2003}. Each telescope's camera was equipped with three $256\times256$ arrays of HgCdTe detectors with $2\arcsec$ pixels and observed simultaneously in the $J$ (1.25 $\mu{\rm m}$), $H$ (1.65 $\mu{\rm m}$), and $K_s$ (2.17 $\mu{\rm m}$) bands. The detectors were sensitive to point sources brighter than about 1 mJy at the $10\sigma$ level, corresponding to limiting (Vega--based) magnitudes of $15.8$, $15.1$, and $14.3$, respectively. Point source photometry is repeatable to better than $10\%$ precision at this level, and the astrometric uncertainty for these sources is less than $0.2\arcsec$. The 2MASS catalogs contain positional and photometric information for $\sim$$5\times10^8$ point sources and $\sim$$2\times10^6$ extended sources. 

\subsection{The Merged Sample \label{merging}}

We have assembled a sample of point sources from the SDSS DR2 catalog \citep{Abazajian2004} using data processed by the photometric pipeline described by \citet{Padmanabhan2007} and \citet{Finkbeiner2004}.  In this reduction pipeline SDSS detections are automatically matched to the nearest 2MASS source within $3\arcsec$ of the astrometric position of the SDSS detection.  To ensure our analysis is not unduly affected by sources with poor photometry, we selected for analysis in this work only those objects which meet the following criteria:

\begin{itemize}
\item{Unblended, unsaturated and accurate 2MASS photometry: $K_s <$ 14.3 (the 2MASS completeness limit), read flag (rd\_flg) $= 2$, blend flag (bl\_flg) $= 1$, and contamination $\&$ confusion flag (cc\_flg) $= 0$ in $J, H$ and $K_s$ \footnote{for a full description of 2MASS flags, see \\ http://www.ipac.caltech.edu/2mass/releases/allsky/doc/explsup.html};}
\item{Unblended, unsaturated, reliable SDSS photometry of stars: $i <$ 21.3 (the SDSS 95\% completeness limit), BLENDED flag $= 0$, DEBLENDED\_AS\_MOVING flag $= 0$, SATURATED flag $= 0$, PRIMARY flag $= 1$, INTERP\_CENTER = 0, EDGE = 0, SATURATED\_CENTER = 0, PSF\_FLUX\_INTERP = 0, object type $=$ STAR \footnote{see \citet{Stoughton2002} for a full description of SDSS flags}.}
\end{itemize}

\begin{figure}
\epsscale{1.2}
\plotone{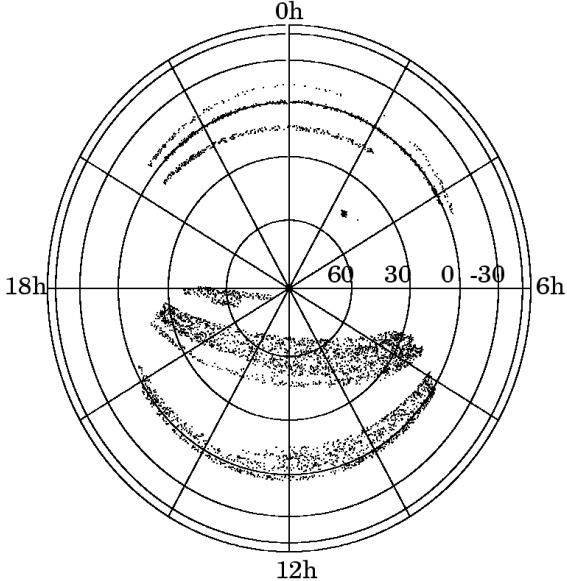}
\caption {\normalsize{The footprint covered by the matched SDSS/2MASS catalog, represented in an equal area polar projection (in RA and Dec coordinates).  For computational efficiency, the footprint is shown using an $\sim 0.5\%$ subset of the full matched SDSS-2MASS dataset.}}
\label{footprint}
\end{figure}

The catalog resulting from these cuts contain 687,150 point sources above the $K_s < 14.3$ effective flux limit.  Each object in the sample has measurements in 8 filters (SDSS $ugriz$, 2MASS $JHK_s$) stretching from 3000\AA~to 2.5$\mu$m; the footprint of the sample shown in Figure \ref{footprint}.  As we seek to study point sources, we adopted PSF magnitudes for SDSS sources, without correcting for Galactic extinction; as the \citet{Schlegel1998} dust map measures the total Galactic extinction along the line of sight, it may over-estimate the true extinction to nearby stars.  In any event, we expect the impact of applying extinction corrections to our sample would be small; as the effect of extinction on infrared colors is small, and the reddening vector is nearly parallel to the stellar locus in SDSS color space \citep{Finkbeiner2004a}, objects with uncorrected extinction will merely be shifted along, and not out of, the standard stellar locus. 

\section{Defining the Standard Stellar Locus \label{locus}}

\subsection{The High Quality Sample \label{highquality}}

To ensure the sample used to determine the basic shape of the typical stellar locus is of high photometric quality, we identify a subset of our catalog that satisfies the following additional quality cuts:

\begin{itemize}
\item{Colors minimally affected by reddening from the interstellar medium: A$_{r} <$ 0.2 as estimated by the dust map of \citet{Schlegel1998};}
\item{Acceptable photometric conditions: SDSS field quality flag (fieldQA) $>$ 0;} 
\item{Isolated sources: SDSS CHILD flag = 0 (When a single photometric detection appears to be made up of multiple overlapping objects, the SDSS photometric pipeline uses a process known as `deblending\footnote{see \url{http://www.sdss.org/dr5/algorithms/deblend.html} for a full discussion of the SDSS deblending algorithm.}' to separate the individual components self-consistently in all filters, and sets the CHILD flag for all components.);}
\item{Minimal chance of spurious catalog matches: SDSS and 2MASS positional separation $<$ 0.6 arcseconds.}
\end{itemize}

These cuts produce a subsample of 311,652 objects that we will refer to as the `high quality' sample.  

\begin{figure*}
\epsscale{1.15}
\plotone{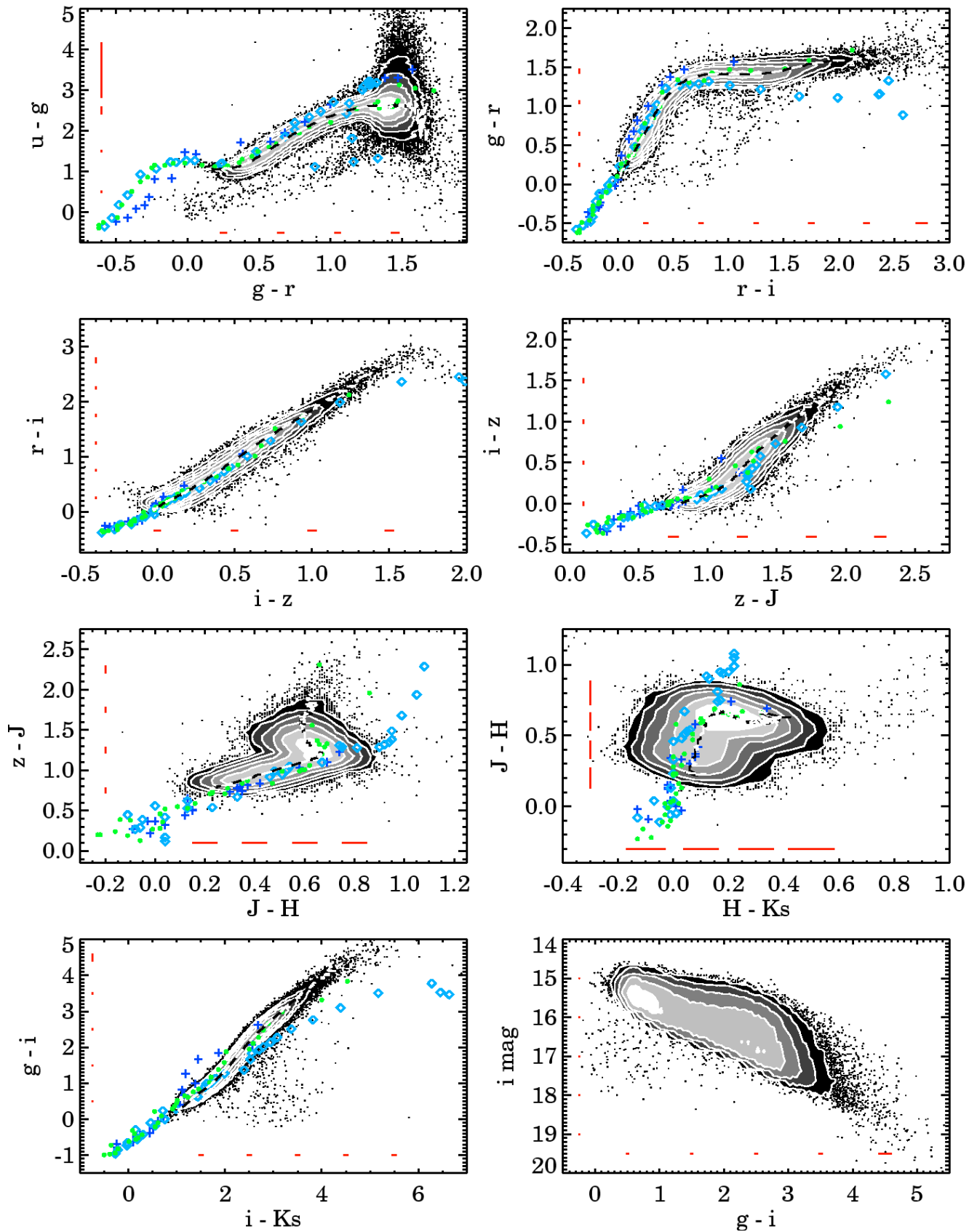}
\caption {\footnotesize{Color-color and color-magnitude diagrams showing the density of objects in the `high quality' SDSS/2MASS sample described in \ref{highquality}.  Black points show individual sources; contours show source density in saturated regions. Contours begin at source densities of 10K and 2K sources per square magnitude in color-color and color-magnitude space respectively; steps between contours indicate an increase in source density by a factor of three.  The black and white dashed line shows the location of the median stellar locus, calculated in \S \ref{medianlocus} and tabulated in Table \ref{tab:locusfit}.  Error bars near the axes of each color-color diagram show the median photometric error as a function of color.  Symbols show the synthetic SDSS/2MASS colors of solar metallicity \citet{Pickles1998} spectral standards, as calculated in \S \ref{typedlocus} and given in Table \ref{tab:solarstars}; dwarf stars are shown as green dots, giants and supergiants are shown as light blue diamonds and dark blue crosses, respectively.}}
\label{colorcolor}
\end{figure*}

Figure \ref{colorcolor} displays the locations of stars in the high quality
sample within color-color and color-magnitude space. We use `native'
magnitudes to construct these diagrams; note that SDSS uses an AB-based magnitude system, while 
2MASS uses a Vega-based magnitude system (\citealp{Finlator2000}, however, provide the $JHK_s$ offsets required 
to place 2MASS magnitudes on an AB system).  The first six panels display color-color diagrams constructed using the seven `adjacent colors' (a phrase which we will use as shorthand for the complete set $u-g$, $g-r$, $r-i$, $i-z$, $z-J$, $J-H$ and $H-K_s$).  The stellar locus is clearly visible in these diagrams, typically extended from the bluest stars (in the lower left corner in all diagrams) to the reddest stars (in the upper right of all diagrams)\footnote{We have chosen to display these color-color diagrams with the bluer color on the y axis and the redder color on the x axis, as is customary for stellar astronomers; this differs from the standard SDSS convention, in which color-color diagrams typically display the bluer color on the x axis and the redder color on the y axis.}.  

The seventh panel of Figure \ref{colorcolor} displays a $g-i$ vs. $i-K_s$ color-color diagram, which samples as broad a range in wavelength as is possible without including $u$ band measurements, which are less reliable for red stars due to their low intrinsic flux and a `red leak' in the SDSS camera.  The red leak stems from flux at 7100 \AA\ passing through the $u$ band filter due to a change in the filter's interference coating under vacuum.  This causes a color-dependent offset to $u$ band photometry of $\sim$0.02 mags for K stars, $\sim$0.06 mags for M0 stars, and 0.3 mags for stars with $r-i \sim$ 1.5.  The effect depends on a star's \textit{instrumental} $u$ and $r$ magnitudes, which are sensitive to airmass, seeing, and the detailed interaction between the $u$ filter in each camera column and the sharp molecular features in the spectra of red stars.  The standard SDSS reduction does not attempt to correct for this effect, resulting in a large $u-g$ dispersion at the red end of the stellar locus (as seen in the $u-g$ vs. $g-r$ color-color diagram in Figure 2), even for a high quality sample such as this one. 

In the eighth panel of Figure \ref{colorcolor} we show the distribution of our sample in $g-i$ vs. $i$ color-magnitude space.  The faint magnitude limit of the sample is clearly apparent, ranging from $i \sim$ 16.3 at $g-i = 1$ to $i \sim$ 17.7 at $g-i = 3$.  This faint limit is a result of selecting only those SDSS sources with well measured 2MASS counterparts, thus requiring that $K_{s} <$ 14.3.  Redder stars are fainter in $i$ for a given $K_s$ magnitude, and thus a constant $K_s$ faint limit results in an effective color-dependent $i$ band faint limit, such that stars with $g-i \sim$ 1 must have $i <$ 16.3 to be detected in 2MASS, while stars with $g-i \sim$ 3 can be more than a magnitude fainter ($i <$ 17.7).

The SDSS catalog contains S/N $=$ 10 detections down to $i=$21.5, and Figure
\ref{colorcolor} demonstrates that our sample is nearly entirely composed of
sources brighter than $i \sim 19$; requiring a 2MASS counterpart has biased
our sample towards the brightest of the stars detected by SDSS. Our
primary goals, however, are to explore the \textit{areas} of color-space populated
by astronomical point sources, not to compare the \textit{relative
  numbers} of objects in those areas; this bias does not seriously compromise
the effectiveness of our study.  It does affect 
our sensitivity to the least luminous, reddest members of the stellar
locus (very low-mass stars and brown dwarfs), which we can observe only in a very small volume centered on the Sun.  To the extent that these objects are extremely hard to
detect within the completeness limits of the two surveys, our high quality
sample does not allow characterization of the behavior of the stellar locus
beyond the location of mid-M type objects. The K band flux limit also preferentially selects
nearby disk stars, resulting in a sample dominated by relatively metal-rich 
stars.

\subsection{The Median Stellar Locus \label{medianlocus}}

In order to identify sources in unusual areas of color-color space, we must first characterize the properties of the stellar locus, where the vast majority of stars are found.  To first order, the observational properties of main sequence stars can be characterized by their photospheric temperature, or T$_{eff}$.  Additionally, most stellar colors become monotonically redder as T$_{eff}$ decreases\footnote{this is not uniformly true, however; $J-H$ becomes bluer with decreasing T$_{eff}$ for M stars, for example.}, allowing a single color to serve as a useful parameterization for the majority of the varience in stellar colors.  

We choose to parameterize the stellar locus as a function of $g-i$ color because it will be applicable to the entire SDSS catalog, not merely the subset with 2MASS counterparts, and because it samples the largest wavelength range possible without relying on shallower $u$ or $z$ measurements.  As can be seen in the lower right hand panel of Figure \ref{colorcolor}, our sample spans more than four magnitudes of $g-i$ color with typical $g-i$ errors $\sim$ 0.03 magnitudes.  The range and precision of the $g-i$ color allow the separation of stars into more than 130 independent $g-i$ bins, in principle a more precise classification than allowed by MK spectral types.  The $g-i$ color most efficiently identifies objects whose effective temperatures (T$_{eff}$) place the peak of their blackbody emission within the wavelength range spanned by the $g$ and $i$ filters, 4000 \AA~to 8200 \AA.  As a result, the $g-i$ color will be most sensitive to selecting objects with 3540 K $<$ T$_{\rm eff} <$ 7200 K, corresponding roughly to spectral types from F to mid-M.  Other colors (such as $i-z$) allow for precise separation of the stellar locus as a function of T$_{eff}$ outside this temperature range.  

\begin{figure}
\epsscale{1.2}
\plotone{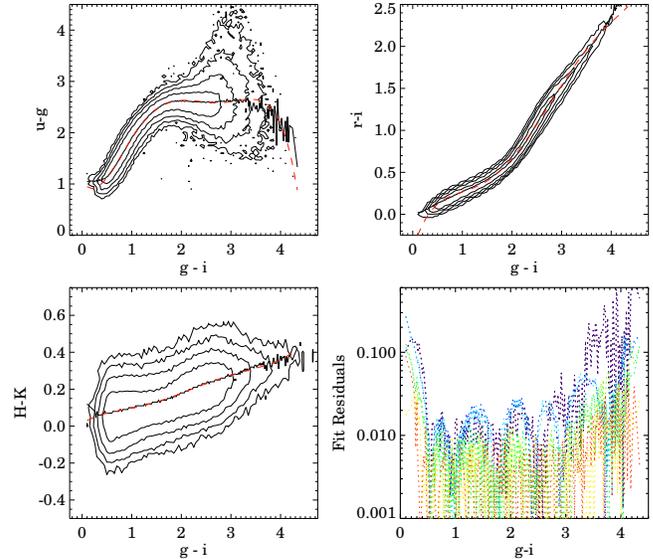}
\caption {\normalsize{First three panels: $u-g$, $r-i$, and $H-K_s$ colors as a function of $g-i$.  Stellar density in each color-color diagram is shown with contours; contour steps indicate a factor of five increase in density.  The median value of each adjacent color as a function of $g-i$ is shown with a solid line; red dashed lines show analytic fits to the median values using the coefficients given in Table \ref{tab:analyticfits}.  Fourth panel: residuals between the median value of each adjacent color and the corresponding analytic fit.  The color of each line represents the mean wavelength of the adjacent color whose residuals are shown, such that $u-g$ residuals are shown in blue and $H-K_s$ residuals are shown in red.}}
\label{colorgi}
\end{figure}

The first three panels of Figure \ref{colorgi} show the $u-g$, $r-i$ and $H-K_s$ colors of the stellar locus as a function of $g-i$.  To parameterize the stellar locus, we have measured the median value of each adjacent color in $g-i$ bins ranging from 0.10 $< g-i <$ 4.34.  The standard bin width (and spacing) is 0.02 magnitudes, with 2 expanded bins of 0.1 magnitudes at each end of the $g-i$ range (necessary to extend the median locus into areas of low stellar density).  We present the median value and dispersion of each adjacent color as a function of $g-i$ in Table \ref{tab:locusfit}.  We use the interquartile width to characterize the width of the locus, as it is a more robust measure of a distribution's width than the standard deviation in the presence of large, non-gaussian outliers.  Following convention, however, we present in Table \ref{tab:locusfit} `psuedo-standard deviations', calculated as 74\% of the interquartile width, using the relation between the interquartile width and standard deviation for a well behaved gaussian distribution.  We do not explicitly include the uncertainty in the median value of each adjacent color, but it can easily be calculated by dividing the pseudo-standard deviation of each bin by the square root of the number of objects in the bin (also given in Table \ref{tab:locusfit}).  The median values of the $u-g$, $r-i$, and $H-K_s$ colors are shown in Figure \ref{colorgi} as a function of $g-i$, with the location of the parameterized locus in color-color space shown in Figure \ref{colorcolor}.

We have fit the adjacent colors of the median stellar locus (covering $0.05 < g-i < 4.4$) using a fifth order polynomial in $g-i$, with the form:
\begin{equation}
\label{colorfit}
color X = \sum_{k=0}^5 A_k (g-i)^k
\end{equation}
The resulting fits are compared to the median $u-g$, $r-i$, and $H-K_s$ colors in the first three panels of Figure \ref{colorgi}, and the coefficients used to produce each fit are presented in Table \ref{tab:analyticfits}.  The last two columns of Table \ref{tab:analyticfits} provides a measure of the accuracy of each fit by presenting the maximum and median offsets between the median stellar locus and each analytic description; residuals between the median adjacent colors and the corresponding analytic fit are shown as a function of $g-i$ in the fourth panel of Figure \ref{colorgi}. These polynomial relations match the actual median behavior well (typical residuals $\leq 0.02$ magnitudes).  As we weighted points along the median locus proportionally to the number of stars in each $g-i$ bin, the fits perform better in densely populated color regions ($0.5 < g-i < 3.0$), and more poorly at the sparsely populated blue and red ends.  

\input{tab2}

We note that while we have restricted our catalog to objects 
identified as point sources by the SDSS reduction pipeline, contamination 
by extragalactic sources or unresolved multiple stars cannot be prevented 
completely, and could bias this measurement of the median stellar locus. 
\citet{Ivezic2002} identify distinct color differences in the 
optical/NIR colors of stars, galaxies, and QSOs (see their Figure 3); 
only 229 of the 311,652 high quality sources (or 0.07\%) analyzed here
lie in the region of color-color space ($i-K_s > 1.5 + 0.75 \times g-i$) occupied 
by extragalactic sources, regardless of their morphological classification.  
This is partially explained by the de facto magnitude limit imposed on our 
sample by requiring 2MASS detections; 98.7\% of the high quality sample is
brighter than $r = 19$, the magnitude at which galaxies begin to outnumber
stars.
 
Unresolved binaries surely represent a larger source of contamination for 
a sample such as this, with estimates of the stellar multiplicity fraction
ranging from 57\% for G type stars \citep{Duquennoy1991a} to 30\% for M 
type stars \citep{Delfosse2004}.  Indeed, \citet{Pourbaix2005} detect
radial velocity variations indicative of binarity in 6\% of SDSS stars with 
repeated spectroscopic observations.  Corrected for the sparse sampling
of the study (most objects were only observed twice), \citet{Pourbaix2005} 
derive a total spectroscopic binary fraction of 18$\%$, consistent with the
findings of a similar study of halo stars by \citet{Carney2003}.  This slightly
overestimates the impact binaries could have in producing systems with odd colors,
however, as the colors of binaries with large luminosity ratios are dominated by 
the primary, while binaries with more closesly matched luminosities will tend
to have similar colors.  Binary systems with a non-main sequence component, 
such as the white-dwarf/M-dwarf pairs identified by \citet{Smolcic2004} and 
\citet{Silvestri2006}, are a notable exception to this generalization, and are
visible in the $u-g$ vs. $g-r$ and $g-r$ vs. $r-i$ color-color diagrams in 
Figure \ref{colorcolor}.  Fortunately, such systems make up less than 0.1\% of 
all stars detected in SDSS photometry \citep{Smolcic2004}, so do not represent a serious source of
contamination and bias for the median stellar locus measured here.

\subsection{Colors as a Function of Spectral Type \label{typedlocus}}

To provide guidance in interpreting the properties of stars within the SDSS/2MASS stellar locus, we have estimated the SDSS/2MASS colors of stars as a function of MK spectral type.  We have used two independent methods: calculating synthetic SDSS/2MASS colors using flux calibrated spectral standards, and assigning spectral types to stars with SDSS spectra and native SDSS photometry.  We describe each of these methods in turn below.

\subsubsection{Synthetic SDSS/2MASS Photometry of Spectral Standards \label{picklecolors}}


\citet{Pickles1998} assembled a grid of flux calibrated spectral type standards with wavelength coverage extending from 1150 \AA\ to 2.5 $\mu$m for solar metallicity stars and 1150 \AA\ to 10,620 \AA\ for a selection of non-solar metallicity stars.  We have calculated `observed' fluxes in Janskys from each spectrum by convolving and integrating the flux transmitted through the SDSS and 2MASS filters\footnote{SDSS and 2MASS filter curves, including the effects of atmospheric transmission, are available at \url{http://www.sdss.org/dr5/instruments/imager/filters/} and \url{http://spider.ipac.caltech.edu/staff/waw/2mass/opt\_cal/index.html}}, and integrating the transmitted flux in Janskys.  

\begin{figure}
\epsscale{1.15}
\plotone{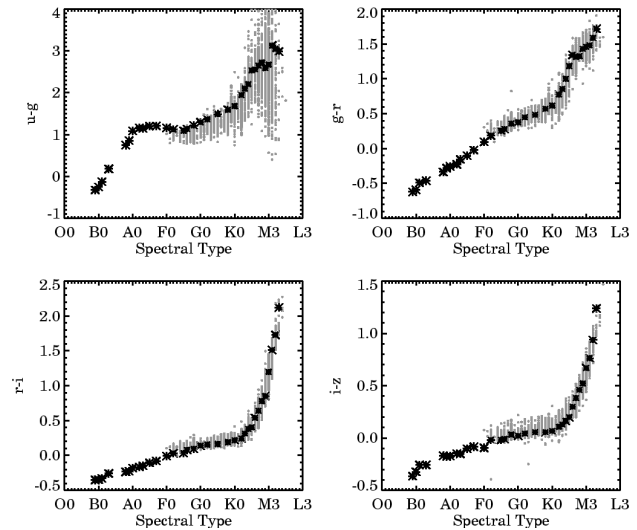}
\caption {\normalsize{The relationship between stellar MK spectral type and color in the SDSS photometric system.  Black asterisks show synthetic SDSS colors calculated for the \citet{Pickles1998} spectral atlas, with grey points indicating the colors of 3443 stars with spectral types assigned to SDSS spectra using the Hammer spectral typing software.  }}
\label{colorspectypefig}
\end{figure}
\begin{figure}
\epsscale{1.15}
\plotone{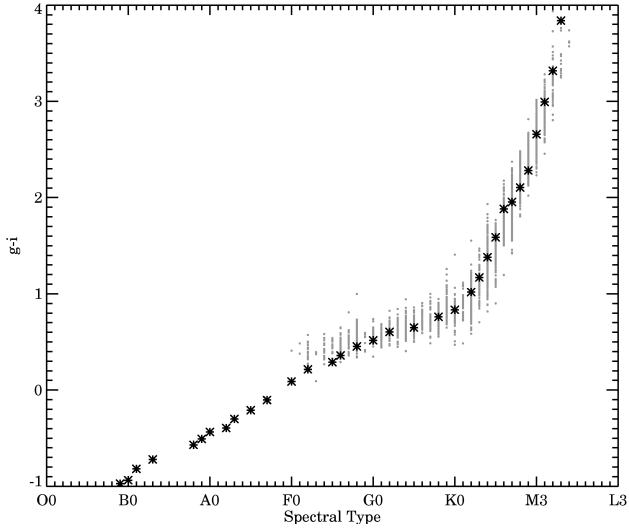}
\caption {\normalsize{The relationship between stellar MK spectral type and $g-i$ color in the SDSS photometric system.  Symbols as in Figure \ref{colorspectypefig}.}}
\label{gispectypefig}
\end{figure}

SDSS magnitudes were calculated from optical fluxes using the survey zero-point flux density (3631 Jy) and the asinh softening parameter (b coefficients) for each filter, and applying offsets of 0.036, -0.012, -0.01, -0.028, and -0.04 to our $ugriz$ magnitudes, respectively, to account for the difference between the SDSS system and a perfect AB system\footnote{see \url{http://www.sdss.org/dr5/algorithms/fluxcal.html} for a detailed description of the flux calibration of SDSS photometry, and \url{http://cosmo.nyu.edu/blanton/kcorrect/} for a description of the SDSS to AB conversion.  Note, however, that the AB offsets assumed here differ slightly from those adopted by \citet{Eisenstein2006}}.  2MASS magnitudes were calculated from near--infrared fluxes using Vega-based zeropoints of 1594, 1024, and 666.7 Jy in the $J$, $H$, and $K_s$ filters respectively \citep{Cohen2003}.  Synthetic SDSS/2MASS colors for the solar-metallicity Pickles standards are presented in Tables \ref{tab:gispectype} and \ref{tab:solarstars}, and shown in Figures \ref{colorspectypefig} and \ref{gispectypefig}; SDSS colors for the non-solar-metallicity Pickles standards are given in Table \ref{tab:nonsolarstars}, as the reduced spectral coverage prevents the calculation of synthetic 2MASS magnitudes for these stars.  With uncertainties of a few percent in the absolute flux calibration of the spectra and the survey zeropoints, the synthetic colors presented in Tables \ref{tab:gispectype} and \ref{tab:solarstars} have characteristic uncertainties of 0.05 magnitudes; as the Pickles spectra have only smoothed flux distributions in the near-infrared, the near-infrared colors of these stars may be somewhat less accurate than the optical colors.

\input{tab4}

\input{tab5}

For completeness, we include in Table \ref{tab:solarstars} absolute magnitudes calculated by \citet{Pickles1998} for each spectrum, transformed onto the 2MASS system via the relations derived by \citet{Carpenter2001} and reported as M$_J$ instead of M$_K$.  We believe these approximate absolute magnitudes may be useful for back of the envelope calculations, particularly for giants and early type (OBA) dwarfs.  More reliable measurements of absolute magnitude as a function of SDSS/2MASS color exist for FGKM dwarf stars; for those stars, \citet{Williams2002}, \citet{Hawley2002}, \citet{West2005}, \citet{Bilir2005a}, \citet{Davenport2006}, \citet{Bochanski2007} and \citet{Golimowski2007} present science-grade empirical measurements, while \citet{Girardi2004} and \citet{Dotter2006} provide stellar models with SDSS and 2MASS photometry.  

\subsubsection{Spectral Typing SDSS Stars \label{HAMMERcolors}}

To provide a check on the accuracy of the synthetic colors described above, we have assigned spectral types to stars with SDSS spectroscopy and native SDSS photometry.  These types were assigned using a custom IDL package, dubbed `the Hammer', whose algorithm is described in full in Appendix A.  In short, the Hammer automatically assigns spectral types to input spectra by measuring a suite of spectral indices and performing a least squares minimization of the residuals between the indices of the target and those measured from spectral type standards.   This code has been made available for community use, and can be downloaded from \anchor{http://www.cfa.harvard.edu/~kcovey/thehammer/}{http://www.cfa.harvard.edu/$\sim$kcovey/thehammer}. 

To measure the median SDSS colors of the MK subclasses, we used the Hammer to assign MK spectral types to 3443 SDSS stellar spectra.  Input SDSS spectra were selected from a subset of SDSS plates that sparsely sample the SDSS color space inhabited by point sources\footnote{This sample is known as the Spectra of Everything sample, contained in part on plates 1062-75, 1077-88, 1090-6, 1101, 1103-7, 1116-7, 1473-6, 1487-8, 1492-7, 1504-5, 1508-9, 1511, 1514-8,1521-3, and 1529, and was originally targeted to test the completeness of the quasar spectroscopic targeting algorithm \citep{Richards2002}.}.  Mean fluxes and S/N ratios were then calculated in three 100 \AA\ bands centered at 4550, 6150, and 8300 \AA\ ; if the wavelength band with the largest mean flux had S/N $<$ 10 per wavelength element, the spectrum was rejected as too noisy for further analysis.  These spectra were classified automatically, and all types were visually confirmed by KRC.  The colors of these stars are shown in Figures \ref{colorspectypefig} and \ref{gispectypefig} as a function of their assigned spectral type, with the associated $g-i$ vs. spectral type relationship presented in Table \ref{tab:gispectype}. Note that this sample contained no stars bluer than $g-i =$ 0.0; equivalently, no stars were assigned spectral types earlier than F0.

\subsubsection{Comparing the color-spectral type relations \label{comparecolorspt}}

The color-spectral type relations obtained from native and synthetic SDSS
photometry agree well, particularly for the $g-r$ and $r-i$ colors that
underly the $g-i$ color used in \S \ref{medianlocus} to parameterize the median
stellar locus.  Residuals between the SDSS colors of
\citet{Pickles1998} standards and SDSS stars assigned the same spectral type
are typically 0.05 magnitudes in $g-r$ and $r-i$ (producing $g-i$ residuals
of $\sim$ 0.1 magnitude), and slightly larger in $i-z$ ( $\sim$ 0.08
magnitudes).  The agreement in $u-g$ is worse, with native SDSS
stars consistently bluer than the \citet{Pickles1998} stars by 0.2 mags
in color. This is easily understood as a metallicity effect; as [Fe/H] $\sim$ -0.75 for typical SDSS stars, they suffer less line blanketing than the solar metallicity Pickles standards, with accordingly bluer $u-g$ colors (Ivezi\'{c} et 
al. 2007, in prep.). 

As the relationship between $g-i$ color and spectral type appears secure to within 0.1 mags in color, Table \ref{tab:gispectype} allows preliminary spectral types to be assigned to SDSS stars based on $g-i$ color alone.  With $g-i$ color increasing by $\sim$0.05 mags between spectral type subclasses, classification accuracies of $\pm 2-3$ subclasses should be achievable for stars with small photometric errors.  

\section{Cataloging Color Outliers}

\subsection{Determining Color Distances}

Having characterized the adjacent colors of the stellar locus from 0.3 to 2.2 $\mu$m, we can now search for color outliers -- objects whose colors place them in areas of color-space well separated from the stellar locus.  In calculating the statistical significance of an object's separation from the standard stellar locus, we account for photometric errors as well as the intrinsic width of the stellar locus.  The catalog enables efficient and robust searches for color outliers by avoiding objects that live in sparsely populated areas of color-color space merely due to large photometric errors, and more interestingly, by aggregating the effects of small offsets in multiple colors.  This alogorithm resembles that used to select quasar candidates from
SDSS imaging for spectroscopic observation, extended to account for an object's
NIR colors \citep[see Appendix A and B of][]{Richards2002}.

We calculate a quantity we dub the `Seven Dimensional Color Distance' (7DCD), which captures the statistical significance of the distance in color-space beween an object (the `target') and a given `locus point' (characterized by colors from a single row in Table \ref{tab:locusfit}) as  
\begin{eqnarray}
\label{colordist}
\textrm{7DCD} = \sum_{k=0}^6 \frac{ (X_k^{targ} - X_k^{locus})^2}{\sigma_{X}^2 (locus) + \sigma_{x}^2}  \qquad \\ \textrm{where} \qquad X_0 = u-g, X_1 = g-r, etc. \nonumber
\end{eqnarray}
In the above equation, the target's $X_k$ color error ($\sigma_{x}$) 
is defined as the quadrature sum of the photometric errors reported 
in either the SDSS or 2MASS catalog in both appropriate filters.  Normalizing
the 7DCD by the target's photometric errors, as well as the observed
width of the stellar locus, which itself includes the effects of photometric 
errors, would `double-count' the ability of photometric error to explain the
separation of a source from the stellar locus; this effect was first detected 
by comparing preliminary 7DCD distributions to synthetic chi-square-based 
distributions, which predicted 7DCD distributions larger than those actually 
observed.  This was an especially troubling fact given that SDSS error estimates 
are, if anything, underestimating errors by $\sim$ 10-20\% \citep{Scranton2005},
which should lead to overestimates of the 7DCD of a given source. 

We account for this effect by including
only the intrinsic width of the stellar locus in the normalization of the color
distance; $\sigma_{X}$ gives the intrinsic width of the stellar
locus in the $X_k$ color at the locus point, calculated by subtracting (in
quadrature) the median color error from the standard deviation of the colors of
 objects in this $g-i$ bin in the high quality sample.  The intrinsic width 
of the stellar locus therefore accounts for the range of colors induced by 
variations in stellar properties (e.g., metallicity; Ivezic et al. 2007) as 
well as instrumental spread not captured in the pipeline error estimates, such 
as that due to the red leak.  It excludes, however, the component of the locus' 
width which is due purely to photometric errors that are well estimated by the
SDSS photometric pipeline, and eliminates the double-counting of the importance of photometric errors in the calculation of the 7DCD.  

Using this algorithm, and adopting minimum errors of 0.03 mags in the SDSS $u-g$, $g-r$, $r-i$, and $i-z$ colors, we have identified the minimum 7DCD and best fit $g-i$ point along the median stellar locus for every object in our matched sample.   
Strictly speaking, however, the minimum 7DCD is not the minimum 
perpendicular distance between a given star and the stellar locus, which requires 
a continuous description of the stellar locus.  As the polynomial fits given 
in \S \ref{medianlocus} possess non-trivial
residuals at extreme colors, we have chosen instead to calculate color distances 
using the discrete tabulation of the stellar locus given in Table \ref{tab:locusfit}.
As Table \ref{tab:locusfit} is finely spaced in $g-i$ color,
the difference between the calculated color distance and the minimum 
perpendicular color distance will be small, particularly for objects with large
color distances.  This procedure also neglects the effects of photometric 
covariance, which \citet{Scranton2005} show can cause SDSS color errors to be 
underestimated by $\sim$20\% for variable stars.  \citet{Scranton2005} 
demonstrate, however, that the effect is negligible for non-variable stars; 
since the majority of our sample is composed of non-variable stars, we expect 
the effects of neglecting covariance to be small.

\begin{figure}
\epsscale{1.2}
\plotone{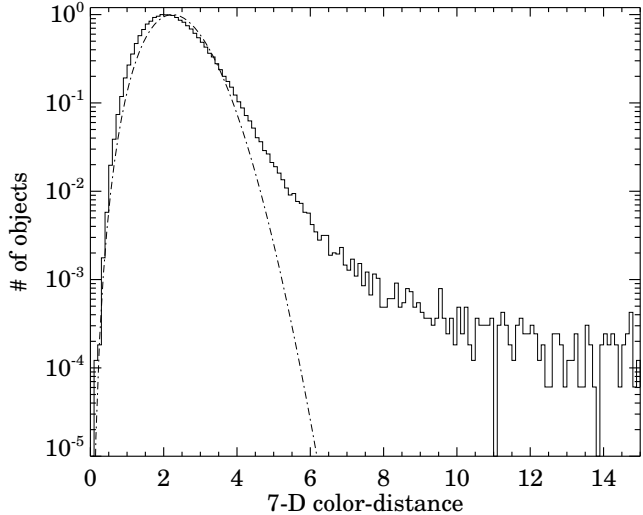}
\caption {\normalsize{7DCDs for objects in the high quality sample (solid line), compared to the chi-square distribution expected for a sample with 6 degrees of freedom (dash-dot).  The sample contains a clear excess of sources with color distances $>$ 6.}}
\label{chisquare}
\end{figure}

A histogram of the 7DCDs in our high quality sample is shown in Figure \ref{chisquare}, with the expected $\chi^2$ distribution shown for comparison.  The $\chi^2$ distribution shown is calculated for a sample with six degrees of freedom, with one degree of freedom for each adjacent color and one degree of freedom removed to account for fitting the object to the closest $g-i$ bin. The observed distribution has a shallower slope than the expected distribution between color distances of three and six -- we attribute this to the intrinsic width of the stellar locus.  Though Equation \ref{colordist} explicitly includes a term to account for the width of the stellar locus, it still implicitly assumes that the stellar locus can be modelled as a centrally concentrated gaussian -- if the stellar locus is significantly non-gaussian, we would expect to see an excess of sources at large color distances.

The slope of 7DCDs changes at color distances
larger than six, becoming noticably shallower.  This change in slope indicates that there are either true
outliers whose distribution in color space differs from objects
within the stellar locus (either for astrophysical reasons, or because of
unrecognized photometric errors), or that our technique for assigning 7DCDs to individual objects is erroneously assigning high color distances to some subset of our catalog.  In the following section we perform a variety of tests to understand the robustness, accuracy, and limitations of our derived 7DCDs.

\subsection{Understanding the Utility and Limitations of Color Distances}

In order to verify that the 7DCD provides an accurate means of identifying point source color outliers, we have conducted a number of tests on the objects within our catalog. A set of tests to probe the sensitivity of the 7DCD to various source characteristics are shown in Figures \ref{gifitqualitytests}-\ref{sdsscdvstotalcd}.  These investigations reveal the following properties of the 7DCDs calculated for the matched SDSS/2MASS sample:

\begin{figure}
\epsscale{1.2}
\plotone{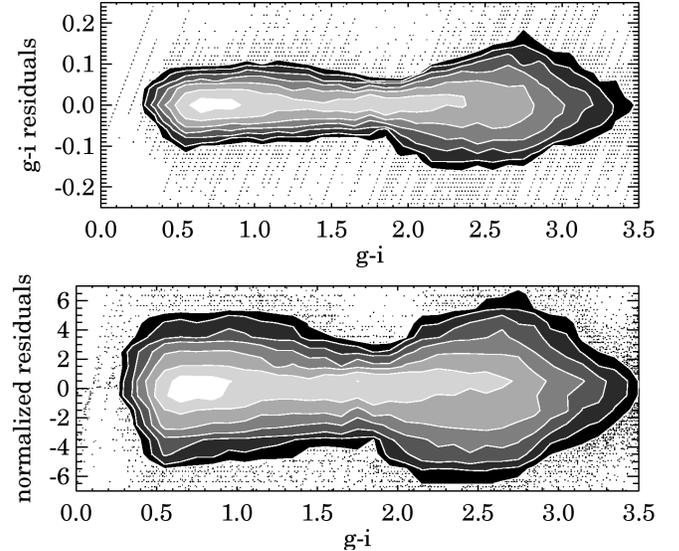}
\caption {\normalsize{Residuals between the $g-i$ colors of the matched sample and the $g-i$ colors of the locus point that minimizes each source's 7DCD, shown as a function of $g-i$ color and before (top) and after (bottom) normalizing by each source's $g-i$ error.  Successive contours indicate an increase in source density by a factor of two.   More than 95\% of the sample have $g-i$ residuals less than 0.1 magnitudes, indicating that the position of normal stars within the SDSS/2MASS stellar locus can be accurately determined by minimizing their 7DCD.}}
\label{gifitqualitytests}
\end{figure}

\begin{itemize}

\item{Minimizing the 7DCD correctly identifies the location of normal stars along the median SDSS/2MASS stellar locus.  As Figure \ref{gifitqualitytests} demonstrates, the $g-i$ color of the best fit locus point matches a star's observed $g-i$ color within 0.1 magnitudes for more than 95\% of the sample.}

\item{Stars with unusual colors are correctly assigned large 7DCDs; Figure \ref{blueredqualitytests} shows that stars in two example regions of unusual color space ($u-g <$ 0.8, $J-K_s >$ 1.2) have median 7DCDs $> 6$.}

\begin{figure}
\epsscale{1.2}
\plotone{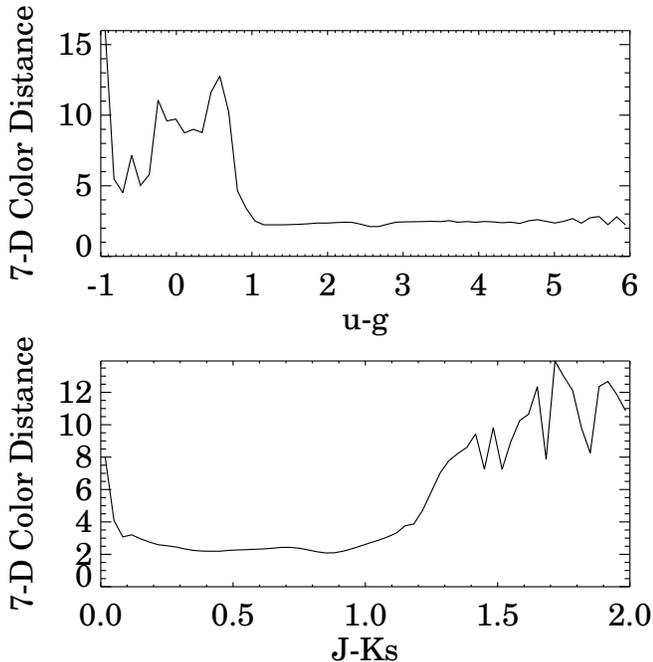}
\caption {\normalsize{Median color distances for sources in bins of $u-g$ and $J-K_s$ color (top and bottom panels, respectively).  Color distances increase sharply for $u-g < 1.0$ and $J-K_s > 1.2$, which are examples of areas of color-space outside the SDSS/2MASS stellar locus.}}
\label{blueredqualitytests}
\end{figure}

\item{The 7DCD is largely insensitive to the $i$ magnitude and $g-i$ color of a source.  Figure \ref{colormagqualitytests} shows that brighter, bluer objects slightly dominate the number of objects in a given color distance bin.  This is consistent, however, with the larger overall fraction of bright, blue objects within our sample (see last panel of Figure \ref{colorcolor}, where the peak stellar density occurs at 15 $< i <$ 16 and $g-i <$ 1).  Aside from this effect, contours are essentially horizontal, implying that to first order color distances derived here are independent of magnitude and $g-i$ color.}

\begin{figure}
\epsscale{1.2}
\plotone{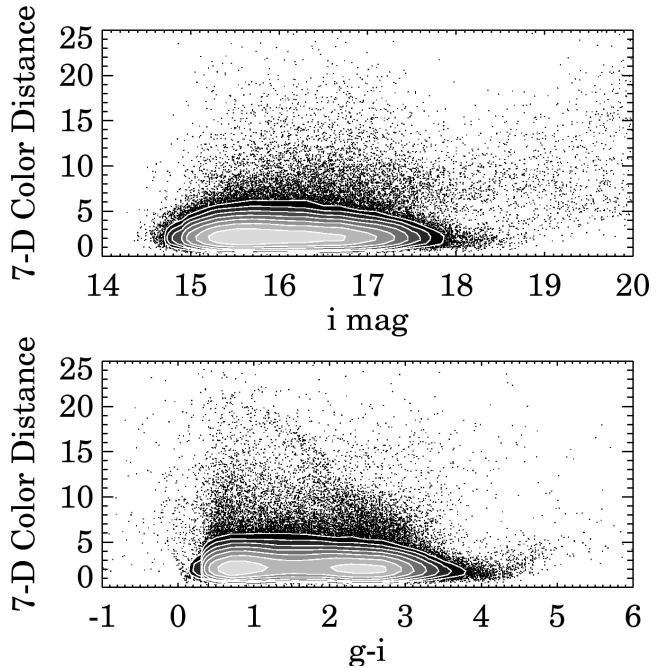}
\caption {\normalsize{7DCDs calculated for the matched SDSS/2MASS sample, displayed as a function of $i$ magnitude (top panel) and $g-i$ color (bottom panel).  Individual stars are shown as black points, and contours indicate source density in areas too crowded for individual points to be distinguished from one another.  Steps between contours indicate a doubling in source density.}}
\label{colormagqualitytests}
\end{figure}

\item{Mismatches between the SDSS-2MASS catalogs generate a population of objects with spuriously high 7DCDs.  This effect is visible in Figure \ref{badmatchqualitytests}, where derived color distances increase with astrometric matching distance (the distance between the positions of an object's SDSS and 2MASS counterparts) past $\sim$ 0.6$\arcsec$.  

\begin{figure}
\epsscale{1.2}
\plotone{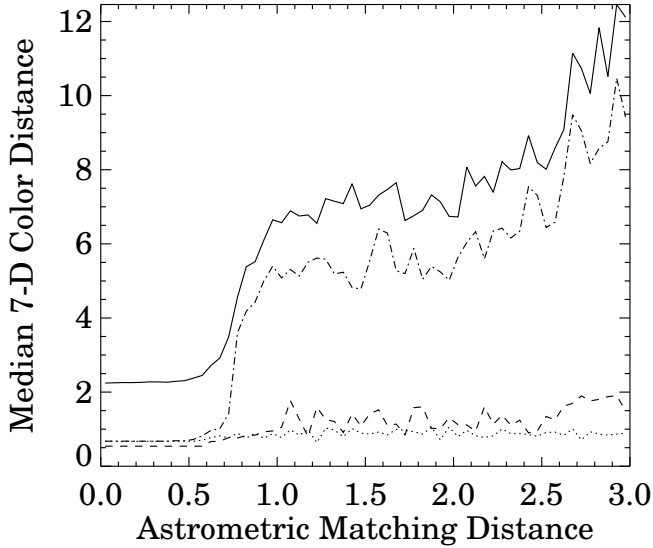}
\caption {\normalsize{Median 7-D color distance (solid line) as a function of the astrometric distance between each matched object's SDSS and 2MASS detections.  Plotted for comparison are the size of the $i-z$ (dashed), $J-H$ (dotted) and $z-J$ (dash-dot) color distance components, also as a function of astrometric distance.  The $i-z$ and $J-H$ color distances are relatively insensitive to astrometric matching distance, as they depend only on an object's colors within a single survey, so are unaffected by incorrect matches between the two surveys.  In contrast, the $z-J$ color distance increases sharply for objects whose SDSS and 2MASS positions are separated by more than 0.6$\arcsec$, as distinct SDSS and 2MASS objects are increasingly spuriously identified as a single matched detection, producing abnormal $z-J$ colors.  }}
\label{badmatchqualitytests}
\end{figure}

Spurious associations generally occur between members of a visual binary, when the SDSS detection of the faint star is
associated with the 2MASS detection of its brighter neighbor, 
resulting in anomalously red $z-J$ colors for the `matched' object.  
Given that SDSS filters make up five of the eight filters used to calculate the colors considered here, that the NIR colors of main-sequence stars change relatively slowly along the main sequence, and that SDSS photometry generally has smaller photometric errors than 2MASS for sources of similar magnitude, mismatches are typically fit to a $g-i$ bin along the stellar locus consistent with their SDSS colors.  This results in relatively small residuals between the colors of the $g-i$ bin identified as the best fit for the object and the colors of that source in a single survey.  The overwhelmingly red $z-J$ color, however, is a very poor match for the $z-J$ color of the object's best fit $g-i$ bin.  This effect is clearly visible in Figure \ref{badmatchqualitytests}, where the single-band $z-J$ color distance clearly increases for sources with astrometric matching distances $> 0.6\arcsec$, with no similar increase detected in the $i-z$ or $J-H$ single-survey color distances.

Mismatches will be significant contaminants for any sample of of sources with large 7DCDs; while sources with astrometric match distances greater than 0.6\arcsec make up only 0.75\% of the SDSS/2MASS sample, they make up 26\% of the sources with 7DCD $>$ 7.}

\item{Objects whose SDSS counterpart is a deblended CHILD produce a disproportionate number of objects with large color distances, as seen in Figure \ref{childqualitytests}; while CHILDren are only 41\% of our sample, they make up 74\% of the sources with 7DCDs greater than 6.  Although the astrometric position of a source may agree in both surveys to within 0.6\arcsec, the SDSS deblended PSF photometry significantly reduces the amount of flux contributed by an object's nearby neighbor, while 2MASS's aperture photometry does not.  This mismatch again produces sources with normal single survey colors and anomalously red $z-J$ colors, spuriously doubling the number of CHILDren assigned large color distances.}

\begin{figure}
\epsscale{1.2}
\plotone{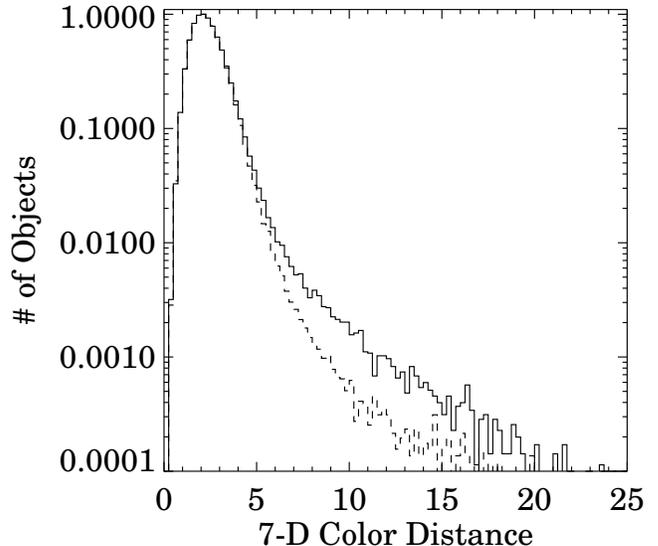}
\caption {\normalsize{7-D Color Distances calculated for objects with (solid) and without (dashed) the SDSS CHILD flag set.  Though the two distributions match well for color distances less than 6, deblended CHILDren produce a clear excess of color outliers at large color distances.}}
\label{childqualitytests}
\end{figure}

\end{itemize}

These tests indicate that 7DCD is a useful tool for identifying point sources in unusual areas of color space, and also enables the best fit $g-i$ color to serve as a simple 1-D parameterization of the properties of typical main-sequence stars.  \textit{The most robust searches for color outliers using 7DCDs, however, should be limited to sources with SDSS/2MASS astrometric matching distances $<$ 0.6$\arcsec$ and CHILD $=$ 0 to filter out objects whose anomalous colors are merely the result of spurious catalog matches.}

As a final test of the utility of the 7DCD, we examine the extent to which each survey contributes to the identification of a source as a color outlier.  Figure \ref{sdsscdvstotalcd} shows that for the point sources included in our sample (high-latitude point sources with high quality detections in both SDSS and 2MASS), SDSS provides the bulk of the leverage for identifying color outliers.  In particular, if each filter contributed an equal amount of signal to the 7DCD, we would expect the typical source to have an SDSS-only color distance equal to 76\% of the total 7DCD.  However, color distances computed from SDSS colors alone are $\geq$ 76\% of the 7DCD for 84.6\% of the sources with 7DCD $>$ 6.  Similarly, the median SDSS-only color distance is 90\% of the median 7DCD for sources with 7DCDs $>$ 6, suggesting that SDSS colors make up most of the signal captured in the 7DCDs of color outliers.  

\begin{figure}
\epsscale{1.2}
\plotone{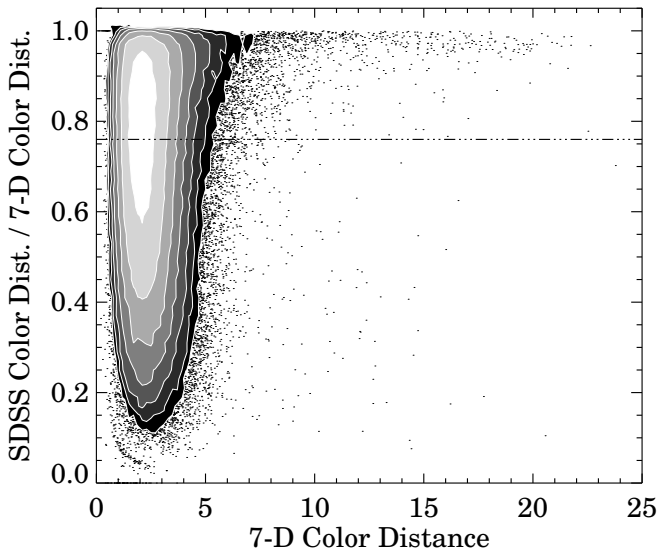}
\caption {\normalsize{The ratio of SDSS-only to 7-D color distance as a function of 7-D color distance, calculated for objects with astrometric matching distances $<$ 0.6\arcsec and CHILD $=$ 0.  Steps between contours indicate a doubling of source densities.  Shown for comparison, as a dash-dot line, is the 0.76 ratio expected if color offsets contribute equally from all filters.  The SDSS-only/7-D color distance ratio is nearly one for high confidence color outliers (7DCDs $>$ 6), indicating that optical/SDSS colors provide the bulk of the leverage for identifying sources with unusual colors.}}
\label{sdsscdvstotalcd}
\end{figure}

As we demonstrate below, the addition of $JHK_s$ photometry can assist in the classification of certain classes of objects identified as color outliers on the basis of SDSS photometry, but it does not appear to identify new classes of outliers which otherwise appear normal in SDSS photometry.  This may change in the near future, as UKIDSS photometry will provide NIR colors with smaller errors and a better match to the dynamic range and spatial resolution of the SDSS, potentially uncovering new classes of faint objects with odd optical-near infrared colors. For now, however, cataloged SDSS-2MASS point sources are most useful for characterizing main sequence stars along the stellar locus, or for studying ``drop-out'' objects, such as late L and T dwarfs, which appear in 2MASS but may be detected only in the SDSS $z$ band (Metchev et al. 2007, submitted).

\subsection{Properties of Identified Color Outliers} 

\begin{figure*}
\epsscale{1.1}
\plotone{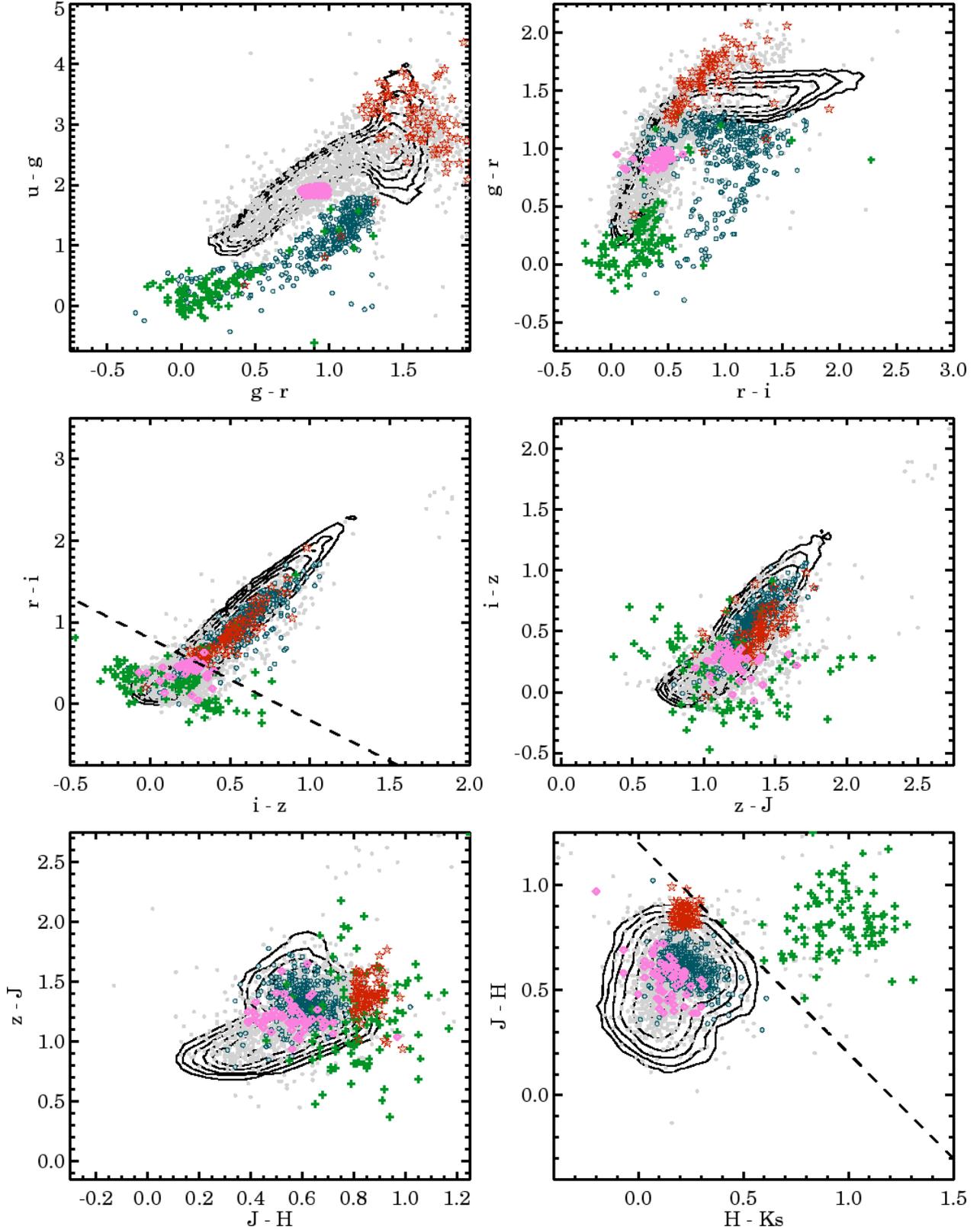}
\caption {\scriptsize{The location in color space of 2117 high quality color outliers in our sample (shown as grey dots)), compared to the location of the SDSS/2MASS stellar locus, shown with black points and contours.  Objects selected as likely QSOs, WDMD pairs, or M giants/Carbon stars are shown as green crosses, blue circles, or red stars, respectively.  Pink diamonds identify a subset of outliers that blend into the margin of the stellar locus; selected on the basis of their $u-g$ and $g-r$ colors, these objects nevertheless show a consistent offset in $r-i$ as well.  Dashed lines indicate the $r-z$ = 0.8 and $J-K$ = 1.2 cuts useful for separating QSOs and WDMD pairs.  }}
\label{multiplot}
\end{figure*}

Having verified the utility of the 7DCD parameter, we have assembled a catalog of 2117 color outliers (0.31\% of the original matched SDSS/2MASS sample) with 7DCDs $>$ 6, astrometric matching distances $<$ 0.6\arcsec, and CHILD $=$ 0.  Motivated largely by the change in slope seen at 7DCD $=$ 6 in Figure \ref{chisquare}, we identify a sample of color outliers with 7DCDs $>$ 6. As Figure \ref{multiplot} shows, the large 7DCD sample makes concentrations of objects with unusual colors, such as QSOs and unresolved white-dwarf/M dwarf pairs (WDMDs), clearly visible.  

Using color constraints we provide preliminary classifications for 25\% of the objects with large 7DCDs.  In particular, a pair of $u-g$ vs. $g-r$ color cuts ( [$u-g <$ 0.7 and $g-r <$ 0.55] or [0.55 $\geq g-r \leq$ 1.3 and $u-g <$ -0.4 + 2 $\times g-r$] ) identify 463 objects within the region of $u-g$ vs. $g-r$ color-color space inhabited by QSOs and WDMDs.  While these objects are easily identified on the basis of their unusual $u-g$ vs. $g-r$ colors, the blue end of the WDMD locus overlaps with the region typically inhabited by QSOs, making the two classes difficult to distinguish photometrically.  As shown in Figure \ref{multiplot}, however, they do possess distinct $J-K_s$ and, to a lesser extent, $r-z$ colors; QSOs typically have $J-K_s >$ 1.2 and $r-z <$ 0.8 and WDMD pairs the opposite.  On the basis of their $J-K_s$ colors, we are able to identify 93 of these outliers as candidate QSOs and 370 as candidate WDMD pairs.

The last class of outliers we identify are 90 objects with NIR colors typical of M giants or carbon stars ($J-H > 0.8$ and $0.15 < H-K_s < 0.3$).  While these sources are most simply identified on the basis of their NIR colors, Figure \ref{multiplot} shows that most are outside the $g-r$ vs. $r-i$ stellar locus, and offset relative to the median $r-i$ vs. $i-z$ locus as well; distinguishing such sources from other objects along the edge of the stellar locus, however, would be difficult on the basis of their optical SDSS photometry alone.  

The majority of the color outliers detected in this catalog, however, appear to be located along the outskirts of the stellar locus.  As an example, we highlight a cluster at $u-g \sim$ 1.9 and $g-r \sim 0.9$ in Figure \ref{multiplot}.  These sources, identified solely on the basis of their $u-g$ and $g-r$ colors, form a coherent, distinctive clump in $g-r$ vs. $r-i$ color color space as well; originally identifed as sources with $g-r$ colors \textit{redward} of most stars with the same $u-g$ color, their $g-r$ colors are consistently \textit{blueward} of sources with similar $r-i$ colors.  A preliminary search of the SDSS spectroscopic database fails to identify any objects with SDSS spectra, so additional spectroscopic programs will be required to identify the cause of the correlated color offsets displayed by these sources.  Their identification, however, demonstrates the utility of the 7DCD for revealing the presence of objects with small, but consistent, offsets in color-color space.

\section{Summary \& Conclusions}

Using a sample of more than 600,000 point sources detected in SDSS and 2MASS,
we have traced the location of main sequence stars through $ugrizJHK_s$
color-color space, parameterizing and tabulating the position and width of the stellar locus as a function of $g-i$ color.  To provide context for this 1-D representation of the stellar locus, we have used synthetic photometry of spectral atlases, as well as analysis of 3000 SDSS stellar spectra by a custom spectral typing pipeline (`The Hammer') to produce estimates of stellar $ugrizJHK_s$ colors and absolute J magnitude (M$_J$) as a function of spectral type.  These measurements will provide guidance for those seeking to interpret the millions of stars detected in SDSS and 2MASS, as well as in future surveys (such as UKIDSS, Pan-STARRS, and Skymapper) making use of similar filter sets.

We have also developed an algorithm to calculate a point source's minimum separation in color space from the stellar locus.  This parameter, which we identify as an object's Seven Dimensional Color Distance (7DCD), accounts for the intrinsic width of the stellar locus as well as photometric errors, and provides a robust identification of objects in unique areas of color space.  Reliability tests reveal the basic utility of the color distance parameter for identifying color outliers, but also identify spurious SDSS/2MASS matches (typically with SDSS/2MASS astrometric separations $>$ 0.6\arcsec or the SDSS CHILD flag set) as a source of erroneously large color distances.  Analysis of a final catalog of 2117 color outliers identified as having color-distances $>$ 6 identifies 370 white-dwarf/M dwarf pairs, 93 QSOs, and 90 M giant/carbon star candidates, and demonstrates how WDMD pairs and QSOs can be distinguished on the basis of their $J-K_s$ and $r-z$ colors.  A group of objects with correlated offsets in both the $u-g$ vs. $g-r$ and $g-r$ vs. $r-i$ color-color spaces is also identified as deserving of subsequent follow-up. Future applications of this 
algorithm to a matched SDSS-UKIDSS catalog may identify additional classes of objects with unusual colors by probing new areas
of color-magnitude space.  

\acknowledgments

The authors would like to thank Coryn Bailer-Jones for a thoughtful referee report; responding to his thoughtful comments and suggestions resulted in considerable improvements to this paper.  Support for this work was provided by NASA through the Spitzer Space Telescope Fellowship Program, through a contract issued by the Jet Propulsion Laboratory, California Institute of Technology under a contract with NASA.  K.R.C also gratefully acknowledges the support of the NASA Graduate Student Researchers Program, which enabled the first stages of this work through grant 80-0273.

Funding for the SDSS and SDSS-II has been provided by the Alfred P. Sloan Foundation, the Participating Institutions, the National Science Foundation, the U.S. Department of Energy, the National Aeronautics and Space Administration, the Japanese Monbukagakusho, the Max Planck Society, and the Higher Education Funding Council for England. The SDSS Web Site is http://www.sdss.org/.

The SDSS is managed by the Astrophysical Research Consortium for the Participating Institutions. The Participating Institutions are the American Museum of Natural History, Astrophysical Institute Potsdam, University of Basel, University of Cambridge, Case Western Reserve University, University of Chicago, Drexel University, Fermilab, the Institute for Advanced Study, the Japan Participation Group, Johns Hopkins University, the Joint Institute for Nuclear Astrophysics, the Kavli Institute for Particle Astrophysics and Cosmology, the Korean Scientist Group, the Chinese Academy of Sciences (LAMOST), Los Alamos National Laboratory, the Max-Planck-Institute for Astronomy (MPIA), the Max-Planck-Institute for Astrophysics (MPA), New Mexico State University, Ohio State University, University of Pittsburgh, University of Portsmouth, Princeton University, the United States Naval Observatory, and the University of Washington. 

 The Two Micron All Sky Survey was a joint project of the University of Massachusetts and the Infrared Processing and Analysis Center (California Institute of Technology). The University of Massachusetts was responsible for the overall management of the project, the observing facilities and the data acquisition. The Infrared Processing and Analysis Center was responsible for data processing, data distribution and data archiving.

This research has made use of NASA's Astrophysics Data System Bibliographic Services, the SIMBAD database, operated at CDS, Strasbourg, France, and the VizieR database of astronomical catalogues \citep{Ochsenbein2000}.  

\appendix

\section{The Hammer -- An IDL-based Spectral Typing Suite}

The Hammer spectral typing algorithm was originally developed for use on late-type SDSS spectra, but has subsequently been modified to allow it to classify spectra in a variety of formats with targets spanning the MK spectral sequence.  In this appendix, we document the Hammer's index set and describe the alogrithm employed to automatically assign spectral types to input target spectra.  As well, we provide a brief discussion of the use of the program, which includes an interactive mode which allows the user to assign final spectral types via visual comparison with a grid of spectral templates.  We conclude this appendix with a discussion of tests of the accuracy of the Hammer, as well as a few caveats concerning its limitations.  Those interested in utilizing the Hammer for their own science goals can download a copy of the code from \anchor{http://www.cfa.harvard.edu/~kcovey/thehammer}{http://www.cfa.harvard.edu/$\sim$kcovey/thehammer}.

\subsection{Spectral Indices}

To estimate the spectral type of an input spectrum, the Hammer measures a set of 26 atomic (H, Ca I, Ca II, Na I, Mg I, Fe I, Rb, Cs) and molecular (Gband, CaH, TiO, VO, CrH) features that are prominent in late type stars, as well as two colors (BlueColor and Color-1) that sample the broadband shape of the SED.

These spectral features are measured with ratios of the mean flux density within different spectral bandpasses, such that:
\begin{eqnarray}
\label{index}
\textrm{Index} = \frac{\textrm{Mean Flux Density}(N)}{\textrm{Mean Flux Density}(D)}
\end{eqnarray}
where N and D are spectral regions bounded by wavelengths given in Table \ref{tab:simpleindices}.  Table \ref{tab:complexindices} summarizes similar information for four indices where the numerator of the spectral index combines information from multiple bandpasses.  For these indices, N is calculated as:
\begin{eqnarray}
\label{multinum}
\textrm{N} = \textrm{N$_1$ Weight} \times \textrm{Mean Flux Density}(N_1) + \textrm{N$_2$ Weight} \times \textrm{Mean Flux Density}(N_2) \nonumber
\end{eqnarray}
where Table \ref{tab:complexindices} gives the wavelength boundaries and weights of N$_1$ and N$_2$.  

\input{tab6}

\input{tab7}

To provide a consistent estimate of the uncertainty for both types of spectral indices, we adopt as a characteristic uncertainty the change in the index induced by the uncertainty in the mean flux density of the denominator.  This index uncertainty, $\sigma_{Index}$, is calculated as
\begin{eqnarray}
\label{indexerr}
\sigma_{Index} = \sqrt{\Big(Index - \big(\frac{\textrm{Mean Flux Density}(N)}{(\textrm{Mean Flux Density}(D) + \sigma_D )} \big) \Big)^2} 
\end{eqnarray}
where $\sigma_D$ is the standard deviation of the flux density divided by the number of pixels used to sample that regime, or
\begin{eqnarray}
\label{derr}
\sigma_D = \frac{ \textrm{Standard Deviation}(D)}{\textrm{n}_{pix}}
\end{eqnarray}

To map out their variation as a function of spectral type, 
we have measured these indices in 594 dwarf standards taken from the 
spectral libraries of \citet{Pickles1998}, \citet{Hawley2002}, 
\citet{Valdes2004}, \citet{Le-Borgne2003}, \citet{Sanchez-Blazquez2006}, 
and \citet{Bochanski2007}, spanning a range in spectral type from O5 to L8.  We measured the median value of each
index as a function of spectral type, linearly interpolating across
gaps in the spectral type grid.  The non-uniform spectral
coverage of these libraries result in some indices being measured
reliably only over a restricted range of spectral types; spectra of the 
reddest stars, for instance, either do not extend to Ca K, or 
are too noisy to produce reliable measurements.  To compensate for
this, when necessary we extended the index values from the earliest and
latest templates which produced reliable measurements to the edges of the full
spectral grid.  Figures \ref{indices1}-\ref{indices3} display the resulting
median spectral type/spectral index relationships, with the values measured from
each individual template also shown for comparison.

\begin{figure}
\epsscale{0.872}
\plotone{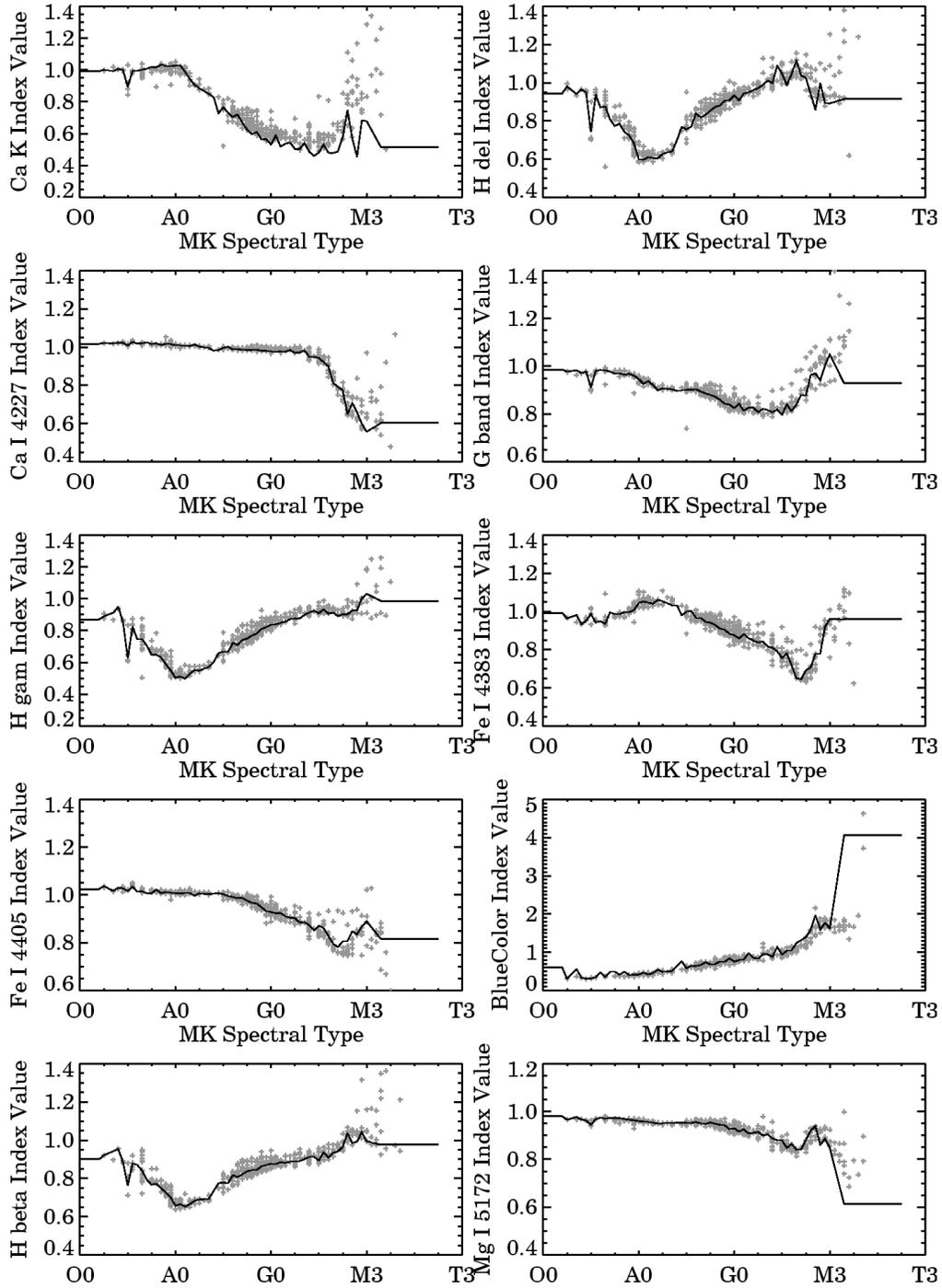}
\caption {\normalsize{Variations of spectral indices as a function of MK spectral type.  Grey crosses denote the spectral index as measured for a single spectral type template; the solid line gives the median index vs. spectral type relation used by the Hammer.}}
\label{indices1}
\end{figure}

\begin{figure}
\epsscale{0.872}
\plotone{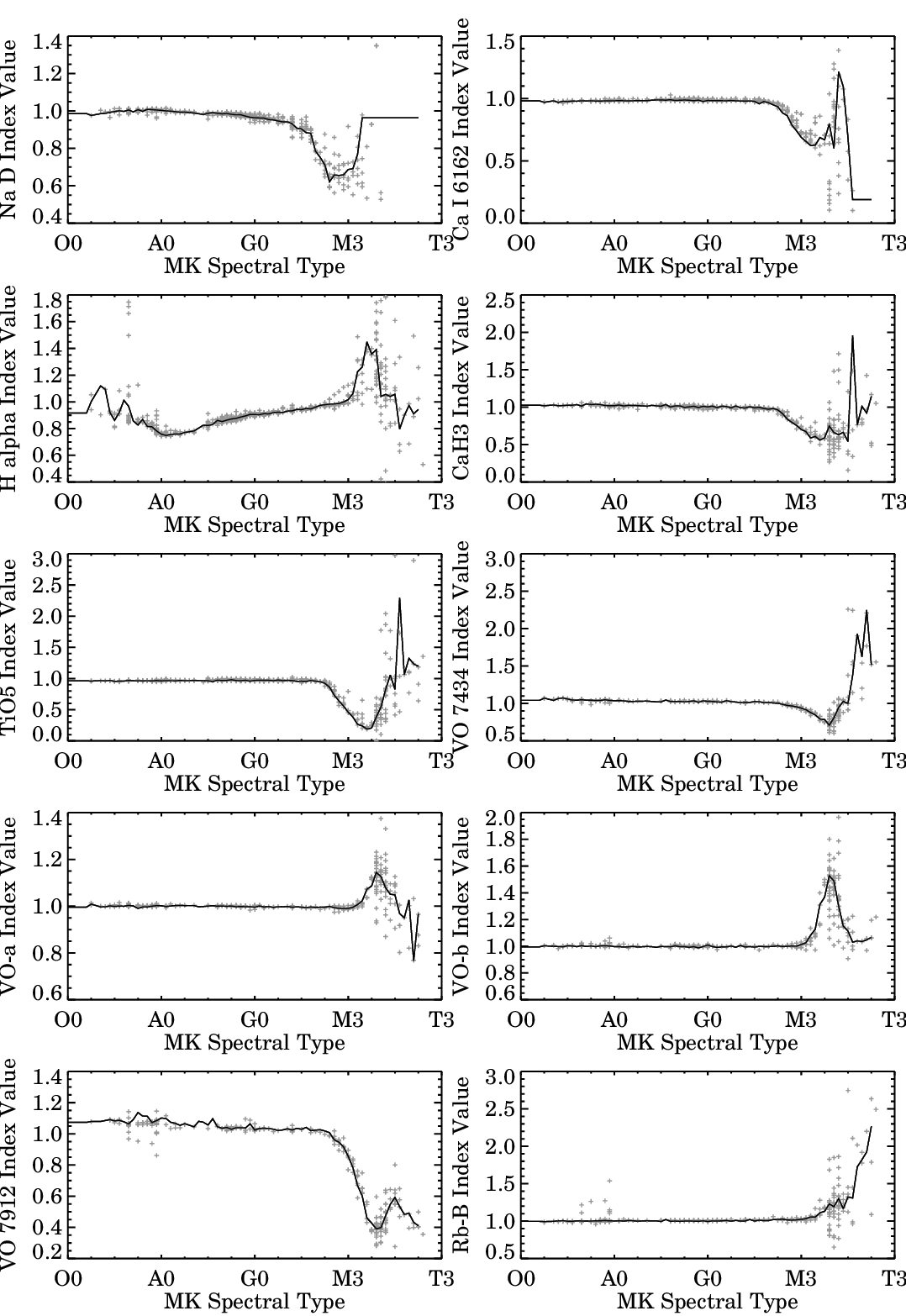}
\caption {\normalsize{Continued from Figure \ref{indices1}}}
\label{indices2}
\end{figure}

\begin{figure}
\epsscale{0.872}
\plotone{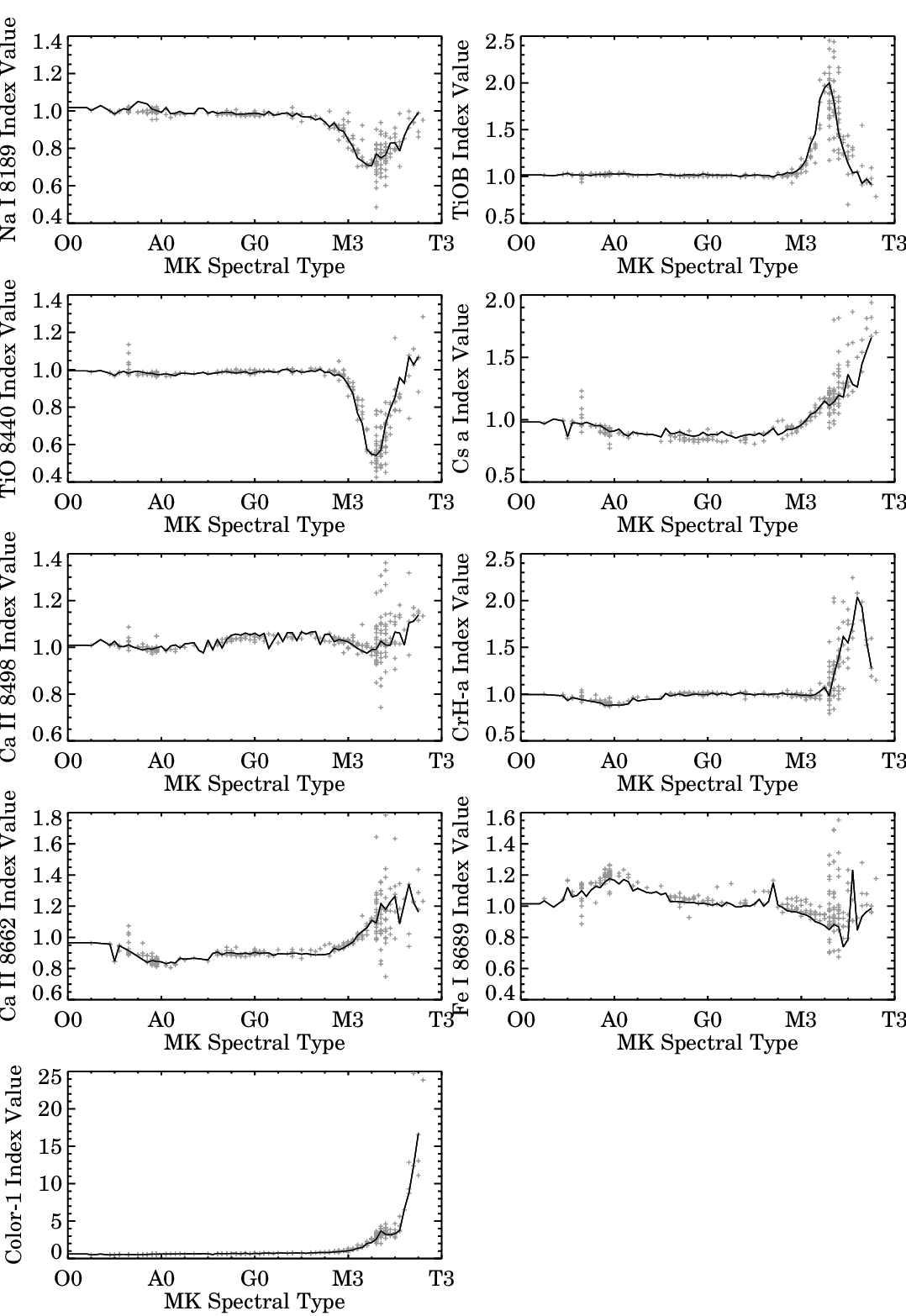}
\caption {\normalsize{Continued from Figure \ref{indices1}}}
\label{indices3}
\end{figure}

\subsection{Automated \& Interactive Spectral Type Determination}

Using these measured spectral type/spectral index relationships, the 
Hammer generates an automatic estimate of the spectral type of an input
target spectrum.  To produce this estimate, the Hammer first measures 
all spectral indices in the above set that are contained within the wavelength 
coverage of the target spectrum.  Spectral indices which are not available in a given 
target spectrum are flagged as being inaccessible, and are excluded from
subsequent analysis.

Each index, however, is most useful within a given
spectral type range, and can reduce the accuracy of the spectral typing routine 
if it contributes an inordinate amount of weight to the goodness of fit parameter.
For this reason, the Hammer attempts to exclude indices that are unlikely to
be useful in constraining the spectral type of a given input star.  To decide which indices
to exclude, the Hammer measures the mean wavelength of the target spectrum between 5000 \AA\
and 7750 \AA\, 
with and without weighting each wavelength by the flux at that wavelength.  
The ratio of these means (weighted:non-weighted) is a crude indicator of the 
shape of a star's SED; late type stars have more flux at longer wavelengths, so possess 
ratios greater than one, while early type stars have more flux at short wavelengths, 
so have ratios less than one.  For the latest type stars ($\sim$ M3 or later; mean wavelength
ratios of 1.03 or larger), the Hammer excludes indices tuned for early type stars 
(Ca K, H $\delta$, Ca I 4227, H $\gamma$, Fe I 4383, Fe I 4404, BlueColor, H $\beta$, Na D,
Ca I 6162, H $\alpha$, Ca II 8498, \& Fe I 8689).  For the earliest type stars ($\sim$ F0 and
earlier, mean wavelength ratios of 0.97 or less), the Hammer excludes indices useful at
cooler temperatures (Ca I 4227, Na D, VO 7434, VO-a, VO-b, VO 7912, Rb-b, TiOB, TiO 8440, Cs-a, Ca II 8498, CrH-a,
Ca II 8662, Fe I 8689, \& color-1).  For intermediate type stars (mean wavelength ratios between 0.97 and 1.03), 
the Hammer uses all available indices except for H $\delta$ and H $\beta$.

The values of the target star's remaining indices are then compared to the values of the template grid to determine the best fit spectral type.  Specifically, the differences in the target's indices and the median values associated with a given spectral type are
normalized by the errors associated with each index in the target spectrum. 
After comparing the target indices with the full grid of indices measured from known standards, the Hammer selects the best
fit spectral type by selecting the spectral type associated with the smallest mean squared error-normalized residuals.  

Once the Hammer has produced an initial estimate of the spectral type of each input spectrum, it enters an 
interactive mode whereby the user can perform a direct visual comparison of the target spectrum to a
grid of spectral templates.  In this mode, the user can smooth the target spectrum for clarity, assign
a final MK spectral type, mark a spectrum as hopelessly noisy (`bad'), or assign it a three character flag for
later identification (the `odd' button).  This routine also allows the user to save the results of interactive typing before the full 
list of input stars have been processed (the `break' button), and saves both the automated and interactive spectral type
assigned to each spectrum.  Though originally developed to process SDSS spectra, 
the Hammer has since been modified to allow it to process spectra in a variety of formats, and also incorporates routines 
written by \citet{West2004} to identify M-type subdwarfs and diagnose stellar magnetic activity; the IDL code for the Hammer can be downloaded from 
\url{http://www.cfa.harvard.edu/~kcovey/thehammer}.

\subsection{Spectral Typing Accuracy \& Caveats}

The robustness of the Hammer's spectral typing algorithm has been tested by measuring the errors in the spectral types it assigns to dwarf templates of known spectral type, synthetically degraded with white Gaussian noise to S/N $\sim$ 5.  These tests indicate that the Hammer's automated spectral types are accurate to within $\pm$ 2 subclasses for K and M type stars, the regime for which the Hammer has been optomised.  At warmer temperatures, the Hammer is somewhat less accurate; typical uncertainties of $\pm$ 4 subtypes are found for A-G stars at S/N $\sim$ 5.  

We note, however, two caveats concerning the utility of the Hammer as a spectral typing engine.  First, the mapping of index strength as a function of spectral type on which the Hammer is constructed has only been calibrated for solar metallicity dwarf stars.  As a result, the Hammer cannot produce accurate spectral types for stars outside this regime of metallicity and luminosity class; spectral types derived for non-solar, non-dwarf stars will be prone to significant, systematic errors.  

Secondly, the Hammer was originally designed to process stellar spectra obtained by SDSS-I at high Galactic latitude.  As the extinction towards these stars are typically very small, and the spectrophotometry produced by the SDSS pipeline is accurate to within a few percent, the Hammer has been designed to make use of the information contained in the slope of the stellar continuum, both in the form of specific indices (BlueColor, Color-1) and in using the SED slope to determine the most useful index set for a given target.  As a result, the Hammer will not produce reliable results for spectra where the intrinsic slope of the stellar continuum has not been preserved, either due to instrumental effects or the presence of significant amounts of extinction.  For spectra such as these, spectral types should be assigned using methods less sensitive to the shape of the stellar continuum.


\clearpage

\LongTables

\input{tab1}

\input{tab3}

\end{document}

%% file: tab2.tex
\begin{deluxetable*}{lcccccccc}
\tablewidth{0pt}
\tabletypesize{\scriptsize}
\tablecaption{Analytic Fits to the Stellar Locus  \label{tab:analyticfits}}
\tablehead{
  \colhead{} &
  \colhead{} &
  \colhead{} &
  \colhead{} &
  \colhead{} &
  \colhead{} &
  \colhead{} &
  \colhead{Max.} &
  \colhead{Med.} \\
  \colhead{color} &
  \colhead{A$_0$} &
  \colhead{A$_1$} &
  \colhead{A$_2$} &
  \colhead{A$_3$} &
  \colhead{A$_4$} &
  \colhead{A$_5$} &
  \colhead{Res.} &
\colhead{Res.} }
\startdata
$u-g$ &  1.0636113 & -1.6267818 &  4.9389572 & -3.2809081 &  0.8725109 & -0.0828035 & 0.6994  & 0.0149 \\
$g-r$ &  0.4290263 & -0.8852323 &  2.0740616 & -1.1091553 &  0.2397461 & -0.0183195 & 0.2702  & 0.0178 \\
$r-i$ & -0.4113500 &  1.8229991 & -1.9989772 &  1.0662075 & -0.2284455 &  0.0172212 & 0.2680  & 0.0164 \\
$i-z$ & -0.2270331 &  0.7794558 & -0.7350749 &  0.3727802 & -0.0735412 &  0.0049808 & 0.1261  & 0.0071 \\
$z-J$ &  0.5908002 &  0.5550226 & -0.0980948 & -0.0744787 &  0.0416410 & -0.0051909 & 0.1294  & 0.0054 \\
$J-H$ &  0.2636025 & -0.2509140 &  0.9660369 & -0.6294004 &  0.1561321 & -0.0134522 & 0.0413  & 0.0042 \\
$H-K_s$ &  0.0134374 &  0.1844403 & -0.1989967 &  0.1300435 & -0.0344786 &  0.0032191 & 0.0900  & 0.0040 \\
\enddata
\end{deluxetable*}

%% file: tab4.tex
\begin{deluxetable}{l|c|cc}
\tablewidth{0pt}
\tabletypesize{\scriptsize}
\tablecaption{$g-i$ as a function of MK Spectral Type  \label{tab:gispectype}}
\tablehead{
  \colhead{Spec.} &
  \colhead{Synth.} &
  \colhead{Spec.} &
\colhead{} \\
  \colhead{Type} &
  \colhead{$g-i$} &
  \colhead{$g-i$} &
\colhead{n$_{spec}$}
}
\startdata
 O5 &  -1.00 &  \nodata &    0\\
 O9 &  -0.97 &  \nodata &    0\\
 B0 &  -0.94 &  \nodata &    0\\
 B1 &  -0.82 &  \nodata &    0\\
 B3 &  -0.72 &  \nodata &    0\\
 B8 &  -0.57 &  \nodata &    0\\
 B9 &  -0.51 &  \nodata &    0\\
 A0 &  -0.44 &  \nodata &    0\\
 A2 &  -0.39 &  \nodata &    0\\
 A3 &  -0.30 &  \nodata &    0\\
 A5 &  -0.21 &  \nodata &    0\\
 A7 &  -0.10 &  \nodata &    0\\
 F0 &   0.09 &     0.41 &    1\\
 F2 &   0.22 &     0.40 &   24\\
 F5 &   0.29 &     0.46 &   16\\
 F6 &   0.36 &     0.46 &   33\\
 F8 &   0.45 &     0.55 &   97\\
 G0 &   0.52 &     0.57 &   53\\
 G2 &   0.60 &     0.60 &   49\\
 G5 &   0.65 &     0.66 &   58\\
 G8 &   0.76 &     0.77 &   95\\
 K0 &   0.83 &     0.81 &   80\\
 K2 &   1.02 &     0.95 &  275\\
 K3 &   1.17 &     1.10 &   88\\
 K4 &   1.38 &     1.31 &  274\\
 K5 &   1.59 &     1.47 &  137\\
 K7 &   1.88 &     1.75 &  148\\
 M0 &   1.95 &     1.98 &  315\\
 M1 &   2.10 &     2.23 &  256\\
 M2 &   2.28 &     2.42 &  303\\
 M3 &   2.66 &     2.63 &  438\\
 M4 &   2.99 &     2.87 &  255\\
 M5 &   3.32 &     3.22 &   57\\
 M6 &   3.84 &     3.39 &   13\\
\enddata
\end{deluxetable}

%% file: tab5.tex
\begin{deluxetable}{lcccccc}
\tablewidth{0pt}
\tabletypesize{\scriptsize}
\tablecaption{Synthetic SDSS Photometry of Pickles (1998) Non-Solar Metallicity Standards  \label{tab:nonsolarstars}}
\tablehead{
  \colhead{Spec.} &
  \colhead{Lum.} &
  \colhead{} &
  \colhead{} &
  \colhead{} &
  \colhead{} &
\colhead{} \\
  \colhead{Type} &
  \colhead{Class} &
  \colhead{$u-g$} &
  \colhead{$g-r$} &
  \colhead{$r-i$} &
  \colhead{$i-z$} &
\colhead{[Fe/H]} }
\startdata
wF5 &   V &    1.03  &    0.26  &    0.05  &   -0.04  &   -0.30 \\
rF6 &   V &    1.20  &    0.33  &    0.10  &    0.00  &    0.30 \\
wF8 &   V &    1.16  &    0.32  &    0.10  &   -0.03  &   -0.60 \\
rF8 &   V &    1.27  &    0.38  &    0.15  &   -0.01  &    0.20 \\
wG0 &   V &    1.28  &    0.37  &    0.13  &    0.03  &   -0.80 \\
rG0 &   V &    1.36  &    0.41  &    0.16  &    0.04  &    0.40 \\
wG5 &   V &    1.30  &    0.47  &    0.13  &    0.03  &   -0.40 \\
rG5 &   V &    1.51  &    0.51  &    0.14  &    0.05  &    0.10 \\
rK0 &   V &    1.83  &    0.67  &    0.19  &    0.09  &    0.50 \\
wG5 & III &    1.69  &    0.65  &    0.21  &    0.10  &   -0.26 \\
rG5 & III &    1.85  &    0.69  &    0.22  &    0.08  &    0.23 \\
wG8 & III &    2.01  &    0.68  &    0.25  &    0.11  &   -0.38 \\
wK0 & III &    1.92  &    0.75  &    0.26  &    0.11  &   -0.33 \\
rK0 & III &    2.29  &    0.81  &    0.27  &    0.13  &    0.18 \\
wK1 & III &    2.20  &    0.82  &    0.27  &    0.13  &   -0.10 \\
rK1 & III &    2.62  &    0.91  &    0.30  &    0.17  &    0.27 \\
wK2 & III &    2.46  &    0.84  &    0.32  &    0.15  &   -0.38 \\
rK2 & III &    2.65  &    0.99  &    0.33  &    0.18  &    0.24 \\
wK3 & III &    2.47  &    0.96  &    0.37  &    0.19  &   -0.36 \\
rK3 & III &    2.92  &    1.10  &    0.36  &    0.21  &    0.27 \\
wK4 & III &    2.91  &    1.17  &    0.42  &    0.26  &   -0.33 \\
rK4 & III &    2.87  &    1.22  &    0.43  &    0.29  &    0.15 \\
rK5 & III &    3.34  &    1.35  &    0.56  &    0.32  &    0.08 \\
\enddata
\end{deluxetable}

%% file: tab6.tex
\begin{deluxetable}{lcccc}
\tablewidth{0pt}
\tabletypesize{\scriptsize}
\tablecaption{Single Numerator Hammer Spectral Indices \label{tab:simpleindices}}
\tablehead{
  \colhead{Spectral} &
  \colhead{N} &
  \colhead{N} &
  \colhead{D} &
  \colhead{D} \\
  \colhead{Feature} &
  \colhead{Start (\AA)} &
  \colhead{End (\AA)} &
  \colhead{Start (\AA)} &
  \colhead{End (\AA)} }
\startdata
Ca K       &  3923.7  &  3943.7  &  3943.7  & 3953.7 \\
H $\delta$ &  4086.7  &  4116.7  &  4136.7  & 4176.7 \\
Ca I 4227  &  4216.7  &  4236.7  &  4236.7  & 4256.7 \\
G band     &  4285.0  &  4315.0  &  4260.0  & 4285.0 \\
H $\gamma$ &  4332.5  &  4347.5  &  4355.0  & 4370.0 \\
Fe I 4383  &  4378.6  &  4388.6  &  4355.0  & 4370.0 \\ 
Fe I 4405  &  4399.8  &  4409.8  &  4414.8  & 4424.8 \\
BlueColor  &  6100.0  &  6300.0  &  4500.0  & 4700.0 \\
H $\beta$  &  4847.0  &  4877.0  &  4817.0  & 4847.0 \\
Mg I 5172  &  5152.7  &  5192.7  &  5100.0  & 5150.0 \\
Na D       &  5880.0  &  5905.0  &  5910.0  & 5935.0 \\
Ca I 6162  &  6150.0  &  6175.0  &  6120.0  & 6145.0 \\
H $\alpha$ &  6548.0  &  6578.0  &  6583.0  & 6613.0 \\
CaH 3      &  6960.0  &  6990.0  &  7042.0  & 7046.0 \\
TiO 5      &  7126.0  &  7135.0  &  7042.0  & 7046.0 \\
VO 7434    &  7430.0  &  7470.0  &  7550.0  & 7570.0 \\
VO 7912    &  7900.0  &  7980.0  &  8100.0  & 8150.0 \\
Na I 8189  &  8177.0  &  8201.0  &  8151.0  & 8175.0 \\
TiO B      &  8400.0  &  8415.0  &  8455.0  & 8470.0 \\
TiO 8440   &  8440.0  &  8470.0  &  8400.0  & 8420.0 \\
Ca II 8498 &  8483.0  &  8513.0  &  8513.0  & 8543.0 \\
CrH-a      &  8580.0  &  8600.0  &  8621.0  & 8641.0 \\
Ca II 8662 &  8650.0  &  8675.0  &  8625.0  & 8650.0 \\
Fe I 8689  &  8684.0  &  8694.0  &  8664.0  & 8674.0 \\
Color-1    &  8900.0  &  9100.0  &  7350.0  & 7550.0 \\
\enddata
\end{deluxetable}

%% file: tab7.tex
\begin{deluxetable*}{lcccccccc}
\tablewidth{0pt}
\tabletypesize{\scriptsize}
\tablecaption{Multiple Numerator Hammer Spectral Indices \label{tab:complexindices}}
\tablehead{
  \colhead{Spectral} &
  \colhead{Num. 1} &
  \colhead{Num. 1} &
  \colhead{Num. 1} &
  \colhead{Num. 2} &
  \colhead{Num. 2} & 
  \colhead{Num. 2} &
  \colhead{Denom.} &
  \colhead{Denom.} \\
  \colhead{Feature} &
  \colhead{Start (\AA)} &
  \colhead{End (\AA)} &
  \colhead{Weight} &
  \colhead{Start (\AA)} &
  \colhead{End (\AA)} & 
  \colhead{Weight} &
  \colhead{Start (\AA)} &
  \colhead{End (\AA)}
}
\startdata
VO-a       &  7350.0  &  7400.0  &  0.5625  & 7510.0  &  7560.0  &  0.4375  & 7420.0   & 7470.0   \\
VO-b       &  7860.0  &  7880.0  &   0.5    & 8080.0  &  8100.0  &  0.5     & 7960.0   & 8000.0   \\
Rb-b       &  7922.6  &  7932.6  &   0.5    & 7962.6  &  7972.6  & 0.5      & 7942.6   & 7952.6   \\
Cs-a       &  8496.1  &  8506.1  &   0.5    & 8536.1  &  8546.1  & 0.5      & 8516.1   & 8526.1   \\
\enddata
\end{deluxetable*}

%% file: tab1.tex
\begin{deluxetable}{lccccccccccccccc}
\tablewidth{0pt}
\tabletypesize{\footnotesize}
\tablecaption{Locus colors as a function of g-i color  \label{tab:locusfit}}
\tablehead{
  \colhead{$g-i$} &
  \colhead{n$_{stars}$} &
  \colhead{$u-g$} &
  \colhead{$\sigma_{u-g}$} &
  \colhead{$g-r$} &
  \colhead{$\sigma_{g-r}$} &
  \colhead{$r-i$} &
  \colhead{$\sigma_{r-i}$} &
  \colhead{$i-z$} &
  \colhead{$\sigma_{i-z}$} &
  \colhead{$z-J$} &
  \colhead{$\sigma_{z-J}$} &
  \colhead{$J-H$} &
  \colhead{$\sigma_{J-H}$} &
  \colhead{$H-K_s$} &
\colhead{$\sigma_{H-K_s}$} }
\startdata
   0.10 &    25 &    1.05  &    0.67  &    0.09  &    0.04  &    0.02  &    0.02  &   -0.03  &    0.06  &    0.76  &    0.19  &    0.23  &    0.38  &    0.12  &    0.51 \\
   0.20 &    70 &    1.06  &    0.10  &    0.18  &    0.03  &    0.03  &    0.03  &   -0.02  &    0.06  &    0.77  &    0.07  &    0.22  &    0.07  &    0.09  &    0.11 \\
   0.26 &    15 &    1.05  &    0.05  &    0.21  &    0.01  &    0.06  &    0.01  &   -0.03  &    0.02  &    0.77  &    0.03  &    0.22  &    0.04  &    0.08  &    0.01 \\
   0.28 &    50 &    1.07  &    0.07  &    0.22  &    0.02  &    0.06  &    0.02  &   -0.03  &    0.03  &    0.77  &    0.04  &    0.24  &    0.05  &    0.05  &    0.08 \\
   0.30 &    49 &    1.07  &    0.07  &    0.24  &    0.03  &    0.06  &    0.03  &    0.00  &    0.04  &    0.79  &    0.04  &    0.23  &    0.06  &    0.08  &    0.10 \\
   0.32 &   132 &    1.05  &    0.10  &    0.25  &    0.03  &    0.08  &    0.02  &    0.00  &    0.04  &    0.78  &    0.04  &    0.25  &    0.06  &    0.06  &    0.09 \\
   0.34 &    86 &    1.09  &    0.09  &    0.26  &    0.02  &    0.09  &    0.02  &    0.00  &    0.04  &    0.80  &    0.04  &    0.25  &    0.07  &    0.06  &    0.07 \\
   0.36 &   212 &    1.07  &    0.10  &    0.28  &    0.03  &    0.09  &    0.02  &    0.00  &    0.03  &    0.80  &    0.04  &    0.26  &    0.05  &    0.06  &    0.07 \\
   0.38 &   178 &    1.08  &    0.07  &    0.29  &    0.03  &    0.09  &    0.03  &    0.01  &    0.04  &    0.81  &    0.04  &    0.28  &    0.05  &    0.06  &    0.08 \\
   0.40 &   442 &    1.07  &    0.09  &    0.30  &    0.02  &    0.10  &    0.03  &    0.01  &    0.03  &    0.81  &    0.04  &    0.28  &    0.06  &    0.06  &    0.07 \\
   0.42 &   383 &    1.10  &    0.07  &    0.32  &    0.02  &    0.11  &    0.02  &    0.01  &    0.03  &    0.82  &    0.04  &    0.29  &    0.05  &    0.07  &    0.07 \\
   0.44 &   903 &    1.10  &    0.08  &    0.33  &    0.03  &    0.11  &    0.02  &    0.02  &    0.04  &    0.82  &    0.04  &    0.29  &    0.05  &    0.07  &    0.07 \\
   0.46 &   706 &    1.11  &    0.07  &    0.34  &    0.02  &    0.12  &    0.02  &    0.02  &    0.03  &    0.83  &    0.04  &    0.30  &    0.05  &    0.08  &    0.07 \\
   0.48 &  1494 &    1.12  &    0.08  &    0.36  &    0.02  &    0.12  &    0.02  &    0.02  &    0.04  &    0.84  &    0.04  &    0.31  &    0.05  &    0.07  &    0.07 \\
   0.50 &  1198 &    1.14  &    0.07  &    0.37  &    0.02  &    0.13  &    0.02  &    0.03  &    0.03  &    0.84  &    0.04  &    0.31  &    0.06  &    0.07  &    0.07 \\
   0.52 &  2351 &    1.15  &    0.08  &    0.38  &    0.02  &    0.14  &    0.02  &    0.03  &    0.03  &    0.85  &    0.04  &    0.32  &    0.06  &    0.07  &    0.07 \\
   0.54 &  1859 &    1.18  &    0.09  &    0.40  &    0.02  &    0.14  &    0.02  &    0.03  &    0.03  &    0.85  &    0.04  &    0.33  &    0.05  &    0.07  &    0.08 \\
   0.56 &  3276 &    1.20  &    0.09  &    0.41  &    0.02  &    0.15  &    0.02  &    0.04  &    0.03  &    0.86  &    0.04  &    0.33  &    0.06  &    0.07  &    0.07 \\
   0.58 &  2325 &    1.22  &    0.09  &    0.43  &    0.02  &    0.15  &    0.02  &    0.04  &    0.03  &    0.87  &    0.04  &    0.34  &    0.05  &    0.07  &    0.07 \\
   0.60 &  3746 &    1.25  &    0.10  &    0.44  &    0.02  &    0.16  &    0.02  &    0.04  &    0.03  &    0.87  &    0.04  &    0.34  &    0.06  &    0.07  &    0.07 \\
   0.62 &  2732 &    1.28  &    0.10  &    0.46  &    0.02  &    0.16  &    0.02  &    0.05  &    0.04  &    0.88  &    0.04  &    0.35  &    0.06  &    0.08  &    0.07 \\
   0.64 &  3831 &    1.31  &    0.11  &    0.47  &    0.03  &    0.17  &    0.02  &    0.05  &    0.03  &    0.89  &    0.04  &    0.35  &    0.05  &    0.08  &    0.07 \\
   0.66 &  2794 &    1.34  &    0.12  &    0.48  &    0.02  &    0.17  &    0.02  &    0.05  &    0.03  &    0.89  &    0.04  &    0.36  &    0.06  &    0.08  &    0.07 \\
   0.68 &  3671 &    1.37  &    0.11  &    0.50  &    0.02  &    0.18  &    0.02  &    0.06  &    0.04  &    0.90  &    0.04  &    0.37  &    0.06  &    0.08  &    0.07 \\
   0.70 &  2747 &    1.40  &    0.12  &    0.51  &    0.02  &    0.19  &    0.02  &    0.06  &    0.03  &    0.91  &    0.04  &    0.38  &    0.06  &    0.08  &    0.07 \\
   0.72 &  3490 &    1.44  &    0.13  &    0.53  &    0.02  &    0.19  &    0.02  &    0.07  &    0.04  &    0.91  &    0.04  &    0.39  &    0.06  &    0.08  &    0.07 \\
   0.74 &  2551 &    1.46  &    0.13  &    0.54  &    0.03  &    0.20  &    0.02  &    0.07  &    0.03  &    0.92  &    0.04  &    0.40  &    0.05  &    0.08  &    0.07 \\
   0.76 &  3475 &    1.49  &    0.13  &    0.56  &    0.02  &    0.20  &    0.02  &    0.07  &    0.04  &    0.93  &    0.04  &    0.40  &    0.06  &    0.09  &    0.08 \\
   0.78 &  2628 &    1.53  &    0.14  &    0.57  &    0.03  &    0.21  &    0.03  &    0.08  &    0.03  &    0.94  &    0.04  &    0.41  &    0.06  &    0.09  &    0.07 \\
   0.80 &  3374 &    1.56  &    0.13  &    0.58  &    0.02  &    0.21  &    0.02  &    0.08  &    0.04  &    0.95  &    0.04  &    0.42  &    0.06  &    0.08  &    0.07 \\
   0.82 &  2555 &    1.60  &    0.13  &    0.60  &    0.02  &    0.22  &    0.03  &    0.09  &    0.04  &    0.95  &    0.04  &    0.43  &    0.06  &    0.09  &    0.07 \\
   0.84 &  3424 &    1.63  &    0.13  &    0.61  &    0.03  &    0.23  &    0.02  &    0.09  &    0.04  &    0.96  &    0.04  &    0.44  &    0.06  &    0.09  &    0.07 \\
   0.86 &  2658 &    1.66  &    0.14  &    0.63  &    0.02  &    0.23  &    0.03  &    0.10  &    0.04  &    0.97  &    0.04  &    0.45  &    0.06  &    0.09  &    0.07 \\
   0.88 &  3335 &    1.69  &    0.13  &    0.64  &    0.03  &    0.24  &    0.03  &    0.10  &    0.03  &    0.97  &    0.04  &    0.45  &    0.06  &    0.09  &    0.07 \\
   0.90 &  2535 &    1.73  &    0.14  &    0.66  &    0.02  &    0.24  &    0.02  &    0.11  &    0.04  &    0.98  &    0.04  &    0.46  &    0.06  &    0.09  &    0.07 \\
   0.92 &  3212 &    1.76  &    0.13  &    0.67  &    0.03  &    0.25  &    0.03  &    0.11  &    0.03  &    0.99  &    0.04  &    0.47  &    0.06  &    0.09  &    0.07 \\
   0.94 &  2550 &    1.80  &    0.14  &    0.68  &    0.03  &    0.25  &    0.02  &    0.11  &    0.03  &    1.00  &    0.05  &    0.48  &    0.05  &    0.10  &    0.07 \\
   0.96 &  3223 &    1.82  &    0.13  &    0.70  &    0.02  &    0.26  &    0.03  &    0.12  &    0.04  &    1.00  &    0.04  &    0.48  &    0.06  &    0.10  &    0.07 \\
   0.98 &  2298 &    1.86  &    0.13  &    0.71  &    0.03  &    0.26  &    0.02  &    0.12  &    0.03  &    1.01  &    0.04  &    0.49  &    0.05  &    0.10  &    0.07 \\
   1.00 &  2879 &    1.90  &    0.13  &    0.73  &    0.03  &    0.27  &    0.03  &    0.12  &    0.04  &    1.01  &    0.04  &    0.49  &    0.05  &    0.10  &    0.07 \\
   1.02 &  2206 &    1.93  &    0.13  &    0.74  &    0.03  &    0.28  &    0.02  &    0.13  &    0.04  &    1.02  &    0.04  &    0.50  &    0.06  &    0.10  &    0.07 \\
   1.04 &  2779 &    1.95  &    0.13  &    0.76  &    0.03  &    0.28  &    0.03  &    0.13  &    0.03  &    1.03  &    0.05  &    0.51  &    0.06  &    0.10  &    0.07 \\
   1.06 &  1929 &    1.98  &    0.13  &    0.77  &    0.03  &    0.29  &    0.02  &    0.13  &    0.04  &    1.03  &    0.04  &    0.51  &    0.05  &    0.10  &    0.07 \\
   1.08 &  2559 &    2.01  &    0.12  &    0.79  &    0.02  &    0.29  &    0.03  &    0.14  &    0.03  &    1.04  &    0.05  &    0.52  &    0.06  &    0.11  &    0.07 \\
   1.10 &  2015 &    2.04  &    0.13  &    0.80  &    0.03  &    0.30  &    0.02  &    0.14  &    0.04  &    1.04  &    0.04  &    0.53  &    0.05  &    0.11  &    0.07 \\
   1.12 &  2373 &    2.07  &    0.12  &    0.82  &    0.03  &    0.30  &    0.03  &    0.15  &    0.04  &    1.05  &    0.05  &    0.53  &    0.06  &    0.11  &    0.07 \\
   1.14 &  1775 &    2.09  &    0.12  &    0.83  &    0.03  &    0.31  &    0.02  &    0.15  &    0.03  &    1.05  &    0.04  &    0.54  &    0.06  &    0.11  &    0.07 \\
   1.16 &  2239 &    2.12  &    0.13  &    0.85  &    0.02  &    0.31  &    0.03  &    0.15  &    0.04  &    1.06  &    0.05  &    0.54  &    0.06  &    0.11  &    0.07 \\
   1.18 &  1870 &    2.14  &    0.11  &    0.86  &    0.03  &    0.32  &    0.03  &    0.16  &    0.04  &    1.06  &    0.04  &    0.55  &    0.06  &    0.11  &    0.07 \\
   1.20 &  2156 &    2.17  &    0.13  &    0.87  &    0.02  &    0.32  &    0.02  &    0.16  &    0.03  &    1.07  &    0.05  &    0.55  &    0.06  &    0.11  &    0.07 \\
   1.22 &  1550 &    2.19  &    0.12  &    0.89  &    0.02  &    0.33  &    0.03  &    0.16  &    0.04  &    1.07  &    0.04  &    0.56  &    0.05  &    0.11  &    0.07 \\
   1.24 &  2093 &    2.22  &    0.12  &    0.90  &    0.03  &    0.33  &    0.02  &    0.17  &    0.04  &    1.07  &    0.04  &    0.56  &    0.06  &    0.11  &    0.07 \\
   1.26 &  1734 &    2.24  &    0.12  &    0.92  &    0.02  &    0.34  &    0.03  &    0.17  &    0.03  &    1.08  &    0.05  &    0.57  &    0.05  &    0.12  &    0.07 \\
   1.28 &  2092 &    2.26  &    0.12  &    0.93  &    0.03  &    0.35  &    0.03  &    0.18  &    0.04  &    1.08  &    0.05  &    0.57  &    0.06  &    0.11  &    0.07 \\
   1.30 &  1520 &    2.28  &    0.11  &    0.94  &    0.02  &    0.35  &    0.03  &    0.18  &    0.04  &    1.09  &    0.04  &    0.58  &    0.05  &    0.12  &    0.06 \\
   1.32 &  1905 &    2.30  &    0.10  &    0.96  &    0.03  &    0.36  &    0.03  &    0.18  &    0.04  &    1.09  &    0.04  &    0.58  &    0.06  &    0.12  &    0.07 \\
   1.34 &  1708 &    2.33  &    0.11  &    0.97  &    0.02  &    0.36  &    0.02  &    0.19  &    0.04  &    1.09  &    0.05  &    0.59  &    0.06  &    0.12  &    0.07 \\
   1.36 &  1985 &    2.35  &    0.10  &    0.99  &    0.02  &    0.37  &    0.03  &    0.19  &    0.03  &    1.10  &    0.04  &    0.59  &    0.05  &    0.12  &    0.07 \\
   1.38 &  1490 &    2.36  &    0.10  &    1.00  &    0.03  &    0.38  &    0.02  &    0.19  &    0.04  &    1.10  &    0.04  &    0.60  &    0.05  &    0.12  &    0.07 \\
   1.40 &  1869 &    2.38  &    0.10  &    1.02  &    0.02  &    0.38  &    0.03  &    0.20  &    0.03  &    1.10  &    0.05  &    0.60  &    0.06  &    0.12  &    0.07 \\
   1.42 &  1821 &    2.40  &    0.10  &    1.03  &    0.03  &    0.39  &    0.03  &    0.20  &    0.04  &    1.11  &    0.04  &    0.60  &    0.05  &    0.13  &    0.07 \\
   1.44 &  1885 &    2.42  &    0.11  &    1.04  &    0.03  &    0.39  &    0.02  &    0.21  &    0.04  &    1.11  &    0.05  &    0.61  &    0.05  &    0.13  &    0.06 \\
   1.46 &  1531 &    2.43  &    0.10  &    1.05  &    0.02  &    0.40  &    0.03  &    0.21  &    0.04  &    1.12  &    0.04  &    0.61  &    0.06  &    0.13  &    0.07 \\
   1.48 &  1810 &    2.45  &    0.10  &    1.07  &    0.03  &    0.41  &    0.02  &    0.21  &    0.04  &    1.12  &    0.04  &    0.61  &    0.05  &    0.13  &    0.07 \\
   1.50 &  1709 &    2.46  &    0.10  &    1.08  &    0.03  &    0.41  &    0.02  &    0.22  &    0.04  &    1.13  &    0.05  &    0.62  &    0.05  &    0.13  &    0.07 \\
   1.52 &  1864 &    2.47  &    0.10  &    1.09  &    0.03  &    0.42  &    0.02  &    0.22  &    0.04  &    1.13  &    0.04  &    0.62  &    0.05  &    0.13  &    0.06 \\
   1.54 &  1508 &    2.49  &    0.10  &    1.11  &    0.03  &    0.43  &    0.03  &    0.23  &    0.04  &    1.13  &    0.04  &    0.63  &    0.05  &    0.14  &    0.07 \\
   1.56 &  1814 &    2.50  &    0.10  &    1.12  &    0.03  &    0.44  &    0.02  &    0.23  &    0.03  &    1.13  &    0.05  &    0.63  &    0.06  &    0.13  &    0.07 \\
   1.58 &  1765 &    2.51  &    0.10  &    1.13  &    0.03  &    0.44  &    0.02  &    0.24  &    0.04  &    1.14  &    0.04  &    0.63  &    0.05  &    0.13  &    0.07 \\
   1.60 &  1996 &    2.53  &    0.10  &    1.15  &    0.02  &    0.45  &    0.03  &    0.24  &    0.04  &    1.14  &    0.05  &    0.64  &    0.05  &    0.14  &    0.07 \\
   1.62 &  1598 &    2.54  &    0.10  &    1.16  &    0.02  &    0.46  &    0.03  &    0.24  &    0.04  &    1.15  &    0.04  &    0.64  &    0.05  &    0.14  &    0.07 \\
   1.64 &  1851 &    2.55  &    0.10  &    1.17  &    0.03  &    0.47  &    0.02  &    0.25  &    0.04  &    1.15  &    0.04  &    0.64  &    0.05  &    0.14  &    0.06 \\
   1.66 &  1863 &    2.56  &    0.10  &    1.18  &    0.03  &    0.47  &    0.02  &    0.25  &    0.04  &    1.15  &    0.05  &    0.65  &    0.05  &    0.14  &    0.07 \\
   1.68 &  1966 &    2.57  &    0.11  &    1.19  &    0.02  &    0.48  &    0.03  &    0.26  &    0.04  &    1.16  &    0.05  &    0.65  &    0.06  &    0.14  &    0.07 \\
   1.70 &  1718 &    2.57  &    0.10  &    1.21  &    0.02  &    0.49  &    0.03  &    0.26  &    0.04  &    1.16  &    0.04  &    0.65  &    0.04  &    0.14  &    0.07 \\
   1.72 &  1843 &    2.58  &    0.11  &    1.22  &    0.02  &    0.50  &    0.03  &    0.27  &    0.04  &    1.16  &    0.04  &    0.65  &    0.06  &    0.15  &    0.06 \\
   1.74 &  1944 &    2.59  &    0.10  &    1.23  &    0.02  &    0.51  &    0.03  &    0.27  &    0.04  &    1.17  &    0.05  &    0.65  &    0.05  &    0.15  &    0.07 \\
   1.76 &  2053 &    2.59  &    0.11  &    1.24  &    0.03  &    0.52  &    0.03  &    0.28  &    0.04  &    1.17  &    0.04  &    0.65  &    0.05  &    0.15  &    0.07 \\
   1.78 &  1848 &    2.60  &    0.11  &    1.25  &    0.03  &    0.53  &    0.03  &    0.28  &    0.04  &    1.17  &    0.04  &    0.66  &    0.05  &    0.15  &    0.06 \\
   1.80 &  1966 &    2.61  &    0.10  &    1.26  &    0.03  &    0.54  &    0.03  &    0.29  &    0.04  &    1.18  &    0.04  &    0.66  &    0.05  &    0.16  &    0.07 \\
   1.82 &  1992 &    2.61  &    0.10  &    1.27  &    0.03  &    0.55  &    0.02  &    0.29  &    0.04  &    1.18  &    0.04  &    0.66  &    0.05  &    0.15  &    0.06 \\
   1.84 &  2069 &    2.61  &    0.11  &    1.28  &    0.03  &    0.55  &    0.03  &    0.30  &    0.04  &    1.18  &    0.04  &    0.66  &    0.05  &    0.16  &    0.07 \\
   1.86 &  1894 &    2.61  &    0.11  &    1.29  &    0.03  &    0.56  &    0.02  &    0.30  &    0.04  &    1.19  &    0.05  &    0.66  &    0.05  &    0.16  &    0.06 \\
   1.88 &  2015 &    2.62  &    0.11  &    1.30  &    0.03  &    0.57  &    0.02  &    0.31  &    0.04  &    1.19  &    0.04  &    0.66  &    0.05  &    0.16  &    0.07 \\
   1.90 &  2200 &    2.61  &    0.10  &    1.31  &    0.03  &    0.59  &    0.03  &    0.32  &    0.04  &    1.19  &    0.04  &    0.67  &    0.05  &    0.16  &    0.07 \\
   1.92 &  2243 &    2.62  &    0.11  &    1.32  &    0.03  &    0.60  &    0.03  &    0.32  &    0.04  &    1.19  &    0.05  &    0.67  &    0.05  &    0.17  &    0.06 \\
   1.94 &  2002 &    2.61  &    0.11  &    1.33  &    0.03  &    0.61  &    0.03  &    0.33  &    0.04  &    1.20  &    0.04  &    0.66  &    0.05  &    0.17  &    0.06 \\
   1.96 &  2329 &    2.62  &    0.12  &    1.34  &    0.02  &    0.62  &    0.03  &    0.33  &    0.04  &    1.20  &    0.04  &    0.66  &    0.05  &    0.17  &    0.07 \\
   1.98 &  2243 &    2.62  &    0.11  &    1.34  &    0.03  &    0.63  &    0.03  &    0.34  &    0.04  &    1.21  &    0.05  &    0.67  &    0.05  &    0.18  &    0.06 \\
   2.00 &  2348 &    2.61  &    0.11  &    1.35  &    0.03  &    0.65  &    0.03  &    0.35  &    0.04  &    1.21  &    0.04  &    0.66  &    0.05  &    0.18  &    0.07 \\
   2.02 &  2568 &    2.62  &    0.12  &    1.36  &    0.03  &    0.66  &    0.03  &    0.36  &    0.04  &    1.21  &    0.05  &    0.67  &    0.05  &    0.18  &    0.06 \\
   2.04 &  2016 &    2.62  &    0.12  &    1.37  &    0.04  &    0.67  &    0.03  &    0.36  &    0.04  &    1.22  &    0.05  &    0.67  &    0.05  &    0.18  &    0.07 \\
   2.06 &  2542 &    2.61  &    0.11  &    1.37  &    0.03  &    0.68  &    0.04  &    0.37  &    0.04  &    1.22  &    0.04  &    0.66  &    0.05  &    0.18  &    0.07 \\
   2.08 &  2499 &    2.61  &    0.13  &    1.38  &    0.04  &    0.70  &    0.03  &    0.38  &    0.04  &    1.22  &    0.04  &    0.66  &    0.06  &    0.19  &    0.06 \\
   2.10 &  2798 &    2.62  &    0.12  &    1.38  &    0.04  &    0.71  &    0.04  &    0.39  &    0.04  &    1.23  &    0.04  &    0.66  &    0.06  &    0.19  &    0.06 \\
   2.12 &  2232 &    2.62  &    0.13  &    1.39  &    0.04  &    0.73  &    0.03  &    0.39  &    0.04  &    1.23  &    0.05  &    0.66  &    0.06  &    0.19  &    0.06 \\
   2.14 &  2664 &    2.61  &    0.13  &    1.39  &    0.03  &    0.74  &    0.04  &    0.40  &    0.04  &    1.23  &    0.04  &    0.66  &    0.06  &    0.19  &    0.07 \\
   2.16 &  2749 &    2.61  &    0.13  &    1.40  &    0.04  &    0.76  &    0.04  &    0.41  &    0.04  &    1.24  &    0.04  &    0.66  &    0.06  &    0.20  &    0.06 \\
   2.18 &  2962 &    2.61  &    0.13  &    1.40  &    0.04  &    0.77  &    0.04  &    0.42  &    0.04  &    1.24  &    0.05  &    0.66  &    0.07  &    0.20  &    0.06 \\
   2.20 &  2403 &    2.61  &    0.13  &    1.41  &    0.04  &    0.79  &    0.04  &    0.43  &    0.04  &    1.25  &    0.04  &    0.65  &    0.06  &    0.20  &    0.06 \\
   2.22 &  2865 &    2.60  &    0.13  &    1.41  &    0.04  &    0.81  &    0.04  &    0.43  &    0.04  &    1.25  &    0.05  &    0.65  &    0.06  &    0.20  &    0.07 \\
   2.24 &  2801 &    2.60  &    0.13  &    1.41  &    0.04  &    0.83  &    0.04  &    0.44  &    0.04  &    1.26  &    0.05  &    0.65  &    0.06  &    0.21  &    0.07 \\
   2.26 &  3046 &    2.60  &    0.13  &    1.41  &    0.04  &    0.85  &    0.04  &    0.45  &    0.04  &    1.26  &    0.04  &    0.64  &    0.07  &    0.21  &    0.06 \\
   2.28 &  2594 &    2.61  &    0.14  &    1.41  &    0.04  &    0.86  &    0.04  &    0.46  &    0.04  &    1.26  &    0.05  &    0.64  &    0.06  &    0.21  &    0.06 \\
   2.30 &  2857 &    2.60  &    0.15  &    1.41  &    0.04  &    0.88  &    0.04  &    0.47  &    0.04  &    1.27  &    0.04  &    0.64  &    0.06  &    0.21  &    0.07 \\
   2.32 &  2853 &    2.60  &    0.15  &    1.42  &    0.04  &    0.90  &    0.04  &    0.48  &    0.04  &    1.28  &    0.04  &    0.64  &    0.06  &    0.21  &    0.07 \\
   2.34 &  3217 &    2.60  &    0.16  &    1.41  &    0.04  &    0.92  &    0.04  &    0.49  &    0.04  &    1.28  &    0.05  &    0.64  &    0.06  &    0.22  &    0.07 \\
   2.36 &  2642 &    2.60  &    0.15  &    1.41  &    0.04  &    0.94  &    0.04  &    0.50  &    0.04  &    1.28  &    0.05  &    0.64  &    0.06  &    0.22  &    0.07 \\
   2.38 &  3011 &    2.60  &    0.16  &    1.41  &    0.05  &    0.96  &    0.04  &    0.51  &    0.04  &    1.29  &    0.04  &    0.64  &    0.07  &    0.22  &    0.07 \\
   2.40 &  2901 &    2.59  &    0.16  &    1.41  &    0.05  &    0.98  &    0.04  &    0.52  &    0.04  &    1.30  &    0.05  &    0.63  &    0.06  &    0.22  &    0.06 \\
   2.42 &  3290 &    2.60  &    0.16  &    1.42  &    0.05  &    1.00  &    0.05  &    0.53  &    0.04  &    1.30  &    0.04  &    0.63  &    0.06  &    0.23  &    0.06 \\
   2.44 &  2771 &    2.60  &    0.17  &    1.41  &    0.05  &    1.02  &    0.05  &    0.54  &    0.04  &    1.31  &    0.04  &    0.62  &    0.07  &    0.23  &    0.07 \\
   2.46 &  2967 &    2.60  &    0.18  &    1.41  &    0.05  &    1.04  &    0.05  &    0.55  &    0.04  &    1.31  &    0.04  &    0.62  &    0.06  &    0.23  &    0.06 \\
   2.48 &  3029 &    2.59  &    0.17  &    1.42  &    0.05  &    1.06  &    0.05  &    0.56  &    0.04  &    1.32  &    0.05  &    0.62  &    0.06  &    0.23  &    0.07 \\
   2.50 &  3278 &    2.59  &    0.17  &    1.41  &    0.05  &    1.08  &    0.05  &    0.57  &    0.04  &    1.32  &    0.05  &    0.62  &    0.06  &    0.23  &    0.06 \\
   2.52 &  2778 &    2.59  &    0.19  &    1.41  &    0.05  &    1.10  &    0.05  &    0.58  &    0.04  &    1.33  &    0.04  &    0.62  &    0.07  &    0.24  &    0.07 \\
   2.54 &  2943 &    2.60  &    0.19  &    1.41  &    0.06  &    1.12  &    0.05  &    0.59  &    0.04  &    1.33  &    0.05  &    0.62  &    0.07  &    0.24  &    0.07 \\
   2.56 &  3096 &    2.60  &    0.21  &    1.41  &    0.06  &    1.14  &    0.05  &    0.60  &    0.04  &    1.34  &    0.04  &    0.62  &    0.07  &    0.23  &    0.07 \\
   2.58 &  3135 &    2.59  &    0.21  &    1.41  &    0.06  &    1.17  &    0.05  &    0.61  &    0.04  &    1.34  &    0.05  &    0.61  &    0.07  &    0.24  &    0.06 \\
   2.60 &  2943 &    2.58  &    0.20  &    1.41  &    0.05  &    1.18  &    0.06  &    0.62  &    0.04  &    1.35  &    0.05  &    0.61  &    0.06  &    0.24  &    0.06 \\
   2.62 &  2853 &    2.60  &    0.21  &    1.41  &    0.05  &    1.20  &    0.05  &    0.63  &    0.04  &    1.36  &    0.05  &    0.61  &    0.06  &    0.24  &    0.06 \\
   2.64 &  2881 &    2.59  &    0.22  &    1.41  &    0.05  &    1.22  &    0.06  &    0.64  &    0.04  &    1.36  &    0.05  &    0.61  &    0.06  &    0.24  &    0.07 \\
   2.66 &  2850 &    2.58  &    0.22  &    1.42  &    0.05  &    1.24  &    0.06  &    0.65  &    0.04  &    1.37  &    0.05  &    0.61  &    0.06  &    0.24  &    0.07 \\
   2.68 &  2716 &    2.60  &    0.24  &    1.42  &    0.05  &    1.26  &    0.05  &    0.66  &    0.04  &    1.37  &    0.05  &    0.60  &    0.07  &    0.25  &    0.07 \\
   2.70 &  2747 &    2.61  &    0.24  &    1.42  &    0.05  &    1.28  &    0.05  &    0.67  &    0.04  &    1.38  &    0.04  &    0.60  &    0.07  &    0.25  &    0.07 \\
   2.72 &  2631 &    2.60  &    0.24  &    1.42  &    0.06  &    1.30  &    0.05  &    0.68  &    0.04  &    1.39  &    0.05  &    0.60  &    0.07  &    0.25  &    0.06 \\
   2.74 &  2480 &    2.61  &    0.27  &    1.42  &    0.06  &    1.31  &    0.05  &    0.69  &    0.04  &    1.40  &    0.05  &    0.60  &    0.06  &    0.25  &    0.07 \\
   2.76 &  2380 &    2.61  &    0.27  &    1.42  &    0.05  &    1.33  &    0.05  &    0.70  &    0.04  &    1.40  &    0.04  &    0.60  &    0.07  &    0.25  &    0.07 \\
   2.78 &  2254 &    2.61  &    0.27  &    1.43  &    0.05  &    1.35  &    0.05  &    0.71  &    0.04  &    1.41  &    0.05  &    0.60  &    0.06  &    0.25  &    0.07 \\
   2.80 &  2036 &    2.62  &    0.28  &    1.43  &    0.06  &    1.36  &    0.05  &    0.72  &    0.04  &    1.41  &    0.05  &    0.60  &    0.06  &    0.26  &    0.06 \\
   2.82 &  1979 &    2.61  &    0.30  &    1.43  &    0.05  &    1.38  &    0.06  &    0.73  &    0.04  &    1.42  &    0.05  &    0.60  &    0.06  &    0.26  &    0.07 \\
   2.84 &  1864 &    2.61  &    0.30  &    1.44  &    0.05  &    1.40  &    0.06  &    0.74  &    0.04  &    1.42  &    0.05  &    0.60  &    0.06  &    0.26  &    0.06 \\
   2.86 &  1634 &    2.62  &    0.30  &    1.44  &    0.06  &    1.42  &    0.05  &    0.75  &    0.04  &    1.43  &    0.05  &    0.60  &    0.06  &    0.26  &    0.07 \\
   2.88 &  1540 &    2.60  &    0.33  &    1.45  &    0.05  &    1.43  &    0.06  &    0.76  &    0.04  &    1.44  &    0.05  &    0.60  &    0.06  &    0.26  &    0.07 \\
   2.90 &  1502 &    2.61  &    0.35  &    1.45  &    0.06  &    1.45  &    0.05  &    0.77  &    0.05  &    1.44  &    0.05  &    0.60  &    0.06  &    0.27  &    0.06 \\
   2.92 &  1303 &    2.65  &    0.33  &    1.45  &    0.05  &    1.46  &    0.05  &    0.78  &    0.04  &    1.45  &    0.06  &    0.60  &    0.06  &    0.27  &    0.07 \\
   2.94 &  1167 &    2.62  &    0.33  &    1.46  &    0.05  &    1.48  &    0.05  &    0.78  &    0.05  &    1.46  &    0.05  &    0.60  &    0.07  &    0.28  &    0.07 \\
   2.96 &  1138 &    2.61  &    0.37  &    1.46  &    0.04  &    1.50  &    0.05  &    0.80  &    0.05  &    1.46  &    0.05  &    0.60  &    0.06  &    0.27  &    0.06 \\
   2.98 &   927 &    2.63  &    0.38  &    1.47  &    0.05  &    1.51  &    0.05  &    0.80  &    0.04  &    1.47  &    0.05  &    0.60  &    0.07  &    0.27  &    0.07 \\
   3.00 &   976 &    2.61  &    0.43  &    1.47  &    0.05  &    1.52  &    0.04  &    0.81  &    0.05  &    1.48  &    0.05  &    0.59  &    0.07  &    0.27  &    0.07 \\
   3.02 &   888 &    2.61  &    0.36  &    1.48  &    0.04  &    1.54  &    0.05  &    0.82  &    0.05  &    1.48  &    0.05  &    0.59  &    0.06  &    0.28  &    0.06 \\
   3.04 &   811 &    2.63  &    0.40  &    1.48  &    0.04  &    1.56  &    0.05  &    0.83  &    0.04  &    1.49  &    0.06  &    0.60  &    0.06  &    0.28  &    0.07 \\
   3.06 &   686 &    2.64  &    0.41  &    1.49  &    0.04  &    1.56  &    0.05  &    0.83  &    0.04  &    1.50  &    0.06  &    0.59  &    0.06  &    0.28  &    0.07 \\
   3.08 &   649 &    2.62  &    0.40  &    1.49  &    0.04  &    1.58  &    0.04  &    0.85  &    0.04  &    1.50  &    0.06  &    0.60  &    0.06  &    0.28  &    0.07 \\
   3.10 &   611 &    2.62  &    0.42  &    1.50  &    0.04  &    1.60  &    0.05  &    0.85  &    0.05  &    1.51  &    0.05  &    0.59  &    0.06  &    0.28  &    0.06 \\
   3.12 &   553 &    2.64  &    0.42  &    1.50  &    0.04  &    1.62  &    0.05  &    0.87  &    0.05  &    1.52  &    0.06  &    0.59  &    0.07  &    0.28  &    0.07 \\
   3.14 &   525 &    2.64  &    0.50  &    1.51  &    0.04  &    1.63  &    0.04  &    0.88  &    0.05  &    1.52  &    0.06  &    0.59  &    0.06  &    0.28  &    0.06 \\
   3.16 &   439 &    2.57  &    0.44  &    1.51  &    0.04  &    1.65  &    0.05  &    0.88  &    0.04  &    1.52  &    0.05  &    0.59  &    0.06  &    0.29  &    0.06 \\
   3.18 &   405 &    2.65  &    0.53  &    1.52  &    0.05  &    1.66  &    0.05  &    0.89  &    0.04  &    1.53  &    0.05  &    0.59  &    0.06  &    0.28  &    0.07 \\
   3.20 &   391 &    2.64  &    0.47  &    1.52  &    0.04  &    1.68  &    0.04  &    0.90  &    0.04  &    1.54  &    0.06  &    0.59  &    0.06  &    0.30  &    0.07 \\
   3.22 &   391 &    2.60  &    0.50  &    1.52  &    0.04  &    1.69  &    0.04  &    0.91  &    0.05  &    1.55  &    0.07  &    0.60  &    0.06  &    0.29  &    0.06 \\
   3.24 &   334 &    2.60  &    0.49  &    1.53  &    0.04  &    1.70  &    0.05  &    0.92  &    0.04  &    1.55  &    0.07  &    0.59  &    0.06  &    0.30  &    0.07 \\
   3.26 &   268 &    2.63  &    0.52  &    1.54  &    0.05  &    1.71  &    0.05  &    0.92  &    0.05  &    1.55  &    0.06  &    0.59  &    0.07  &    0.30  &    0.06 \\
   3.28 &   267 &    2.65  &    0.53  &    1.54  &    0.04  &    1.73  &    0.04  &    0.95  &    0.05  &    1.57  &    0.06  &    0.58  &    0.06  &    0.31  &    0.07 \\
   3.30 &   242 &    2.60  &    0.57  &    1.54  &    0.04  &    1.75  &    0.04  &    0.94  &    0.05  &    1.57  &    0.06  &    0.60  &    0.05  &    0.30  &    0.07 \\
   3.32 &   244 &    2.66  &    0.60  &    1.54  &    0.04  &    1.77  &    0.05  &    0.95  &    0.06  &    1.58  &    0.06  &    0.60  &    0.06  &    0.31  &    0.07 \\
   3.34 &   197 &    2.68  &    0.61  &    1.55  &    0.05  &    1.79  &    0.05  &    0.97  &    0.05  &    1.58  &    0.06  &    0.59  &    0.07  &    0.30  &    0.07 \\
   3.36 &   195 &    2.51  &    0.61  &    1.56  &    0.04  &    1.80  &    0.04  &    0.96  &    0.05  &    1.59  &    0.06  &    0.59  &    0.06  &    0.31  &    0.06 \\
   3.38 &   177 &    2.61  &    0.58  &    1.56  &    0.05  &    1.81  &    0.06  &    0.98  &    0.04  &    1.60  &    0.05  &    0.59  &    0.07  &    0.31  &    0.07 \\
   3.40 &   158 &    2.50  &    0.63  &    1.56  &    0.04  &    1.83  &    0.04  &    0.99  &    0.05  &    1.60  &    0.06  &    0.60  &    0.05  &    0.32  &    0.06 \\
   3.42 &   108 &    2.43  &    0.86  &    1.56  &    0.05  &    1.85  &    0.05  &    1.00  &    0.04  &    1.62  &    0.07  &    0.60  &    0.06  &    0.32  &    0.07 \\
   3.44 &   130 &    2.52  &    0.58  &    1.57  &    0.05  &    1.86  &    0.05  &    1.01  &    0.06  &    1.62  &    0.06  &    0.59  &    0.07  &    0.32  &    0.06 \\
   3.46 &   114 &    2.58  &    0.77  &    1.57  &    0.05  &    1.88  &    0.05  &    1.03  &    0.05  &    1.63  &    0.05  &    0.60  &    0.06  &    0.32  &    0.07 \\
   3.48 &   110 &    2.50  &    0.79  &    1.58  &    0.05  &    1.90  &    0.05  &    1.02  &    0.06  &    1.62  &    0.07  &    0.58  &    0.05  &    0.33  &    0.07 \\
   3.50 &    91 &    2.46  &    0.67  &    1.58  &    0.03  &    1.92  &    0.04  &    1.03  &    0.06  &    1.63  &    0.04  &    0.60  &    0.06  &    0.33  &    0.06 \\
   3.52 &    99 &    2.53  &    0.58  &    1.59  &    0.05  &    1.93  &    0.06  &    1.04  &    0.07  &    1.64  &    0.06  &    0.60  &    0.05  &    0.32  &    0.06 \\
   3.54 &    83 &    2.41  &    0.61  &    1.59  &    0.05  &    1.94  &    0.05  &    1.06  &    0.04  &    1.67  &    0.06  &    0.59  &    0.05  &    0.33  &    0.06 \\
   3.56 &   104 &    2.43  &    0.61  &    1.59  &    0.06  &    1.97  &    0.05  &    1.07  &    0.06  &    1.66  &    0.05  &    0.59  &    0.07  &    0.33  &    0.07 \\
   3.58 &    66 &    2.58  &    0.64  &    1.59  &    0.04  &    1.99  &    0.04  &    1.06  &    0.05  &    1.67  &    0.07  &    0.58  &    0.06  &    0.34  &    0.05 \\
   3.60 &    67 &    2.47  &    0.56  &    1.59  &    0.04  &    2.01  &    0.04  &    1.08  &    0.05  &    1.66  &    0.07  &    0.60  &    0.07  &    0.33  &    0.06 \\
   3.62 &    57 &    2.73  &    0.76  &    1.59  &    0.05  &    2.03  &    0.05  &    1.09  &    0.05  &    1.69  &    0.07  &    0.60  &    0.07  &    0.34  &    0.07 \\
   3.64 &    45 &    2.43  &    0.76  &    1.59  &    0.06  &    2.04  &    0.06  &    1.10  &    0.06  &    1.67  &    0.09  &    0.59  &    0.07  &    0.33  &    0.06 \\
   3.66 &    52 &    2.43  &    0.67  &    1.61  &    0.04  &    2.05  &    0.04  &    1.10  &    0.04  &    1.69  &    0.07  &    0.58  &    0.05  &    0.35  &    0.05 \\
   3.68 &    48 &    2.39  &    0.81  &    1.60  &    0.07  &    2.07  &    0.06  &    1.13  &    0.04  &    1.68  &    0.06  &    0.61  &    0.06  &    0.31  &    0.06 \\
   3.70 &    42 &    2.20  &    0.79  &    1.62  &    0.07  &    2.08  &    0.07  &    1.13  &    0.07  &    1.72  &    0.08  &    0.60  &    0.05  &    0.33  &    0.07 \\
   3.72 &    44 &    2.27  &    0.87  &    1.60  &    0.08  &    2.11  &    0.07  &    1.16  &    0.05  &    1.71  &    0.05  &    0.60  &    0.05  &    0.30  &    0.08 \\
   3.74 &    38 &    2.46  &    0.65  &    1.61  &    0.05  &    2.12  &    0.06  &    1.14  &    0.05  &    1.70  &    0.05  &    0.61  &    0.04  &    0.34  &    0.04 \\
   3.76 &    39 &    2.27  &    0.75  &    1.61  &    0.08  &    2.15  &    0.07  &    1.17  &    0.05  &    1.70  &    0.07  &    0.60  &    0.07  &    0.35  &    0.07 \\
   3.78 &    27 &    2.55  &    0.72  &    1.61  &    0.04  &    2.16  &    0.04  &    1.18  &    0.05  &    1.70  &    0.09  &    0.63  &    0.07  &    0.35  &    0.06 \\
   3.80 &    34 &    2.47  &    0.69  &    1.59  &    0.09  &    2.20  &    0.10  &    1.18  &    0.07  &    1.73  &    0.08  &    0.61  &    0.07  &    0.33  &    0.06 \\
   3.82 &    23 &    2.33  &    0.41  &    1.61  &    0.04  &    2.20  &    0.03  &    1.19  &    0.05  &    1.76  &    0.06  &    0.64  &    0.04  &    0.32  &    0.06 \\
   3.84 &    39 &    2.46  &    0.90  &    1.65  &    0.07  &    2.19  &    0.06  &    1.22  &    0.05  &    1.77  &    0.05  &    0.62  &    0.05  &    0.34  &    0.07 \\
   3.86 &    27 &    1.99  &    0.53  &    1.62  &    0.07  &    2.24  &    0.05  &    1.24  &    0.05  &    1.80  &    0.06  &    0.62  &    0.05  &    0.37  &    0.07 \\
   3.88 &    19 &    2.52  &    0.84  &    1.64  &    0.06  &    2.23  &    0.06  &    1.25  &    0.05  &    1.78  &    0.04  &    0.60  &    0.06  &    0.34  &    0.06 \\
   3.90 &    19 &    2.37  &    0.78  &    1.62  &    0.07  &    2.27  &    0.09  &    1.22  &    0.03  &    1.81  &    0.05  &    0.63  &    0.07  &    0.34  &    0.07 \\
   3.92 &    16 &    1.75  &    0.66  &    1.62  &    0.09  &    2.29  &    0.07  &    1.27  &    0.06  &    1.84  &    0.10  &    0.59  &    0.05  &    0.34  &    0.04 \\
   3.94 &    15 &    2.66  &    0.85  &    1.67  &    0.10  &    2.26  &    0.08  &    1.26  &    0.07  &    1.78  &    0.04  &    0.63  &    0.04  &    0.39  &    0.09 \\
   3.96 &    13 &    2.30  &    1.07  &    1.66  &    0.07  &    2.29  &    0.07  &    1.24  &    0.07  &    1.85  &    0.10  &    0.67  &    0.04  &    0.31  &    0.05 \\
   3.98 &    17 &    2.09  &    0.53  &    1.64  &    0.10  &    2.33  &    0.10  &    1.27  &    0.04  &    1.81  &    0.05  &    0.59  &    0.10  &    0.36  &    0.13 \\
   4.00 &    17 &    2.14  &    1.10  &    1.66  &    0.03  &    2.34  &    0.03  &    1.28  &    0.06  &    1.84  &    0.04  &    0.63  &    0.08  &    0.38  &    0.07 \\
   4.02 &    16 &    2.28  &    0.66  &    1.62  &    0.06  &    2.39  &    0.07  &    1.30  &    0.07  &    1.80  &    0.09  &    0.62  &    0.06  &    0.36  &    0.07 \\
   4.04 &    20 &    2.08  &    0.90  &    1.66  &    0.10  &    2.38  &    0.09  &    1.28  &    0.07  &    1.85  &    0.07  &    0.59  &    0.04  &    0.37  &    0.07 \\
   4.06 &    15 &    1.84  &    0.66  &    1.56  &    0.12  &    2.49  &    0.13  &    1.36  &    0.10  &    1.84  &    0.09  &    0.60  &    0.07  &    0.37  &    0.04 \\
   4.08 &    12 &    2.07  &    0.66  &    1.65  &    0.07  &    2.43  &    0.09  &    1.30  &    0.09  &    1.86  &    0.05  &    0.60  &    0.04  &    0.38  &    0.06 \\
   4.10 &    16 &    2.23  &    1.11  &    1.61  &    0.07  &    2.49  &    0.07  &    1.32  &    0.10  &    1.88  &    0.10  &    0.62  &    0.07  &    0.33  &    0.06 \\
   4.12 &    13 &    1.82  &    0.70  &    1.63  &    0.08  &    2.48  &    0.07  &    1.36  &    0.05  &    1.90  &    0.10  &    0.61  &    0.05  &    0.38  &    0.05 \\
   4.14 &     8 &    2.31  &    1.31  &    1.64  &    0.18  &    2.49  &    0.12  &    1.34  &    0.05  &    1.86  &    0.06  &    0.62  &    0.12  &    0.34  &    0.02 \\
   4.16 &    15 &    1.81  &    0.72  &    1.67  &    0.10  &    2.48  &    0.07  &    1.37  &    0.03  &    1.93  &    0.06  &    0.62  &    0.06  &    0.37  &    0.05 \\
   4.18 &     8 &    2.14  &    0.58  &    1.69  &    0.13  &    2.49  &    0.10  &    1.37  &    0.05  &    1.97  &    0.10  &    0.60  &    0.05  &    0.39  &    0.05 \\
   4.24 &    29 &    2.09  &    0.95  &    1.66  &    0.12  &    2.54  &    0.10  &    1.39  &    0.10  &    1.92  &    0.11  &    0.63  &    0.06  &    0.36  &    0.08 \\
   4.34 &    32 &    1.33  &    0.75  &    1.68  &    0.13  &    2.64  &    0.14  &    1.48  &    0.13  &    1.96  &    0.15  &    0.63  &    0.07  &    0.42  &    0.05 \\
\enddata
\end{deluxetable}

%% file: tab3.tex
\begin{deluxetable}{lccccccccc}
\tablewidth{0pt}
\tabletypesize{\scriptsize}
\tablecaption{Synthetic SDSS/2MASS Photometry of Pickles (1998) Solar Metallicity Standards  \label{tab:solarstars}}
\tablehead{
  \colhead{Spec.} &
  \colhead{Lum.} &
  \colhead{$u-g$} &
  \colhead{$g-r$} &
  \colhead{$r-i$} &
  \colhead{$i-z$} &
  \colhead{$z-J$} &
  \colhead{$J-H$} &
  \colhead{$H-K_s$} &
\colhead{M$_J$} \\
  \colhead{Type} &
  \colhead{Class} &
  \colhead{} &
  \colhead{} &
  \colhead{} &
  \colhead{} &
  \colhead{} &
  \colhead{} &
  \colhead{} &
\colhead{} }
\startdata
O5 &   V &   -0.39  &   -0.62  &   -0.37  &   -0.35  &    0.20  &   -0.23  &   -0.13  &   -4.50 \\
O9 &   V &   -0.32  &   -0.62  &   -0.35  &   -0.36  &    0.24  &   -0.16  &   -0.11  &   -3.81 \\
B0 &   V &   -0.25  &   -0.59  &   -0.35  &   -0.32  &    0.19  &   -0.11  &   -0.04  &   -2.80 \\
B1 &   V &   -0.13  &   -0.49  &   -0.33  &   -0.26  &    0.13  &   -0.06  &   -0.03  &   -2.03 \\
B3 &   V &    0.18  &   -0.46  &   -0.26  &   -0.26  &    0.20  &   -0.22  &   -0.08  &   -1.55 \\
B5/7 &   V &    0.51  &   -0.39  &   -0.23  &   -0.23  &    0.27  &   -0.01  &   -0.02  &   -1.37 \\
B8 &   V &    0.75  &   -0.33  &   -0.23  &   -0.17  &    0.28  &    0.02  &    0.02  &   -1.21 \\
B9 &   V &    0.86  &   -0.28  &   -0.23  &   -0.18  &    0.36  &    0.03  &   -0.01  &   -0.53 \\
A0 &   V &    1.09  &   -0.25  &   -0.18  &   -0.17  &    0.46  &    0.02  &   -0.03  &    0.43 \\
A2 &   V &    1.16  &   -0.23  &   -0.17  &   -0.15  &    0.38  &   -0.10  &   -0.01  &    1.17 \\
A3 &   V &    1.15  &   -0.16  &   -0.15  &   -0.15  &    0.39  &   -0.14  &   -0.02  &    1.25 \\
A5 &   V &    1.20  &   -0.10  &   -0.11  &   -0.10  &    0.55  &    0.09  &    0.01  &    1.38 \\
A7 &   V &    1.21  &   -0.02  &   -0.08  &   -0.08  &    0.53  &    0.13  &    0.03  &    1.73 \\
F0 &   V &    1.16  &    0.10  &   -0.01  &   -0.09  &    0.59  &    0.16  &    0.02  &    2.43 \\
F2 &   V &    1.12  &    0.19  &    0.03  &   -0.02  &    0.52  &    0.04  &   -0.00  &    2.63 \\
F5 &   V &    1.09  &    0.26  &    0.03  &   -0.02  &    0.71  &    0.22  &    0.00  &    2.62 \\
F6 &   V &    1.14  &    0.28  &    0.08  &   -0.01  &    0.72  &    0.23  &    0.01  &    2.90 \\
F8 &   V &    1.22  &    0.36  &    0.10  &    0.03  &    0.74  &    0.23  &    0.00  &    2.98 \\
G0 &   V &    1.30  &    0.38  &    0.14  &    0.02  &    0.71  &    0.28  &    0.00  &    3.18 \\
G2 &   V &    1.37  &    0.45  &    0.16  &    0.04  &    0.84  &    0.35  &    0.00  &    3.45 \\
G5 &   V &    1.49  &    0.49  &    0.16  &    0.06  &    0.78  &    0.29  &    0.03  &    3.54 \\
G8 &   V &    1.60  &    0.57  &    0.19  &    0.06  &    0.87  &    0.37  &    0.04  &    3.85 \\
K0 &   V &    1.68  &    0.62  &    0.22  &    0.07  &    0.84  &    0.47  &    0.07  &    4.17 \\
K2 &   V &    1.95  &    0.78  &    0.24  &    0.11  &    0.99  &    0.51  &    0.07  &    4.50 \\
K3 &   V &    2.10  &    0.85  &    0.32  &    0.13  &    0.99  &    0.58  &    0.01  &    4.94 \\
K4 &   V &    2.21  &    1.00  &    0.38  &    0.16  &    1.01  &    0.59  &    0.10  &    5.21 \\
K5 &   V &    2.54  &    1.18  &    0.40  &    0.20  &    1.06  &    0.62  &    0.10  &    5.45 \\
K7 &   V &    2.56  &    1.34  &    0.54  &    0.30  &    1.01  &    0.60  &    0.11  &    5.77 \\
M0 &   V &    2.65  &    1.31  &    0.64  &    0.38  &    1.29  &    0.68  &    0.15  &    5.72 \\
M1 &   V &    2.72  &    1.33  &    0.78  &    0.46  &    1.20  &    0.67  &    0.25  &    6.04 \\
M2 &   V &    2.59  &    1.43  &    0.85  &    0.52  &    1.33  &    0.69  &    0.15  &    6.23 \\
M2.5 &   V &    2.74  &    1.47  &    1.01  &    0.63  &    1.32  &    0.66  &    0.18  &    6.51 \\
M3 &   V &    2.67  &    1.46  &    1.20  &    0.67  &    1.37  &    0.65  &    0.20  &    6.77 \\
M4 &   V &    3.13  &    1.48  &    1.51  &    0.76  &    1.56  &    0.63  &    0.23  &    7.06 \\
M5 &   V &    3.05  &    1.59  &    1.73  &    0.94  &    1.96  &    0.86  &    0.24  &    7.39 \\
M6 &   V &    2.99  &    1.72  &    2.12  &    1.24  &    2.31  &    0.66  &    0.32  &    7.72 \\
O8 & III &   -0.35  &   -0.58  &   -0.38  &   -0.36  &    0.12  &    0.04  &   -0.07  &   -4.67 \\
B1/2 & III &   -0.14  &   -0.53  &   -0.32  &   -0.28  &    0.27  &   -0.08  &   -0.13  &   -3.65 \\
B3 & III &    0.18  &   -0.48  &   -0.25  &   -0.25  &    0.17  &    0.04  &    0.00  &   -2.71 \\
B5 & III &    0.42  &   -0.42  &   -0.23  &   -0.24  &    0.29  &   -0.06  &   -0.01  &   -2.16 \\
B9 & III &    0.93  &   -0.33  &   -0.23  &   -0.17  &    0.39  &   -0.05  &    0.01  &   -0.76 \\
A0 & III &    1.07  &   -0.22  &   -0.20  &   -0.16  &    0.42  &    0.04  &    0.00  &   -0.08 \\
A3 & III &    1.24  &   -0.12  &   -0.17  &   -0.15  &    0.45  &   -0.11  &   -0.05  &    0.10 \\
A5 & III &    1.23  &   -0.08  &   -0.10  &   -0.10  &    0.53  &    0.13  &   -0.01  &    0.17 \\
A7 & III &    1.27  &   -0.01  &   -0.09  &   -0.08  &    0.56  &    0.00  &   -0.02  &    0.51 \\
F0 & III &    1.27  &    0.05  &   -0.05  &   -0.02  &    0.54  &    0.23  &    0.00  &    0.56 \\
F2 & III &    1.18  &    0.23  &    0.01  &   -0.03  &    0.62  &    0.13  &   -0.01  &    0.85 \\
F5 & III &    1.21  &    0.24  &    0.05  &   -0.02  &    0.67  &    0.33  &    0.00  &    0.83 \\
G0 & III &    1.49  &    0.48  &    0.15  &    0.05  &    0.92  &    0.46  &   -0.00  &    0.78 \\
G5 & III &    1.86  &    0.68  &    0.22  &    0.08  &    0.99  &    0.48  &    0.03  &    1.82 \\
G8 & III &    2.02  &    0.75  &    0.24  &    0.11  &    0.95  &    0.51  &    0.04  &    1.72 \\
K0 & III &    2.20  &    0.75  &    0.27  &    0.12  &    1.01  &    0.53  &    0.05  &    1.67 \\
K1 & III &    2.34  &    0.85  &    0.29  &    0.15  &    1.05  &    0.54  &    0.06  &    0.31 \\
K2 & III &    2.48  &    0.93  &    0.34  &    0.18  &    1.10  &    0.67  &    0.04  &   -0.63 \\
K3 & III &    2.71  &    1.02  &    0.35  &    0.18  &    1.31  &    0.74  &    0.16  &   -1.13 \\
K4 & III &    3.10  &    1.23  &    0.43  &    0.27  &    1.28  &    0.81  &    0.16  &   -2.17 \\
K5 & III &    3.23  &    1.27  &    0.56  &    0.34  &    1.29  &    0.75  &    0.17  &   -3.01 \\
M0 & III &    3.17  &    1.28  &    0.65  &    0.38  &    1.29  &    0.90  &    0.13  &   -3.49 \\
M1 & III &    3.16  &    1.29  &    0.73  &    0.43  &    1.32  &    0.92  &    0.12  &   -3.71 \\
M2 & III &    3.19  &    1.32  &    0.82  &    0.47  &    1.35  &    0.94  &    0.18  &   -3.92 \\
M3 & III &    3.13  &    1.27  &    1.01  &    0.58  &    1.38  &    0.95  &    0.17  &   -4.15 \\
M4 & III &    3.01  &    1.22  &    1.29  &    0.73  &    1.49  &    0.95  &    0.20  &   -4.51 \\
M5 & III &    2.68  &    1.13  &    1.64  &    0.93  &    1.68  &    0.99  &    0.22  &   -4.75 \\
M6 & III &    2.43  &    1.11  &    1.99  &    1.18  &    1.94  &    1.05  &    0.22  &   -5.28 \\
M7 & III &    1.82  &    1.15  &    2.36  &    1.58  &    2.29  &    1.08  &    0.22  &   -5.64 \\
M8 & III &    1.33  &    1.33  &    2.45  &    2.03  &    2.93  &    1.07  &    0.26  &   -6.13  \\
M9 & III &    1.24  &    1.16  &    2.37  &    2.01  &    3.17 &     0.97 &     0.3   &   -6.6 \\
M10 & III &    1.12  &    0.89  &    2.58  &    2.22  &   3.22  &    0.83  &    0.4  &    -7.04\\
B0 &   I &   -0.23  &   -0.50  &   -0.28  &   -0.31  &    0.22  &   -0.02  &   -0.13  &   -6.59 \\
B1 &   I &   -0.14  &   -0.42  &   -0.26  &   -0.34  &    0.27  &   -0.09  &   -0.09  &   -6.23 \\
B3 &   I &    0.08  &   -0.36  &   -0.27  &   -0.28  &    0.37  &    0.00  &    0.01  &   -6.28 \\
B5 &   I &    0.17  &   -0.30  &   -0.18  &   -0.19  &    0.32  &    0.04  &   -0.01  &   -6.19 \\
B8 &   I &    0.37  &   -0.27  &   -0.17  &   -0.13  &    0.37  &   -0.03  &    0.03  &   -6.64 \\
A0 &   I &    0.81  &   -0.23  &   -0.16  &   -0.10  &    0.44  &    0.12  &   -0.02  &   -6.62 \\
A2 &   I &    0.83  &   -0.11  &   -0.14  &   -0.13  &    0.51  &    0.15  &   -0.02  &   -6.87 \\
F0 &   I &    1.47  &   -0.02  &   -0.03  &   -0.03  &    0.50  &    0.15  &   -0.01  &   -7.17 \\
F5 &   I &    1.43  &    0.06  &   -0.00  &   -0.03  &    0.65  &    0.13  &   -0.02  &   -7.37 \\
F8 &   I &    1.72  &    0.37  &    0.03  &   -0.03  &    0.73  &    0.30  &    0.06  &   -7.56 \\
G0 &   I &    1.73  &    0.58  &    0.10  &   -0.02  &    0.74  &    0.35  &    0.08  &   -7.67 \\
G2 &   I &    1.96  &    0.68  &    0.14  &   -0.01  &    0.79  &    0.34  &   -0.01  &   -7.68 \\
G5 &   I &    2.22  &    0.82  &    0.17  &    0.03  &    0.84  &    0.42  &    0.09  &   -7.74 \\
G8 &   I &    2.65  &    1.00  &    0.27  &    0.04  &    0.78  &    0.33  &    0.03  &   -7.82 \\
K2 &   I &    3.21  &    1.27  &    0.40  &    0.17  &    0.82  &    0.37  &    0.08  &   -8.06 \\
K3 &   I &    3.30  &    1.38  &    0.47  &    0.17  &    1.03  &    0.58  &    0.08  &   -8.17 \\
K4 &   I &    3.30  &    1.47  &    0.60  &    0.38  &    1.23  &    0.74  &    0.21  &   -8.40 \\
M2 &   I &    3.51  &    1.57  &    1.05  &    0.55  &    1.10  &    0.69  &    0.34  &  -10.04 \\
\enddata
\end{deluxetable}

%% file: ms.bbl
\begin{thebibliography}{63}
\expandafter\ifx\csname natexlab\endcsname\relax\def\natexlab#1{#1}\fi

\bibitem[{{Abazajian} {et~al.}(2004){Abazajian}, {Adelman-McCarthy},
  {Ag{\"u}eros}, {Allam}, {Anderson}, {Anderson}, {Annis}, {Bahcall}, {Baldry},
  {Bastian}, {Berlind}, {Bernardi}, {Blanton}, {Bochanski}, {Boroski},
  {Briggs}, {Brinkmann}, {Brunner}, {Budav{\'a}ri}, {Carey}, {Carliles},
  {Castander}, {Connolly}, {Csabai}, {Doi}, {Dong}, {Eisenstein}, {Evans},
  {Fan}, {Finkbeiner}, {Friedman}, {Frieman}, {Fukugita}, {Gal}, {Gillespie},
  {Glazebrook}, {Gray}, {Grebel}, {Gunn}, {Gurbani}, {Hall}, {Hamabe},
  {Harris}, {Harris}, {Harvanek}, {Heckman}, {Hendry}, {Hennessy}, {Hindsley},
  {Hogan}, {Hogg}, {Holmgren}, {Ichikawa}, {Ichikawa}, {Ivezi{\'c}}, {Jester},
  {Johnston}, {Jorgensen}, {Kent}, {Kleinman}, {Knapp}, {Kniazev}, {Kron},
  {Krzesinski}, {Kunszt}, {Kuropatkin}, {Lamb}, {Lampeitl}, {Lee}, {Leger},
  {Li}, {Lin}, {Loh}, {Long}, {Loveday}, {Lupton}, {Malik}, {Margon},
  {Matsubara}, {McGehee}, {McKay}, {Meiksin}, {Munn}, {Nakajima}, {Nash},
  {Neilsen}, {Newberg}, {Newman}, {Nichol}, {Nicinski}, {Nieto-Santisteban},
  {Nitta}, {Okamura}, {O'Mullane}, {Ostriker}, {Owen}, {Padmanabhan},
  {Peoples}, {Pier}, {Pope}, {Quinn}, {Richards}, {Richmond}, {Rix}, {Rockosi},
  {Schlegel}, {Schneider}, {Scranton}, {Sekiguchi}, {Seljak}, {Sergey},
  {Sesar}, {Sheldon}, {Shimasaku}, {Siegmund}, {Silvestri}, {Smith}, {Smol{\v
  c}i{\'c}}, {Snedden}, {Stebbins}, {Stoughton}, {Strauss}, {SubbaRao},
  {Szalay}, {Szapudi}, {Szkody}, {Szokoly}, {Tegmark}, {Teodoro}, {Thakar},
  {Tremonti}, {Tucker}, {Uomoto}, {Vanden Berk}, {Vandenberg}, {Vogeley},
  {Voges}, {Vogt}, {Walkowicz}, {Wang}, {Weinberg}, {West}, {White}, {Wilhite},
  {Xu}, {Yanny}, {Yasuda}, {Yip}, {Yocum}, {York}, {Zehavi}, {Zibetti}, \&
  {Zucker}}]{Abazajian2004}
{Abazajian}, K. {et~al.} 2004, \aj, 128, 502

\bibitem[{{Adelman-McCarthy} {et~al.}(2007){Adelman-McCarthy}, {Ag{\"u}eros},
  {Allam}, {Anderson}, {Anderson}, {Annis}, {Bahcall}, {Baldry}, {Barentine},
  {Berlind}, {Bernardi}, {Blanton}, {Boroski}, {Brewington}, {Brinchmann},
  {Brinkmann}, {Brunner}, {Budav{\'a}ri}, {Carey}, {Carr}, {Castander},
  {Connolly}, {Csabai}, {Czarapata}, {Dalcanton}, {Doi}, {Dong}, {Eisenstein},
  {Evans}, {Fan}, {Finkbeiner}, {Friedman}, {Frieman}, {Fukugita}, {Gillespie},
  {Glazebrook}, {Gray}, {Grebel}, {Gunn}, {Gurbani}, {de Haas}, {Hall},
  {Harris}, {Harvanek}, {Hawley}, {Hayes}, {Hendry}, {Hennessy}, {Hindsley},
  {Hirata}, {Hogan}, {Hogg}, {Holmgren}, {Holtzman}, {Ichikawa}, {Ivezi{\'c}},
  {Jester}, {Johnston}, {Jorgensen}, {Juri{\'c}}, {Kent}, {Kleinman}, {Knapp},
  {Kniazev}, {Kron}, {Krzesinski}, {Kuropatkin}, {Lamb}, {Lampeitl}, {Lee},
  {Leger}, {Lin}, {Long}, {Loveday}, {Lupton}, {Margon},
  {Mart{\'{\i}}nez-Delgado}, {Mandelbaum}, {Matsubara}, {McGehee}, {McKay},
  {Meiksin}, {Munn}, {Nakajima}, {Nash}, {Neilsen}, {Newberg}, {Newman},
  {Nichol}, {Nicinski}, {Nieto-Santisteban}, {Nitta}, {O'Mullane}, {Okamura},
  {Owen}, {Padmanabhan}, {Pauls}, {Peoples}, {Pier}, {Pope}, {Pourbaix},
  {Quinn}, {Richards}, {Richmond}, {Rockosi}, {Schlegel}, {Schneider},
  {Schroeder}, {Scranton}, {Seljak}, {Sheldon}, {Shimasaku}, {Smith}, {Smol{\v
  c}i{\'c}}, {Snedden}, {Stoughton}, {Strauss}, {SubbaRao}, {Szalay},
  {Szapudi}, {Szkody}, {Tegmark}, {Thakar}, {Tucker}, {Uomoto}, {Vanden Berk},
  {Vandenberg}, {Vogeley}, {Voges}, {Vogt}, {Walkowicz}, {Weinberg}, {West},
  {White}, {Xu}, {Yanny}, {Yocum}, {York}, {Zehavi}, {Zibetti}, \&
  {Zucker}}]{Adelman-McCarthy2007}
{Adelman-McCarthy}, J.~K. {et~al.} 2007, \apjs

\bibitem[{{Ag{\"u}eros} {et~al.}(2005){Ag{\"u}eros}, {Ivezi{\'c}}, {Covey},
  {Obri{\'c}}, {Hao}, {Walkowicz}, {West}, {Vanden Berk}, {Lupton}, {Knapp},
  {Gunn}, {Richards}, {Bochanski}, {Brooks}, {Claire}, {Haggard}, {Kaib},
  {Kimball}, {Gogarten}, {Seth}, \& {Solontoi}}]{Agueros2005}
{Ag{\"u}eros}, M.~A. {et~al.} 2005, \aj, 130, 1022

\bibitem[{{Anderson} {et~al.}(2007){Anderson}, {Margon}, {Voges}, {Plotkin},
  {Syphers}, {Haggard}, {Collinge}, {Meyer}, {Strauss}, {Ag{\"u}eros}, {Hall},
  {Homer}, {Ivezi{\'c}}, {Richards}, {Richmond}, {Schneider}, {Stinson},
  {Vanden Berk}, \& {York}}]{Anderson2007}
{Anderson}, S.~F. {et~al.} 2007, \aj, 133, 313

\bibitem[{{Bilir} {et~al.}(2005){Bilir}, {Karaali}, \& {Tun{\c
  c}el}}]{Bilir2005a}
{Bilir}, S., {Karaali}, S., \& {Tun{\c c}el}, S. 2005, Astronomische
  Nachrichten, 326, 321

\bibitem[{{Bochanski} {et~al.}(2007){Bochanski}, {West}, {Hawley}, \&
  {Covey}}]{Bochanski2007}
{Bochanski}, J.~J., {West}, A.~A., {Hawley}, S.~L., \& {Covey}, K.~R. 2007,
  \aj, 133, 531

\bibitem[{{Burgasser} {et~al.}(1999){Burgasser}, {Kirkpatrick}, {Brown},
  {Reid}, {Gizis}, {Dahn}, {Monet}, {Beichman}, {Liebert}, {Cutri}, \&
  {Skrutskie}}]{Burgasser1999}
{Burgasser}, A.~J. {et~al.} 1999, \apjl, 522, L65

\bibitem[{{Carney} {et~al.}(2003){Carney}, {Latham}, {Stefanik}, {Laird}, \&
  {Morse}}]{Carney2003}
{Carney}, B.~W., {Latham}, D.~W., {Stefanik}, R.~P., {Laird}, J.~B., \&
  {Morse}, J.~A. 2003, \aj, 125, 293

\bibitem[{{Carpenter}(2001)}]{Carpenter2001}
{Carpenter}, J.~M. 2001, \aj, 121, 2851

\bibitem[{{Cohen} {et~al.}(2003){Cohen}, {Wheaton}, \& {Megeath}}]{Cohen2003}
{Cohen}, M., {Wheaton}, W.~A., \& {Megeath}, S.~T. 2003, \aj, 126, 1090

\bibitem[{{Cutri} {et~al.}(2003){Cutri}, {Skrutskie}, {van Dyk}, {Beichman},
  {Carpenter}, {Chester}, {Cambresy}, {Evans}, {Fowler}, {Gizis}, {Howard},
  {Huchra}, {Jarrett}, {Kopan}, {Kirkpatrick}, {Light}, {Marsh}, {McCallon},
  {Schneider}, {Stiening}, {Sykes}, {Weinberg}, {Wheaton}, {Wheelock}, \&
  {Zacarias}}]{Cutri2003}
{Cutri}, R.~M. {et~al.} 2003, VizieR Online Data Catalog, 2246, 0

\bibitem[{{Davenport} {et~al.}(2006){Davenport}, {West}, {Matthiesen},
  {Schmieding}, \& {Kobelski}}]{Davenport2006}
{Davenport}, J.~R.~A., {West}, A.~A., {Matthiesen}, C.~K., {Schmieding}, M., \&
  {Kobelski}, A. 2006, \pasp, 118, 1679

\bibitem[{{Delfosse} {et~al.}(2004){Delfosse}, {Beuzit}, {Marchal}, {Bonfils},
  {C.~Perrier}, {S{\'e}gransan}, {Udry}, {Mayor}, \&
  {Forveille}}]{Delfosse2004}
{Delfosse}, X. {et~al.} 2004, in ASP Conf. Ser. 318: Spectroscopically and
  Spatially Resolving the Components of the Close Binary Stars, 166--174

\bibitem[{Dotter \& Chaboyer(2006)}]{Dotter2006}
Dotter, A.~L., \& Chaboyer, B. 2006, in preparation

\bibitem[{{Duquennoy} \& {Mayor}(1991)}]{Duquennoy1991a}
{Duquennoy}, A., \& {Mayor}, M. 1991, \aap, 248, 485

\bibitem[{{Eisenstein} {et~al.}(2006){Eisenstein}, {Liebert}, {Harris},
  {Kleinman}, {Nitta}, {Silvestri}, {Anderson}, {Barentine}, {Brewington},
  {Brinkmann}, {Harvanek}, {Krzesi{\'n}ski}, {Neilsen}, {Long}, {Schneider}, \&
  {Snedden}}]{Eisenstein2006}
{Eisenstein}, D.~J. {et~al.} 2006, \apjs, 167, 40

\bibitem[{{Finkbeiner} {et~al.}(2004{\natexlab{a}}){Finkbeiner}, {Padmanabhan},
  {Schlegel}, {Carr}, {Gunn}, {Rockosi}, {Sekiguchi}, {Lupton}, {Knapp},
  {Ivezi{\'c}}, {Blanton}, {Hogg}, {Adelman-McCarthy}, {Annis}, {Hayes},
  {Kinney}, {Long}, {Seljak}, {Strauss}, {Yanny}, {Ag{\"u}eros}, {Allam},
  {Anderson}, {Bahcall}, {Baldry}, {Bernardi}, {Boroski}, {Briggs},
  {Brinkmann}, {Brunner}, {Budav{\'a}ri}, {Castander}, {Covey}, {Csabai},
  {Doi}, {Dong}, {Eisenstein}, {Fan}, {Friedman}, {Fukugita}, {Gillespie},
  {Grebel}, {Gurbani}, {de Haas}, {Harris}, {Hendry}, {Hennessy}, {Jester},
  {Johnston}, {Jorgensen}, {Juri{\'c}}, {Kent}, {Kniazev}, {Krzesi{\'n}ski},
  {Leger}, {Lin}, {Loveday}, {Mannery}, {Mart{\'{\i}}nez-Delgado}, {McGehee},
  {Meiksin}, {Munn}, {Neilsen}, {Newman}, {Nitta}, {Pauls}, {Quinn}, {Rafikov},
  {Richards}, {Richmond}, {Schneider}, {Schroeder}, {Shimasaku}, {Siegmund},
  {Smith}, {Snedden}, {Stebbins}, {Szalay}, {Szokoly}, {Tegmark}, {Tucker},
  {Uomoto}, {Vanden Berk}, {Weinberg}, {West}, {Yasuda}, {Yocum}, {York}, \&
  {Zehavi}}]{Finkbeiner2004}
{Finkbeiner}, D.~P. {et~al.} 2004{\natexlab{a}}, \aj, 128, 2577

\bibitem[{{Finkbeiner} {et~al.}(2004{\natexlab{b}}){Finkbeiner}, {Schlegel},
  {Gunn}, {Hogg}, {Ivezic}, {Knapp}, \& {Lupton}}]{Finkbeiner2004a}
{Finkbeiner}, D.~P., {Schlegel}, D.~J., {Gunn}, J.~E., {Hogg}, D.~W., {Ivezic},
  Z., {Knapp}, G.~R., \& {Lupton}, R.~H. 2004{\natexlab{b}}, in Bulletin of the
  American Astronomical Society, Vol.~36, Bulletin of the American Astronomical
  Society, 770--+

\bibitem[{{Finlator} {et~al.}(2000){Finlator}, {Ivezi{\'c}}, {Fan}, {Strauss},
  {Knapp}, {Lupton}, {Gunn}, {Rockosi}, {Anderson}, {Csabai}, {Hennessy},
  {Hindsley}, {McKay}, {Nichol}, {Schneider}, {Smith}, {York}, \& {the SDSS
  Collaboration}}]{Finlator2000}
{Finlator}, K. {et~al.} 2000, \aj, 120, 2615

\bibitem[{{Fukugita} {et~al.}(1996){Fukugita}, {Ichikawa}, {Gunn}, {Doi},
  {Shimasaku}, \& {Schneider}}]{Fukugita1996}
{Fukugita}, M., {Ichikawa}, T., {Gunn}, J.~E., {Doi}, M., {Shimasaku}, K., \&
  {Schneider}, D.~P. 1996, \aj, 111, 1748

\bibitem[{{Girardi} {et~al.}(2004){Girardi}, {Grebel}, {Odenkirchen}, \&
  {Chiosi}}]{Girardi2004}
{Girardi}, L., {Grebel}, E.~K., {Odenkirchen}, M., \& {Chiosi}, C. 2004, \aap,
  422, 205

\bibitem[{{Golimowski} {et~al.}(2007){Golimowski}, {Henry}, \&
  {Reid}}]{Golimowski2007}
{Golimowski}, D.~A., {Henry}, T.~J., \& {Reid}, I.~N. 2007, SDSS parallaxes (in
  prep.)

\bibitem[{{Gunn} {et~al.}(1998){Gunn}, {Carr}, {Rockosi}, {Sekiguchi}, {Berry},
  {Elms}, {de Haas}, {Ivezi{\'c}}, {Knapp}, {Lupton}, {Pauls}, {Simcoe},
  {Hirsch}, {Sanford}, {Wang}, {York}, {Harris}, {Annis}, {Bartozek},
  {Boroski}, {Bakken}, {Haldeman}, {Kent}, {Holm}, {Holmgren}, {Petravick},
  {Prosapio}, {Rechenmacher}, {Doi}, {Fukugita}, {Shimasaku}, {Okada}, {Hull},
  {Siegmund}, {Mannery}, {Blouke}, {Heidtman}, {Schneider}, {Lucinio}, \&
  {Brinkman}}]{Gunn1998}
{Gunn}, J.~E. {et~al.} 1998, \aj, 116, 3040

\bibitem[{{Gunn} {et~al.}(2006){Gunn}, {Siegmund}, \& {Mannery}}]{Gunn2006}
{Gunn}, J.~E., {Siegmund}, W.~A., \& {Mannery}, E.~J. 2006, ArXiv Astrophysics
  e-prints

\bibitem[{{Hawley} {et~al.}(2002){Hawley}, {Covey}, {Knapp}, {Golimowski},
  {Fan}, {Anderson}, {Gunn}, {Harris}, {Ivezi{\'c}}, {Long}, {Lupton},
  {McGehee}, {Narayanan}, {Peng}, {Schlegel}, {Schneider}, {Spahn}, {Strauss},
  {Szkody}, {Tsvetanov}, {Walkowicz}, {Brinkmann}, {Harvanek}, {Hennessy},
  {Kleinman}, {Krzesinski}, {Long}, {Neilsen}, {Newman}, {Nitta}, {Snedden}, \&
  {York}}]{Hawley2002}
{Hawley}, S.~L. {et~al.} 2002, \aj, 123, 3409

\bibitem[{{Hogg} {et~al.}(2001){Hogg}, {Finkbeiner}, {Schlegel}, \&
  {Gunn}}]{Hogg2001}
{Hogg}, D.~W., {Finkbeiner}, D.~P., {Schlegel}, D.~J., \& {Gunn}, J.~E. 2001,
  \aj, 122, 2129

\bibitem[{{Ivezi{\'c}} {et~al.}(2002){Ivezi{\'c}}, {Becker}, {Blanton}, {Fan},
  {Finlator}, {Gunn}, {Hall}, {Kim}, {Knapp}, {Loveday}, {Lupton}, {Menou},
  {Narayanan}, {Richards}, {Rockosi}, {Schlegel}, {Schneider}, {Strateva},
  {Strauss}, {vanden Berk}, {Voges}, {Yanny}, \& {The SDSS
  Collaboration}}]{Ivezic2002}
{Ivezi{\'c}}, {\v Z}. {et~al.} 2002, in ASP Conf. Ser. 284: AGN Surveys, 137--+

\bibitem[{{Ivezi{\'c}} {et~al.}(2004){Ivezi{\'c}}, {Lupton}, {Schlegel},
  {Boroski}, {Adelman-McCarthy}, {Yanny}, {Kent}, {Stoughton}, {Finkbeiner},
  {Padmanabhan}, {Rockosi}, {Gunn}, {Knapp}, {Strauss}, {Richards},
  {Eisenstein}, {Nicinski}, {Kleinman}, {Krzesinski}, {Newman}, {Snedden},
  {Thakar}, {Szalay}, {Munn}, {Smith}, {Tucker}, \& {Lee}}]{Ivezic2004}
{Ivezi{\'c}}, {\v Z}. {et~al.} 2004, Astronomische Nachrichten, 325, 583

\bibitem[{{Ivezi{\'c}} {et~al.}(2005){Ivezi{\'c}}, {Vivas}, {Lupton}, \&
  {Zinn}}]{Ivezic2005}
{Ivezi{\'c}}, {\v Z}., {Vivas}, A.~K., {Lupton}, R.~H., \& {Zinn}, R. 2005,
  \aj, 129, 1096

\bibitem[{{Kaiser} {et~al.}(2002){Kaiser}, {Aussel}, {Burke}, {Boesgaard},
  {Chambers}, {Chun}, {Heasley}, {Hodapp}, {Hunt}, {Jedicke}, {Jewitt},
  {Kudritzki}, {Luppino}, {Maberry}, {Magnier}, {Monet}, {Onaka}, {Pickles},
  {Rhoads}, {Simon}, {Szalay}, {Szapudi}, {Tholen}, {Tonry}, {Waterson}, \&
  {Wick}}]{Kaiser2002}
{Kaiser}, N. {et~al.} 2002, in Survey and Other Telescope Technologies and
  Discoveries. Edited by Tyson, J. Anthony; Wolff, Sidney. Proceedings of the
  SPIE, Volume 4836, pp. 154-164 (2002)., ed. J.~A. {Tyson} \& S.~{Wolff},
  154--164

\bibitem[{{Keller} {et~al.}(2007){Keller}, {Schmidt}, {Bessell}, {Conroy},
  {Francis}, {Granlund}, {Kowald}, {Oates}, {Martin-Jones}, {Preston},
  {Tisserand}, {Vaccarella}, \& {Waterson}}]{Keller2007}
{Keller}, S.~C. {et~al.} 2007, ArXiv Astrophysics e-prints

\bibitem[{{Kirkpatrick} {et~al.}(1999){Kirkpatrick}, {Reid}, {Liebert},
  {Cutri}, {Nelson}, {Beichman}, {Dahn}, {Monet}, {Gizis}, \&
  {Skrutskie}}]{Kirkpatrick1999}
{Kirkpatrick}, J.~D. {et~al.} 1999, \apj, 519, 802

\bibitem[{{Kurucz}(1979)}]{Kurucz1979}
{Kurucz}, R.~L. 1979, \apjs, 40, 1

\bibitem[{{Le Borgne} {et~al.}(2003)}]{Le-Borgne2003}
{Le Borgne}, J.-F., {et~al.} 2003, \aap, 402, 433

\bibitem[{{Lupton} {et~al.}(2001){Lupton}, {Gunn}, {Ivezi{\'c}}, {Knapp},
  {Kent}, \& {Yasuda}}]{Lupton2001}
{Lupton}, R., {Gunn}, J.~E., {Ivezi{\'c}}, Z., {Knapp}, G.~R., {Kent}, S., \&
  {Yasuda}, N. 2001, in ASP Conf. Ser. 238: Astronomical Data Analysis Software
  and Systems X, 269--+

\bibitem[{{Margon} {et~al.}(2002){Margon}, {Anderson}, {Harris}, {Strauss},
  {Knapp}, {Fan}, {Schneider}, {Vanden Berk}, {Schlegel}, {Deutsch},
  {Ivezi{\'c}}, {Hall}, {Williams}, {Davidsen}, {Brinkmann}, {Csabai}, {Hayes},
  {Hennessy}, {Kinney}, {Kleinman}, {Lamb}, {Long}, {Neilsen}, {Nichol},
  {Nitta}, {Snedden}, \& {York}}]{Margon2002}
{Margon}, B. {et~al.} 2002, \aj, 124, 1651

\bibitem[{{Newman} {et~al.}(2004){Newman}, {Long}, {Snedden}, {Kleinman},
  {Nitta}, {Harvanek}, {Krzesinski}, {Brewington}, {Barentine}, {Neilsen}, \&
  {Schlegel}}]{Newman2004}
{Newman}, P.~R. {et~al.} 2004, in Ground-based Instrumentation for Astronomy.
  Edited by Alan F. M. Moorwood and Iye Masanori. Proceedings of the SPIE,
  Volume 5492, pp. 533-544 (2004)., ed. A.~F.~M. {Moorwood} \& M.~{Iye},
  533--544

\bibitem[{{Obri{\'c}} {et~al.}(2006){Obri{\'c}}, {Ivezi{\'c}}, {Best},
  {Lupton}, {Tremonti}, {Brinchmann}, {Ag{\"u}eros}, {Knapp}, {Gunn},
  {Rockosi}, {Schlegel}, {Finkbeiner}, {Ga{\'c}e{\v s}a}, {Smol{\v c}i{\'c}},
  {Anderson}, {Voges}, {Juri{\'c}}, {Siverd}, {Steinhardt}, {Jagoda},
  {Blanton}, \& {Schneider}}]{Obric2006}
{Obri{\'c}}, M. {et~al.} 2006, \mnras, 370, 1677

\bibitem[{{Ochsenbein} {et~al.}(2000){Ochsenbein}, {Bauer}, \&
  {Marcout}}]{Ochsenbein2000}
{Ochsenbein}, F., {Bauer}, P., \& {Marcout}, J. 2000, \aaps, 143, 23

\bibitem[{{Padmanabhan} {et~al.}(2007){Padmanabhan}, {Schlegel}, {Finkbeiner},
  {Barentine}, {Blanton}, {Brewington}, {Gunn}, {Harvanek}, {Hogg}, {Ivezic},
  {Johnston}, {Kent}, {Kleinman}, {Knapp}, {Krzesinski}, {Long}, {Neilsen},
  {Nitta}, {Loomis}, {Lupton}, {Roweis}, {Snedden}, {Strauss}, \&
  {Tucker}}]{Padmanabhan2007}
{Padmanabhan}, N. {et~al.} 2007, ArXiv Astrophysics e-prints

\bibitem[{{Pickles}(1998)}]{Pickles1998}
{Pickles}, A.~J. 1998, \pasp, 110, 863

\bibitem[{{Pier} {et~al.}(2003){Pier}, {Munn}, {Hindsley}, {Hennessy}, {Kent},
  {Lupton}, \& {Ivezi{\'c}}}]{Pier2003}
{Pier}, J.~R., {Munn}, J.~A., {Hindsley}, R.~B., {Hennessy}, G.~S., {Kent},
  S.~M., {Lupton}, R.~H., \& {Ivezi{\'c}}, {\v Z}. 2003, \aj, 125, 1559

\bibitem[{{Pourbaix} {et~al.}(2005){Pourbaix}, {Knapp}, {Szkody}, {Ivezi{\'c}},
  {Kleinman}, {Long}, {Snedden}, {Nitta}, {Harvanek}, {Krzesinski},
  {Brewington}, {Barentine}, {Neilsen}, \& {Brinkmann}}]{Pourbaix2005}
{Pourbaix}, D. {et~al.} 2005, \aap, 444, 643

\bibitem[{{Raymond} {et~al.}(2003){Raymond}, {Szkody}, {Hawley}, {Anderson},
  {Brinkmann}, {Covey}, {McGehee}, {Schneider}, {West}, \&
  {York}}]{Raymond2003}
{Raymond}, S.~N. {et~al.} 2003, \aj, 125, 2621

\bibitem[{{Richards} {et~al.}(2002){Richards}, {Fan}, {Newberg}, {Strauss},
  {Vanden Berk}, {Schneider}, {Yanny}, {Boucher}, {Burles}, {Frieman}, {Gunn},
  {Hall}, {Ivezi{\'c}}, {Kent}, {Loveday}, {Lupton}, {Rockosi}, {Schlegel},
  {Stoughton}, {SubbaRao}, \& {York}}]{Richards2002}
{Richards}, G.~T. {et~al.} 2002, \aj, 123, 2945

\bibitem[{{Sanchez-Blazquez} {et~al.}(2006){Sanchez-Blazquez}, {Peletier},
  {Jimenez-Vicente}, {Cardiel}, {Cenarro}, {Falcon-Barroso}, {Gorgas}, {Selam},
  \& {Vazdekis}}]{Sanchez-Blazquez2006}
{Sanchez-Blazquez}, P. {et~al.} 2006, ArXiv Astrophysics e-prints

\bibitem[{{Schlegel} {et~al.}(1998){Schlegel}, {Finkbeiner}, \&
  {Davis}}]{Schlegel1998}
{Schlegel}, D.~J., {Finkbeiner}, D.~P., \& {Davis}, M. 1998, \apj, 500, 525

\bibitem[{{Schneider} {et~al.}(2007){Schneider}, {Hall}, {Richards}, {Strauss},
  {Vanden Berk}, {Anderson}, {Brandt}, {Fan}, {Jester}, {Gray}, {Gunn},
  {SubbaRao}, {Thakar}, {Stoughton}, {Szalay}, {Yanny}, {York}, {Bahcall},
  {Barentine}, {Blanton}, {Brewington}, {Brinkmann}, {Brunner}, {Castander},
  {Csabai}, {Frieman}, {Fukugita}, {Harvanek}, {Hogg}, {Ivezic}, {Kent},
  {Kleinman}, {Knapp}, {Kron}, {Krzesinski}, {Long}, {Lupton}, {Nitta}, {Pier},
  {Saxe}, {Shen}, {Snedden}, {Weinberg}, \& {Wu}}]{Schneider2007}
{Schneider}, D.~P. {et~al.} 2007, ArXiv e-prints, 704

\bibitem[{Scranton {et~al.}(2005)}]{Scranton2005}
Scranton, R., {et~al.} 2005

\bibitem[{{Silvestri} {et~al.}(2006){Silvestri}, {Hawley}, {West}, {Szkody},
  {Bochanski}, {Eisenstein}, {McGehee}, {Schmidt}, {Smith}, {Wolfe}, {Harris},
  {Kleinman}, {Liebert}, {Nitta}, {Barentine}, {Brewington}, {Brinkmann},
  {Harvanek}, {Krzesi{\'n}ski}, {Long}, {Neilsen}, {Schneider}, \&
  {Snedden}}]{Silvestri2006}
{Silvestri}, N.~M. {et~al.} 2006, \aj, 131, 1674

\bibitem[{{Skrutskie} {et~al.}(1997){Skrutskie}, {Schneider}, {Stiening},
  {Strom}, {Weinberg}, {Beichman}, {Chester}, {Cutri}, {Lonsdale}, {Elias},
  {Elston}, {Capps}, {Carpenter}, {Huchra}, {Liebert}, {Monet}, {Price}, \&
  {Seitzer}}]{Skrutskie1997}
{Skrutskie}, M.~F. {et~al.} 1997, in ASSL Vol. 210: The Impact of Large Scale
  Near-IR Sky Surveys, 25--+

\bibitem[{{Smith} {et~al.}(2002){Smith}, {Tucker}, {Kent}, {Richmond},
  {Fukugita}, {Ichikawa}, {Ichikawa}, {Jorgensen}, {Uomoto}, {Gunn}, {Hamabe},
  {Watanabe}, {Tolea}, {Henden}, {Annis}, {Pier}, {McKay}, {Brinkmann}, {Chen},
  {Holtzman}, {Shimasaku}, \& {York}}]{Smith2002}
{Smith}, J.~A. {et~al.} 2002, \aj, 123, 2121

\bibitem[{{Smol{\v c}i{\'c}} {et~al.}(2004){Smol{\v c}i{\'c}}, {Ivezi{\'c}},
  {Knapp}, {Lupton}, {Pavlovski}, {Iliji{\'c}}, {Schlegel}, {Smith}, {McGehee},
  {Silvestri}, {Hawley}, {Rockosi}, {Gunn}, {Strauss}, {Fan}, {Eisenstein}, \&
  {Harris}}]{Smolcic2004}
{Smol{\v c}i{\'c}}, V. {et~al.} 2004, \apjl, 615, L141

\bibitem[{{Stoughton} {et~al.}(2002){Stoughton}, {Lupton}, {Bernardi},
  {Blanton}, {Burles}, {Castander}, {Connolly}, {Eisenstein}, {Frieman},
  {Hennessy}, {Hindsley}, {Ivezi{\'c}}, {Kent}, {Kunszt}, {Lee}, {Meiksin},
  {Munn}, {Newberg}, {Nichol}, {Nicinski}, {Pier}, {Richards}, {Richmond},
  {Schlegel}, {Smith}, {Strauss}, {SubbaRao}, {Szalay}, {Thakar}, {Tucker},
  {Vanden Berk}, {Yanny}, {Adelman}, {Anderson}, {Anderson}, {Annis},
  {Bahcall}, {Bakken}, {Bartelmann}, {Bastian}, {Bauer}, {Berman},
  {B{\"o}hringer}, {Boroski}, {Bracker}, {Briegel}, {Briggs}, {Brinkmann},
  {Brunner}, {Carey}, {Carr}, {Chen}, {Christian}, {Colestock}, {Crocker},
  {Csabai}, {Czarapata}, {Dalcanton}, {Davidsen}, {Davis}, {Dehnen},
  {Dodelson}, {Doi}, {Dombeck}, {Donahue}, {Ellman}, {Elms}, {Evans}, {Eyer},
  {Fan}, {Federwitz}, {Friedman}, {Fukugita}, {Gal}, {Gillespie}, {Glazebrook},
  {Gray}, {Grebel}, {Greenawalt}, {Greene}, {Gunn}, {de Haas}, {Haiman},
  {Haldeman}, {Hall}, {Hamabe}, {Hansen}, {Harris}, {Harris}, {Harvanek},
  {Hawley}, {Hayes}, {Heckman}, {Helmi}, {Henden}, {Hogan}, {Hogg}, {Holmgren},
  {Holtzman}, {Huang}, {Hull}, {Ichikawa}, {Ichikawa}, {Johnston}, {Kauffmann},
  {Kim}, {Kimball}, {Kinney}, {Klaene}, {Kleinman}, {Klypin}, {Knapp},
  {Korienek}, {Krolik}, {Kron}, {Krzesi{\'n}ski}, {Lamb}, {Leger},
  {Limmongkol}, {Lindenmeyer}, {Long}, {Loomis}, {Loveday}, {MacKinnon},
  {Mannery}, {Mantsch}, {Margon}, {McGehee}, {McKay}, {McLean}, {Menou},
  {Merelli}, {Mo}, {Monet}, {Nakamura}, {Narayanan}, {Nash}, {Neilsen},
  {Newman}, {Nitta}, {Odenkirchen}, {Okada}, {Okamura}, {Ostriker}, {Owen},
  {Pauls}, {Peoples}, {Peterson}, {Petravick}, {Pope}, {Pordes}, {Postman},
  {Prosapio}, {Quinn}, {Rechenmacher}, {Rivetta}, {Rix}, {Rockosi}, {Rosner},
  {Ruthmansdorfer}, {Sandford}, {Schneider}, {Scranton}, {Sekiguchi}, {Sergey},
  {Sheth}, {Shimasaku}, {Smee}, {Snedden}, {Stebbins}, {Stubbs}, {Szapudi},
  {Szkody}, {Szokoly}, {Tabachnik}, {Tsvetanov}, {Uomoto}, {Vogeley}, {Voges},
  {Waddell}, {Walterbos}, {Wang}, {Watanabe}, {Weinberg}, {White}, {White},
  {Wilhite}, {Wolfe}, {Yasuda}, {York}, {Zehavi}, \& {Zheng}}]{Stoughton2002}
{Stoughton}, C. {et~al.} 2002, \aj, 123, 485

\bibitem[{{Strauss} {et~al.}(1999){Strauss}, {Fan}, {Gunn}, {Leggett},
  {Geballe}, {Pier}, {Lupton}, {Knapp}, {Annis}, {Brinkmann}, {Crocker},
  {Csabai}, {Fukugita}, {Golimowski}, {Harris}, {Hennessy}, {Hindsley},
  {Ivezi{\'c} }, {Kent}, {Lamb}, {Munn}, {Newberg}, {Rechenmacher},
  {Schneider}, {Smith}, {Stoughton}, {Tucker}, {Waddell}, \&
  {York}}]{Strauss1999}
{Strauss}, M.~A. {et~al.} 1999, \apjl, 522, L61

\bibitem[{{Tucker} {et~al.}(2006){Tucker}, {Kent}, {Richmond}, {Annis},
  {Smith}, {Allam}, {Rodgers}, {Stute}, {Adelman-McCarthy}, {Brinkmann}, {Doi},
  {Finkbeiner}, {Fukugita}, {Goldston}, {Greenway}, {Gunn}, {Hendry}, {Hogg},
  {Ichikawa}, {Ivezi{\'c}}, {Knapp}, {Lampeitl}, {Lee}, {Lin}, {McKay},
  {Merrelli}, {Munn}, {Neilsen}, {Newberg}, {Richards}, {Schlegel},
  {Stoughton}, {Uomoto}, \& {Yanny}}]{Tucker2006}
{Tucker}, D.~L. {et~al.} 2006, Astronomische Nachrichten, 327, 821

\bibitem[{{Valdes} {et~al.}(2004){Valdes}, {Gupta}, {Rose}, {Singh}, \&
  {Bell}}]{Valdes2004}
{Valdes}, F., {Gupta}, R., {Rose}, J.~A., {Singh}, H.~P., \& {Bell}, D.~J.
  2004, \apjs, 152, 251

\bibitem[{{Warren} {et~al.}(2007)}]{Warren2007}
{Warren}, S.~J., {et~al.} 2007

\bibitem[{{West}(2007)}]{West2007}
{West}, A.~A. 2007, submitted

\bibitem[{{West} {et~al.}(2004){West}, {Hawley}, {Walkowicz}, {Covey},
  {Silvestri}, {Raymond}, {Harris}, {Munn}, {McGehee}, {Ivezi{\'c}}, \&
  {Brinkmann}}]{West2004}
{West}, A.~A. {et~al.} 2004, \aj, 128, 426

\bibitem[{West {et~al.}(2005)West, Walkowicz, \& Hawley}]{West2005}
West, A.~A., Walkowicz, L.~M., \& Hawley, S.~L. 2005

\bibitem[{{Williams} {et~al.}(2002){Williams}, {Golimowski}, {Uomoto}, {Reid},
  {Henry}, {Dieterich}, {Jue}, {Long}, {Neilsen}, {Spahn}, \&
  {Walkowicz}}]{Williams2002}
{Williams}, C.~C. {et~al.} 2002, Bulletin of the American Astronomical Society,
  34, 1292

\bibitem[{{York} {et~al.}(2000){York}, {Adelman}, {Anderson}, {Anderson},
  {Annis}, {Bahcall}, {Bakken}, {Barkhouser}, {Bastian}, {Berman}, {Boroski},
  {Bracker}, {Briegel}, {Briggs}, {Brinkmann}, {Brunner}, {Burles}, {Carey},
  {Carr}, {Castander}, {Chen}, {Colestock}, {Connolly}, {Crocker}, {Csabai},
  {Czarapata}, {Davis}, {Doi}, {Dombeck}, {Eisenstein}, {Ellman}, {Elms},
  {Evans}, {Fan}, {Federwitz}, {Fiscelli}, {Friedman}, {Frieman}, {Fukugita},
  {Gillespie}, {Gunn}, {Gurbani}, {de Haas}, {Haldeman}, {Harris}, {Hayes},
  {Heckman}, {Hennessy}, {Hindsley}, {Holm}, {Holmgren}, {Huang}, {Hull},
  {Husby}, {Ichikawa}, {Ichikawa}, {Ivezi{\'c}}, {Kent}, {Kim}, {Kinney},
  {Klaene}, {Kleinman}, {Kleinman}, {Knapp}, {Korienek}, {Kron}, {Kunszt},
  {Lamb}, {Lee}, {Leger}, {Limmongkol}, {Lindenmeyer}, {Long}, {Loomis},
  {Loveday}, {Lucinio}, {Lupton}, {MacKinnon}, {Mannery}, {Mantsch}, {Margon},
  {McGehee}, {McKay}, {Meiksin}, {Merelli}, {Monet}, {Munn}, {Narayanan},
  {Nash}, {Neilsen}, {Neswold}, {Newberg}, {Nichol}, {Nicinski}, {Nonino},
  {Okada}, {Okamura}, {Ostriker}, {Owen}, {Pauls}, {Peoples}, {Peterson},
  {Petravick}, {Pier}, {Pope}, {Pordes}, {Prosapio}, {Rechenmacher}, {Quinn},
  {Richards}, {Richmond}, {Rivetta}, {Rockosi}, {Ruthmansdorfer}, {Sandford},
  {Schlegel}, {Schneider}, {Sekiguchi}, {Sergey}, {Shimasaku}, {Siegmund},
  {Smee}, {Smith}, {Snedden}, {Stone}, {Stoughton}, {Strauss}, {Stubbs},
  {SubbaRao}, {Szalay}, {Szapudi}, {Szokoly}, {Thakar}, {Tremonti}, {Tucker},
  {Uomoto}, {Vanden Berk}, {Vogeley}, {Waddell}, {Wang}, {Watanabe},
  {Weinberg}, {Yanny}, \& {Yasuda}}]{York2000}
{York}, D.~G. {et~al.} 2000, \aj, 120, 1579

\end{thebibliography}
